\documentclass[useAMS,usenatbib]{mn2e}

\usepackage{graphicx}
\usepackage{times}
\usepackage{natbib}
\newif\ifAMStwofonts
\AMStwofontstrue

\newcommand{\be}{\begin{equation}}
\newcommand{\ee}{\end{equation}}
\newcommand{\ba}{\begin{eqnarray}}
\newcommand{\ea}{\end{eqnarray}}
\newcommand{\brr}{\begin{array}}
\newcommand{\err}{\end{array}}
\newcommand{\bc}{\begin{center}}
\newcommand{\ec}{\end{center}}
\newcommand{\hm}{\,h^{-1}{\rm Mpc}}

\newcommand{\msun}{\,h^{-1}M_\odot}

\newcommand{\rvir}{\mbox{$R_{\rmn{vir}}$}}

\newcommand{\vel}{\,{\rm km\,s^{-1}}}

\newcommand{\hder}[2]{\textrm{d}\,{#1}/\textrm{d}{#2}}
\newcommand{\der}[2]{\frac{\textrm{d}\,#1}{\textrm{d}#2}}

\newcommand{\mincir}{\raise
  -2.truept\hbox{\rlap{\hbox{$\sim$}}\raise5.truept \hbox{$<$}\ }}
\newcommand{\magcir}{\raise
  -2.truept\hbox{\rlap{\hbox{$\sim$}}\raise5.truept \hbox{$>$}\ }}
\newcommand{\siml}{\raise
  -2.truept\hbox{\rlap{\hbox{$\sim$}}\raise5.truept \hbox{$<$}\ }}
\newcommand{\simg}{\raise
  -2.truept\hbox{\rlap{\hbox{$\sim$}}\raise5.truept \hbox{$>$}\ }}

\newcommand{\gd}{{\small GADGET-2~}}

\newcommand{\refsec}[1]{Sec. (\ref{s:#1})}

\title[Chemical enrichment of galaxy clusters] {Chemical
  enrichment of galaxy clusters from hydrodynamical
  simulations} \author[Tornatore et al.] {L. Tornatore$^{1,3}$,
  S. Borgani$^{2,3,4}$, K. Dolag$^{5}$ \& F. Matteucci$^{2,4}$
  \\~\\
  $^1$ SISSA -- International School for Advanced Studies, via Beirut
  4, I-34100 Trieste, Italy (torna@sissa.it)\\
  $^2$ Dipartimento di Astronomia dell'Universit\`a di Trieste, via
  Tiepolo 11, I-34131 Trieste, Italy (borgani@oats.inaf.it)\\
  $^3$ INFN -- Istituto Nazionale di Fisica Nucleare, Trieste, Italy\\
  $^4$ INAF -- Istituto Nazionale di Astrofisica, Trieste, Italy\\
  $^5$ Max-Planck-Institut f\"ur Astrophysik, Karl-Schwarzschild Strasse
  1, Garching bei M\"unchen, Germany (kdolag,charlot@mpa-garching.mpg.de)\\
}

\begin{document}
\label{firstpage}
\maketitle

\begin{abstract}
  We present cosmological hydrodynamical simulations of galaxy
  clusters aimed at studying the process of metal enrichment of the
  intra--cluster medium (ICM). These simulations have been performed
  by implementing a detailed model of chemical evolution in the
  Tree-SPH \gd code. This model allows us to follow the metal release
  from SNII, SNIa and AGB stars, by properly accounting for the
  lifetimes of stars of different mass, as well as to change the
  stellar initial mass function (IMF), the lifetime function and the
  stellar yields. As such, our implementation of chemical evolution
  represents a powerful instrument to follow the cosmic history of
  metal production. The simulations presented here have been performed
  with the twofold aim of checking numerical effects, as well as the
  impact of changing the model of chemical evolution and the efficiency
  of stellar feedback. In general, we find that the distribution of
  metals produced by SNII are more clumpy than for product of
  low--mass stars, as a consequence of the different time--scales over
  which they are released. Using a standard Salpeter IMF produces a
  radial profile of Iron abundance which is in fairly good agreement
  with observations available out to $\simeq 0.6R_{500}$. This result
  holds almost independent of the numerical scheme adopted to
  distribute metals around star--forming regions. The mean age of
  enrichment of the ICM corresponds to redshift $z\sim 0.5$, which
  progressively increases outside the virial region. Increasing
  resolution, we improve the description of a diffuse high--redshift
  enrichment of the inter--galactic medium (IGM). This turns into a
  progressively more efficient enrichment of the cluster outskirts,
  while having a smaller impact at $R\mincir 0.5R_{500}$. As for the
  effect of the model of chemical evolution, we find that changing the
  IMF has the strongest impact. Using an IMF, which is top--heavier
  than the Salpeter one, provides a larger Iron abundance, possibly in
  excess of the observed level, also significantly increasing the
  [O/Fe] relative abundance. Our simulations always show an excess of
  low--redshift star formation and, therefore, of the abundance of
  Oxygen in central cluster regions, at variance with
  observations. This problem is not significantly ameliorated by
  increasing the efficiency of the stellar feedback.
\end{abstract}


% introduction
%\input{sect.1}
\section{Introduction}

Observations of the power spectrum of the CMB anisotropies, combined
with those of the cosmic distance scales, and the statistical
properties of cosmic structures are now providing a convincing
validation of the standard cosmological scenario and placing quite
stringent constraints on the value of the cosmological parameters
\citep[see][ for a recent review, and references
therein]{2006Natur.440.1137S}. As for the study of the formation
process of cosmic structures, this allows us to disentangle the
effects of cosmic evolution from those induced by the astrophysical
processes which determine the observational properties of galaxies and
of the diffuse baryons. Star formation, feedback of energy and metals
from explosions of supernovae (SN) and from Active Galactic Nuclei
(AGN) play a fundamental role in determining the thermo-dynamical and
chemo-dynamical status of cosmic baryons.

In this respect, clusters of galaxies play a very important role.
Thanks to the high density and temperature reached by the gas trapped
in their potential wells, they are the ideal signposts where to trace
the past history of the inter--galactic medium (IGM). Observations in
the X--ray band with the Chandra and XMM--Newton satellites are
providing invaluable information on the thermodynamical properties of
the intra--cluster medium (ICM; see \citealt{2002ARA&A..40..539R} and
\citealt{2005RvMP...77..207V}, for reviews). These observations
highlight that non--gravitational sources of energy, such as energy
feedback from SN and AGN have played an important role in determining
the ICM physical properties.

At the same time, spatially resolved X--ray spectroscopy permits to
measure the equivalent width of emission lines associated to
transitions of heavily ionized elements and, therefore, to trace the
pattern of chemical enrichment \citep[e.g.,][for a
review]{2004cgpc.symp..123M}. In turn, this information, as obtainable
from X--ray observations, are inextricably linked to the history of
formation and evolution of the galaxy population \citep[e.g.][, and
references therein]{1997ApJ...488...35R,2002NewA....7..227P}, as
inferred from observational in the optical band. For instance,
\cite{2004A&A...419....7D} have shown that cool core clusters are
characterized by a significant central enhancement of the Iron
abundance in the ICM, which closely correlates with the magnitude of
the Brightest Cluster Galaxies (BCGs) and the temperature of the
cluster. This demonstrates that a fully coherent description of the
evolution of cosmic baryons in the condensed stellar phase and in the
diffuse hot phase requires properly accounting for the mechanisms of
production and release of both energy and metals.

In this framework, semi--analytical models of galaxy formation
provide a flexible tool to explore the space of parameters which
describe a number of dynamical and astrophysical processes. In their
most recent formulation, such models are coupled to dark matter (DM)
cosmological simulations, to trace the merging history of the halos
where galaxy formation takes place, and include a treatment of metal
production from type-Ia and type-II supernovae (SNIa and SNII,
hereafter;
\citealt{2004MNRAS.349.1101D,2005MNRAS.358.1247N,2006astro.ph.10805M}),
so as to properly address the study of the chemical enrichment of the
ICM.
%, and of energy feedback from AGN
%\citep[e.g.,][]{2006MNRAS.365...11C,2006MNRAS.370..645B}. In their
%standard application, semi--analytical models (SAM) are coupled to
%cosmological simulations, which include only dark matter (DM), to
%trace the merging history of the DM halos where galaxy formation takes
%place. While this approach allows one to make predictions on the
%overall level of chemical enrichment of the ICM, the lack of an
%explicit description of gas dynamics does not allow to account for
%the effect of a number of processes (e.g., local gas motions,
%ram--pressure stripping), which play a key role in determining the
%distribution of metals.
%
%In order to partly remedy to these limitations,
\cite{2006MNRAS.368.1540C} recently applied an alternative 
%hybrid
approach, in which non--radiative SPH cluster simulations are used to
trace at the same time the formation history of DM halos and the
dynamics of the gas. In this approach, metals are produced by SAM
galaxies and then suitably assigned to gas particles, thereby
providing a chemo--dynamical description of the
ICM. \cite{2006A&A...452..795D} used hydrodynamical simulations, which
include simple prescriptions for gas cooling, star formation and
feedback, to address the specific role played by ram--pressure
stripping in determining the distribution of metals.

While these approaches offer obvious advantages with respect to
standard semi--analytical models, they still do not provide a fully
self--consistent picture, where chemical enrichment is the outcome of
the process of star formation, associated to the cooling of the gas
infalling in high density regions, as described in the numerical
hydrodynamical treatment.  In this sense, a fully self--consistent
approach requires that the simulations must include the processes of
gas cooling, star formation and evolution, along with the
corresponding feedback in energy and metals.

A number of authors have presented hydrodynamical simulations for the
formation of cosmic structures, which include treatments of the
chemical evolution at different levels of complexity, using both
Eulerian and SPH codes. Starting from the pioneering work by
\cite{1992A&A...265..465T}, a number of chemo--dynamical models based
on Eulerian codes have been presented \citep[e.g.,
][]{1998ApJ...496..155S,2001MNRAS.322..800R} with the aim of studying
the metallicity evolution of galaxies. Although not in a cosmological
framework, these analyses generally include detailed models of
chemical evolution, thereby accounting for the metal production from
SNIa, SNII and intermediate-- and low--mass
stars. \cite{1996A&A...315..105R} presented SPH simulations of the
Galaxy, forming in a isolated halo, by following Iron and Oxygen
production from SNII and SNIa stars, also accounting for the effect of
stellar lifetimes. \cite{2001MNRAS.325...34M} presented a detailed
analysis of chemo--dynamical SPH simulations, aimed at studying both
numerical stability of the results and the enrichment properties of
galactic objects in a cosmological context.
\cite{2002MNRAS.330..821L} discussed a statistical approach to follow
metal production in SPH simulations, which have a large number of star
particles, showing applications to simulations of a disc--like galaxy
and of a galaxy cluster. \cite{2003MNRAS.340..908K} carried out
cosmological chemo--dynamical simulations of elliptical galaxies, with
an SPH code, by including the contribution from SNIa and SNII, also
accounting for stellar lifetimes. \cite{2003MNRAS.339.1117V} applied
the method presented by \cite{2002MNRAS.330..821L} to an exteded set
of simulated galaxy clusters. This analysis showed that profiles of
the Iron abundance are steeper than the observed
ones. \cite{2004MNRAS.349L..19T} presented results from a first
implementation of a chemical evolution model in the {\tt GADGET-2}
code \citep{2005MNRAS.364.1105S}, also including the contribution from
intermediate and low mass stars. Using an earlier version of the code
presented in this paper, they studied the effect of changing the
prescription for the stellar initial mass function (IMF) and of the
feedback efficiency on the ICM enrichment in Iron, Oxygen and
Silicon. A similar analysis has been presented by
\cite{2006MNRAS.371..548R}, who also considered the effect of varying
the IMF and the feedback efficiency on the enrichment pattern of the
ICM.  \cite{2005MNRAS.364..552S} presented another implementation of a
model of chemical enrichment in the {\tt GADGET-2} code, coupled to a
self--consistent model for star formation and feedback \citep[see
also][]{2006MNRAS.371.1125S}. In their model, which was applied to
study the enrichment of galaxies, they included the contribution from
SNIa and SNII, assuming that all long--lived stars die after a fixed
delay time.

In this paper we present in detail a novel implementation of chemical
evolution in the Tree+SPH \gd code \citep{SP01.1,2005MNRAS.364.1105S},
which largely improves that originally introduced by
\cite{2003MNRAS.339..289S} (SH03 hereafter). The model by SH03 assumes
that feedback in energy and metals is provided only by SNII, by
assuming a Salpeter initial mass function \citep[IMF;
][]{1955ApJ...121..161S}, under the instantaneous recycling
approximation (IRA; i.e. stars exploding at the same time of their
formation). Furthermore, no detailed stellar yields are taken into
account, so that the original code described a global metallicity,
without following the production of individual elements. Finally,
radiative losses of the gas are described by a cooling function
computed at zero metallicity.

As a first step to improve with respect to this description, we
properly include life-times for stars of different masses, so as to
fully account for the time--delay between star formation and release
of energy and metals. Furthermore, we account for the contribution of
SNII, SNIa and low and intermediate mass stars to the production of
metals, while only SNII and SNIa contribute to energy feedback. The
contributions from different stars are consistently computed for any
choice of the IMF. Also, radiative losses are computed by accounting
for the dependence of the cooling function on the gas local
metallicity. Accurate stellar yields are included so as to follow in
detail the production of different heavy elements. The code
implementation of chemical evolution is build in the \gd structure in
an efficient way, so that its overhead in terms of computational cost
is always very limited.

%In a previous paper \citep{2004MNRAS.349L..19T} we presented
%preliminary results from an earlier version of this code. 
In the following of this paper, we will discuss in detail the effect
that parameters related both to numerics and to the model of chemical
evolution have on the pattern and history of the ICM chemical
enrichment. While observational results on the ICM enrichment will be
used to guide the eye, we will not perform here a detailed comparison
with observations, based on a statistical ensemble of simulated
clusters. In a forthcoming paper, we will compare different
observational constraints on the ICM metal content with results from
an extended set of simulated clusters.

The plan of the paper is as follows. In Section 2 we describe the
implementation of the models of chemical evolution. After providing a
description of the star formation algorithm and of its numerical
implementation, we will discuss in detail the ingredients of the model
of chemical evolution, finally reviewing the model of feedback through
galactic winds, as implemented by SH03. In Section 3 we will present
the results of our simulations. This presentation will be divided in
three parts. The first one will focus on the role played by numerical
effects related to the prescription adopted to distribute metals
around star forming regions (Sect. 3.1). The second part will
concentrate on the numerical effects related to mass and force
resolution (Sect. 3.2), while the third part (Sect. 3.3) describes in
detail how changing IMF, yields, feedback strength and stellar
life-times affects the resulting chemical enrichment of the ICM. The
readers not interested in the discussion of the numerical issues can
skip Sect. 3.1 and 3.2. We will critically discuss our results and
draw our main conclusions in Section 4.

%%%%%%%%%%%%%%%%%%%%%%%%%%%%%%%%%%%%%%%%%%%%%%%%%%%%%%%%%%%%%%%%%%%%%%%%%%%%%
\section{The simulation code}
%%%%%%%%%%%%%%%%%%%%%%%%%%%%%%%%%%%%%%%%%%%%%%%%%%%%%%%%%%%%%%%%%%%%%%%%%%%%%
\label{s:NumericalMethod}
We use the TreePM-SPH \gd code \citep{SP01.1,2005MNRAS.364.1105S} as
the starting point for our implementation of chemical evolution in
cosmological hydrodynamical simulations. The \gd code contains a fully
adaptive time--stepping, an explicit entropy--conserving formulation
of the SPH scheme \citep{2002MNRAS.333..649S}, heating from a uniform
evolving UV background \citep{1996ApJ...461...20H}, radiative cooling
from a zero metallicity cooling function, a sub--resolution model for
star formation in multi--phase interstellar medium (SH03), and a
phenomenological model for feedback from galactic ejecta powered by
explosions of SNII. Chemical enrichment was originally implemented by
accounting only for the contribution of the SNII expected for a
Salpeter IMF \citep{1955ApJ...121..161S}, under the instantaneous
recycling approximation (IRA) using global stellar yields.

As we will describe in this Section, we have improved this simplified
model along the following lines.
\begin{description} 
\item[(a)] We include the contributions of SNIa, SNII and AGB stars
  to the chemical enrichment, while both SNIa and SNII contributes
  to thermal feedback.
\item[(b)] We account for the age of different stellar populations,
  so that metals and energy are released over different time--scales
  by stars of different mass.
\item[(c)] We allow for different Initial Mass Functions (IMFs), so as
  to check the effect of changing its shape both on the stellar
  populations and on the properties of the diffuse gas.
\item[(d)] Different choices for stellar yields from SNII, SNIa and
  PNe are considered.
\item[(e)] Different schemes to distribute SN ejecta around star
  forming regions are considered, so as to check in detail the effect
  of changing the numerical treatment of metal and energy spreading.
\end{description}

%%%%%%%%%%%%%%%%%%%%%%%%%%%%%%%%%%%%%%%%%%%%%%%%%%%%%%%%%%%%%%%%%%%%%%%%%%%%%
\subsection{The star formation model}
%%%%%%%%%%%%%%%%%%%%%%%%%%%%%%%%%%%%%%%%%%%%%%%%%%%%%%%%%%%%%%%%%%%%%%%%%%%%%
\label{s:sf_in_gadget}
In the original \gd code, SH03 modeled the star formation process
through an effective description of the inter-stellar medium (ISM).
In this model, the ISM is described as an ambient hot gas containing
cold clouds, which provide the reservoir of star formation, the two
phases being in pressure equilibrium. The density of the cold and of
the hot phase represents an average over small regions of the ISM,
within which individual molecular clouds cannot be resolved by
simulations sampling cosmological volumes. 

In this description, baryons can exist in three phases: hot gas,
clouds and stars.  The mass fluxes between the these phases are
regulated by three physical processes: {\it(1)} hot gas cools and
forms cold clouds through radiative cooling; {\it(2)} stars are formed
from the clouds at a rate given a Schmidt Law; {\it(3)} stars explode,
thereby restoring mass and energy to the hot phase, and evaporating
clouds with an efficiency, which scales with the gas density. Under
the assumption that the time--scale to reach equilibrium is much
shorter than other timescales, the energy from SNe also sets the
equilibrium temperature of the hot gas in the star--formation regions.

The original \gd code only accounts for energy from SNII, that are
supposed to promptly explode, with no delay time from the star
formation episode. Therefore, the specific energy available per unit
mass of stars formed is \(\epsilon_{SNe} = e_{SNe} \times n_{SN}^{II}
\). Here, the energy produced by a single SN explosion is assumed to
be $e_{SNe}=10^{51}$~ergs, while the number of stars per solar mass
ending in SNII for a Salpeter IMF (\citealt{1955ApJ...121..161S}, S55
hereafter) is $n_{SN}^{II}=0.0074$~M$_\odot^{-1}$.

In the effective model by SH03, a gas particle is flagged as
star--forming whenever its density exceeds a given density--threshold,
above which that particle is treated as multi--phase. Once the clouds
evaporation efficiency and the star--formation (SF) timescale are
specified, the value of the threshold is self--consistently computed
by requiring {\it(1)} that the temperature of the hot phase at that
density coincides with the temperature, $T_{ti}$, at which thermal
instability sets on, and {\it(2)} that the specific {\it effective}
energy (see eq. [11] of SH03) of the gas changes in a continuous way
when crossing that threshold. Accordingly, the value of the density
threshold for star formation depends on the value of the cooling
function at $T_{ti}$, on the characteristic time--scale for star
formation, and on the energy budget from SNII. For reference, SH03
computed this threshold to correspond to $n_H\simeq 0.14$~cm$^{-3}$ for the
number density of hydrogen atoms in a gas of primordial composition.

In the simulations presented in this paper, we adopt the above
effective model of star formation from a multi-phase medium.  However,
in implementing a more sophisticated model of chemical evolution we
want to account explicitly for stellar life--times, thereby avoiding
the approximation of instantaneous recycling, as well as including
the possibility to change the IMF and the yields. Therefore, while the
general structure of the SH03 effective model is left unchanged, we
have substantially modified it in several aspects. Here below we
describe the key features that we have implemented in the code, while
we postpone further technical details to \refsec{chemical_evolution}.

\begin{description}
\item[(1)] The amount of metals and energy produced by each star
  particle during the evolution are self--consistently computed for
  different choices of the IMF. In principle, the code also allows one
  to treat an IMF which changes with time and whose shape depends on
  the local conditions (e.g., metallicity) of the star--forming
  gas. This feature is not used in the simulations that we will
  discuss in this paper.
\item[(2)] As in the original SH03 model, self--regulation of star
  formation is achieved by assuming that the energy of short--living
  stars is promptly available, while all the other stars die according
  to their lifetimes. We define as short living all the stars with
  mass $\ge M_{SL}$, where $M_{SL}$ must be considered as a parameter
  whose value ranges from the minimum mass of core--collapse SNe (we
  assume $8M_\odot$), up to the maximum mass where the IMF is
  computed. This allows us to parametrize the feedback strength in the
  self--regulation of the star formation process. We emphasize that
  the above mass threshold for short living stars is only relevant for
  the energy available to self--regulate star formation, while metal
  production takes place by accounting for life--times, also for stars
  with mass $\ge M_{SL}$. In the following runs we set
  $M_{SL}=8M_\odot$, thus assuming that all the energy from SNII is
  used for the self--regulation of star formation. 
\item[(3)] We include the contribution of metals to the cooling
  function. To this purpose, our implementation of cooling proceeds as
  follows. The original cooling function provided in the \gd code is
  used to account for the photo-ionization equilibrium of Hydrogen and
  Helium, while the tables by \cite{1993ApJS...88..253S} are used to
  account for the contribution of metals to the cooling function. We
  note that the cooling tables by \cite{1993ApJS...88..253S} assume
  the relative proportions of different metal species to be
  solar. Including more refined cooling rates, which depend explicitly
  on the individual abundances of different metal species, involves a
  straightforward modification of the code. Due to the lack of an
  explicit treatment of metal diffusion, a heavily enriched gas
  particle does not share its metal content with any neighbor metal
  poor particle. This may cause a spurious noise in the cooling
  process, in the sense that close particles may have heavily
  different cooling rates, depending on how different is their
  metallicity. To overcome this problem, we decided to smooth gas
  metallicity using the same kernel used for the computation of the
  hydrodynamical forces (i.e., a B--spline kernel using 64 neighbors),
  but only for the purpose of the computation of the cooling
  function. Therefore, while each particle retains its metal mass, its
  cooling rate is computed by accounting also for the enrichment of
  the surrounding gas particles.
\item[(4)] A self--consistent computation of the density threshold for
  star formation implies introducing a complex interplay between
  different ingredients. Firstly, changing the IMF changes the amount
  of energy available from short--living stars, in such a way that the
  threshold increases with this energy. Secondly, including the
  dependence of the cooling function on the local metallicity causes
  the density threshold to decrease for more enriched star forming
  gas. In the following, we fix the value of this threshold at $n_H =
  0.1$~cm$^{-3}$ in terms of the number density of hydrogen atoms, a
  value that has been adopted in a number of previous studies of star
  formation in hydrodynamical simulations
  \citep[e.g.,][]{1993PhDT........65S,1993MNRAS.265..271N,1996ApJS..105...19K,2002MNRAS.330..113K},
  and which is comparable to that, $n_H = 0.14$~cm$^{-3}$, computed by
  SH03 in their effective model.  We defer to a future analysis the
  discussion of a fully self--consistent model to compute the star
  formation threshold.
\item[(5)] The computation of the fraction of clouds proceeds exactly
  as in the original effective model (see eq.[18] in SH03), except
  that we explicitly include the dependence of the cooling function
  on the local gas metallicity, and the SN energy used for the
  computation of the pressure equilibrium is consistently computed for
  a generic IMF, as described above.
\end{description}

%%%%%%%%%%%%%%%%%%%%%%%%%%%%%%%%%%%%%%%%%%%%%%%%%%%%%%%%%%%%%%%%%%%%%%%%%%%%%
\subsection{The numerical implementation of star formation}
%%%%%%%%%%%%%%%%%%%%%%%%%%%%%%%%%%%%%%%%%%%%%%%%%%%%%%%%%%%%%%%%%%%%%%%%%%%%%

In order to the define the rule to transform star--forming gas
particles into collisionless star particles we closely follow the
implementation by SH03 of the algorithm originally developed by
\cite{1996ApJS..105...19K}. This algorithm describes the
formation of star particles as a stochastic process, rather than as a
``smooth'' continuous process. Basically, at a given time the star
formation rate of a multi--phase gas particle is computed using a
Schmidt-type law \citep{1959ApJ...129..243S}:
\be
\dot m_\star = xm/t_\star\,.
\label{eq:mst}
\ee
Here, $x$ is the fraction of gas in cold clouds, so that $xm$ is the
mass of cold clouds providing the reservoir for star formation. Within
the effective star formation model by SH03, the star formation
time--scale, $t_\star({\rho})$, is computed as
\be
t_\star(\rho) = t_0^*(\rho/\rho_{th})^{-1/2}\,,
\label{eq:tst}
\ee
where $t_0^*$ is a parameter of the model, while $\rho_{th}$ is the
density threshold for star formation, that we defined above. SH03 showed
that the value of $t_0^*$ should be chosen so as to reproduce the
observed relation between the disc--averaged star formation per unit
area and the gas surface density \citep{1998ApJ...498..541K}. Following
\cite{2003MNRAS.339..312S}, we assume $t_0^*=1.5$ Gyr, and checked
that with this value we reproduce the Kennicutt law within the
observational uncertainties.

Following eq.(\ref{eq:mst}), the stellar mass expected to form in a
given time interval $\Delta t$ is
\be 
m_\star = m \left\{1-\exp\left(-\frac{x\Delta
t}{t_\star}\right) \right\}\,.  
\label{eq:mst1}
\ee
Within the stochastic approach to star formation, we define the number
of stellar generations, $N_*$, as the number of star particles, which
are generated by a single gas particles. Therefore, each star
particle will be created with mass
\be
\label{eq:star_mass}
m_{*,0} = m_{g,0}/N_\star \,,
\ee
where $m_{g,0}$ is the initial mass of the gas particles.
Within this approach, a star particle is created once a
random number drawn in the interval $[0,1]$ falls below the
probability
\be 
p = \frac{m_{g,0}}{m_{*,0}}\left[1-\exp\left(-\frac{x\Delta
t}{t_\star}\right) \right]\,.
\ee
After the occurrence of a large enough number of star formation
events, the stochastic star formation history will converge to the
continuous one. In case a gas particle already spawned $(N_g^\star-1)$
star particles, then it is entirely converted into a star particle.
As we shall discuss in Section \refsec{stev}, star particles are
allowed, in our implementation of chemical evolution, to restore part
of their mass to the gas phase, as a consequence of stellar mass
loss. Since this restored mass is assigned to surrounding gas
particles, the latter have masses which can change during the
evolution. Therefore, in the eq.(\ref{eq:star_mass}) the actual mass
$m_g$ replaces the initial mass $m_{g,0}$, which is assumed to be the
same for all gas particles. As a consequence, star particles can have
different masses due to {\em (a)} their mass loss, and {\em (b)} the
mass of their parent gas particles.

Clearly, the larger the number of generations, the closer is the
stochastic representation to the continuous description of star
formation. In the simulations presented in this paper we will use
$N_*=3$ as a reference value for the number of stellar generations, 
%while we will check the effect of increasing it to $N_*=12$.
while we checked that there is no appreciable variation of the final
results when increasing it to $N_*=12$.

%%%%%%%%%%%%%%%%%%%%%%%%%%%%%%%%%%%%%%%%%%%%%%%%%%%%%%%%%%%%%%%%%%%%%%%%%%%%%
\subsection{The chemical evolution model}
%%%%%%%%%%%%%%%%%%%%%%%%%%%%%%%%%%%%%%%%%%%%%%%%%%%%%%%%%%%%%%%%%%%%%%%%%%%%%
\label{s:chemical_evolution}
Due to the stochastic representation of the star formation process,
each star particle must be treated as a simple stellar population
(SSP), i.e. as an ensemble of coeval stars having the same initial
metallicity. Every star particle carries all the physical information
(e.g. birth time $t_b$, initial metallicity and mass), which are
needed to calculate the evolution of the stellar populations, that
they represent, once the lifetime function (see Section
\ref{s:lifet}), the IMF (see Section \ref{s:imf}) and the yields (see
Section \ref{s:yields}) for SNe and AGB stars have been
specified. Therefore, we can compute for every star particle at any
given time $t>t_b$ how many stars are dying as SNII and SNIa, and how
many stars undergo the AGB phase, according to the equations of
chemical evolution that we will discuss in \refsec{stev} below. The
accuracy with which chemical evolution is followed is set by defining
suitable ``chemical'' time--steps. These time--steps are adaptively
computed during the evolution by fixing the percentage of SNe of each
type, which explode within each time step. In our simulations, we
choose this fraction to be 10 per cent for SNII and 2 per cent for
SNIa. As such, these time--steps depend both on the choice of the IMF
and on the life--time function.

In the following, we assume that SNIa arises from stars belonging to
binary systems, having mass in the range 0.8--8~$M_\odot$
\citep{1983A&A...118..217G}, while SNII arise from stars with mass
$>8M_\odot$. Besides SNe, which release energy and metals, we also
account for the mass loss by the stars in the AGB phase. They
contribute to metal production, but not to the energy feedback, and
are identified with those stars, not turning into SNIa, in the mass
range 0.8--8~$M_\odot$.

In summary, the main ingredients that define the model of chemical
evolution, as we implemented in the code, are the following: {\em (a)}
the SNe explosion rates, {\em (b)} the adopted lifetime function, {\em
  (c)} the adopted yields and {\em (d)} the IMF which fixes the number
of stars of a given mass. We describe each of these ingredients here
in the following.

As we shall discuss in \refsec{metspr}, once produced by a star
particle, metals are spread over surrounding particles according to a
suitable kernel.

%SB. Section already revised by Luca & Stefano.
\subsubsection{The equations of chemical evolution}
\label{s:stev}
We describe here the equations for the evolution of the rates of
SNIa, SNII and AGB stars, along with their respective metal
production. We provide here a short description of the basic results
and of the equations, which are actually solved in our simulations,
while we refer to the textbook by \cite{2003ceg..book.....M} for a
detailed discussion.

Let $\tau(m)$ be defined as the life--time function, i.e. the age at
which a star of mass $m$ dies. Accordingly, the the rate of explosions
of SNIa reads
\ba
R_{SNIa}(t) & = & -\der{m(t)}{t}\bigg|_{m_2\equiv
  \tau^{-1}(t)}\\ \nonumber
& \times & 24\,m_2^2\,A\int_{M_{Bm}}^{M_{BM}}\phi(m_B)\frac{1}{m_B^3}dm_B\,. 
\label{eq:my_snia_rate}
\ea
where the term in the first line is the mass of stars dying at the
time $t$, $\tau^{-1}(t)$ is the inverse of the lifetime function
$\tau(m)$, $\phi(m)$ is the IMF and $A$ is the fraction of stars in
binary systems of that particular type to be progenitors of SNIa. The
integral is over the mass $m_B$ of the binary system, which runs in
the range of the minimum and maximum allowed for the progenitor binary
system, $M_{Bm}$ and $M_{BM}$, respectively. Following
\cite{1983A&A...118..217G} and \cite{1986A&A...154..279M}, we assume
$A=0.1$, $M_{Bm}=3M_\odot$ and
$M_{BM}=16M_\odot$. \cite{1995A&A...304...11M} applied a model of
chemical enrichment of the ICM and found that $A=0.1$ was required to
reproduce the observed Iron enrichment, by assuming a Scalo IMF
\citep{1986FCPh...11....1S}. Changing the IMF would in principle
require to change the value of $A$. Since our model of ICM enrichment
is quite different from that by \cite{1995A&A...304...11M}, we prefer
here to fix the value of $A$ and check the agreement with observations
for different IMFs, rather than adjusting by hand its value case by
case. Eq.(\ref{eq:my_snia_rate}) holds under the assumption of
impulsive star formation. Indeed, since each star particle is
considered as a SSP, the associated star formation history, $\psi(t)$,
is a delta--function, $\delta(t-t_0)$, centered on the formation time
$t_0$.

As for the SNII and the low and intermediate mass stars stars, the
rate is given by
\be
R_{SNII|AGB}(t)=\phi(m(t)) \times \left( -\frac{d\,m(t)}{d\, t}\right)\,.
\label{eq:my_snii_rate}
\ee
We note that the above expression must be multiplied by a factor of
$(1-A)$ for AGB rates if the interested mass $m(t)$ falls in the same
range of masses which is relevant for the secondary stars of SNIa
binary systems. 

The release of energy and chemical elements by stars (binary systems
in case of SNIa) of a given mass is obtained by multiplying the above
rates by the yields $p_{Z_i}(m, Z)$, which give the mass of the
element $i$ produced by a star of mass $m$ and initial metallicity
$Z$. Then, the equation which describes the evolution of the mass
$\rho_i(t)$ for the element $i$, holding for a generic form of the
star formation history $\psi(t)$, reads: \be
\begin{array}{l}
\dot{\rho}_i(t)=-\psi(t)Z_i(t) + \\
\\
A\int_{M_{Bm}}^{M_{BM}}\phi(m)\left[\int_{\mu_{min}}^{0.5}
f(\mu)\psi(t-\tau_{m_2})p_{Z_i}(m, Z)\,d\mu \right]\, dm + \\
\\
(1-A)\int_{M_{Bm}}^{M_{BM}}
\psi(t-\tau(m))p_{Z_i}(m, Z)\varphi(m)\,dm + \\
\\
\int_{M_L}^{M_{Bm}}
\psi(t-\tau(m))p_{Z_i}(m, Z)\varphi(m)\,dm + \\
\\
\int_{M_{BM}}^{M_{U}}
\psi(t-\tau(m))p_{Z_i}(m, Z)\varphi(m)\,dm.
\label{eq:stev}
\end{array}
\ee 
In the above equation, $M_L$ and $M_U$ are the minimum and maximum
mass of a star, respectively. In the following we use $M_L=0.1M_\odot$
and $M_U=100M_\odot$. The term in the first line of eq.(\ref{eq:stev})
accounts for the metals which are locked up in stars. The term in the
second line accounts for metal ejection contributed by SNIa. Here we
have explicitly written the inner integral that accounts for all the
possible mass ratios $\mu = m_2/(m_1+m_2)$ between the secondary star
mass and the total mass; $\mu_{\rm min}$ is the minimum value of $\mu$
and $f(\mu)$ is the corresponding distribution function. The terms on
the third and fourth lines describe the enrichment by mass--loss from
intermediate and low mass stars, while the last line accounts for
ejecta by SNII.

The $\mu$ distribution function is assumed to be
\be
f(\mu)=2^{1+\gamma}(1+\gamma)\mu^\gamma
\ee
where $\gamma=2$. This functional form of $f(\mu)$ has been derived
from statistical studies of the stellar population in the solar
neighborhood \citep{1980IAUS...88...15T,2001ApJ...558..351M}. The
value of $\mu_{\rm min}$ is calculated for a binary system of mass
$M_B$ as
\be \mu_{\rm min}={\rm max}\left(\frac{m_2}{m_B},
\frac{m_B-0.5M_{BM}}{m_B} \right).  
\ee
Taking the impulsive star--formation, the terms in eq.(\ref{eq:stev})
must be recast in the form that we actually use for calculating the rates.

In order to solve eqs.(\ref{eq:my_snia_rate}), (\ref{eq:my_snii_rate})
and (\ref{eq:stev}) in the \gd code we proceed as follows. At the
beginning of each run, two tables, one for SNII and one for SNIa and
low and intermediate mass stars, are built to specify at what delay
times the chemical evolution should be calculated. The accuracy of
these ``chemical time--steps'' is set by two run-time parameters that
specify what fraction of stars must be evolved at each step.
Accordingly, during each global time-step of the code only a small
fraction (typically few percent) of all stars is processed.  This
implementation of chemical evolution is quite efficient in terms of
computational cost, especially when the number of stars grows. We
verified that using $N_*=3$ for the number of stellar generations, the
overhead associated to the chemical evolution part amounts only to
$\mincir 10$ per cent of the total computational cost for a typical
simulation.

\subsubsection{The lifetime function}
\label{s:lifet}
In our reference run, we use the function given by
\cite{1993ApJ...416...26P} (PM hereafter),
\be
\label{eq:PM_lifetimes}
\tau(m) = \left\{\begin{array}{ll}
10^{[(1.34 - \sqrt{1.79 - 0.22 (7.76 -\log(m))}) / 0.11 ]
  - 9 } \\
{\rm for}\,\,m \le 6.6~{\rm M}_\odot\\
\\
1.2\, m^{-1.85} + 0.003 \,\,\,\,\, {\rm otherwise} \\
\end{array}\right.
\ee
Furthermore, we also consider the lifetime function originally proposed by
\cite{1989A&A...210..155M} (MM hereafter), and extrapolated by
\cite{1997ApJ...477..765C} to very high ($> 60 M_\odot$) and very low
($< 1.3 M_\odot$) masses:
\be
\label{eq:MM_lifetimes}
\tau(m)=\left\{\begin{array}{ll}
10^{ -0.6545 \log m + 1} & m \le 1.3~{\rm M}_\odot\\
\vspace{-0.2cm}&\\
10^{ -3.7 \log m + 1.351} & 1.3 < m \le 3~{\rm M}_\odot\\
\vspace{-0.2cm}&\\
10^{ -2.51 \log m + 0.77} & 3 < m \le 7~{\rm M}_\odot\\
\vspace{-0.2cm}&\\
10^{ -1.78 \log m + 0.17} & 7 < m \le 15~{\rm M}_\odot\\
\vspace{-0.2cm}&\\
10^{ -0.86 \log m - 0.94} & 15 < m \le 53~{\rm M}_\odot\\
\vspace{-0.3cm}&\\
1.2 \times m^{-1.85}+ 0.003 & otherwise 
\end{array} \right.
\ee
We refer to the paper by \cite{2005A&A...430..491R} for a detailed
discussion on the effects of different lifetime functions on the
chemical enrichment model of the Milky Way.

\begin{figure}
\centerline{
\includegraphics[width=8.5cm]{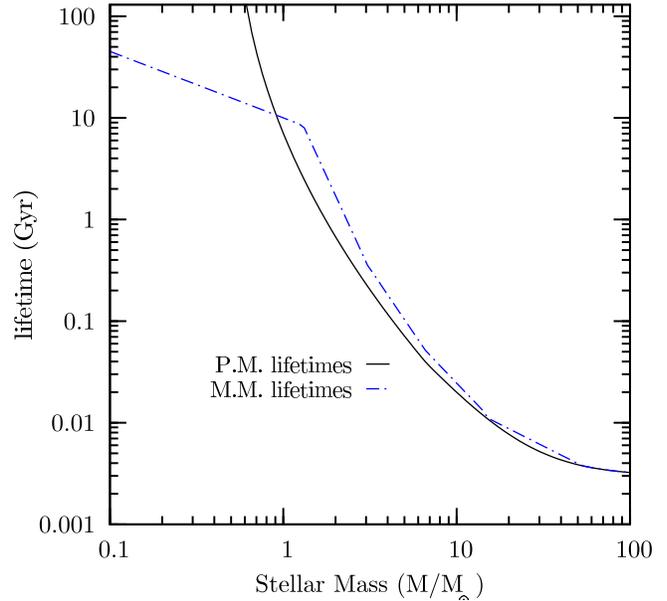}}
\caption{The dependence of the lifetime functions on the stellar mass.
  The solid and the dashed lines are for the lifetimes by
  \protect\cite{1993ApJ...416...26P} and by
  \protect\cite{1989A&A...210..155M}, respectively.}
\label{fi:lifet}
\end{figure}

A comparison between the life--time functions of
eqs.(\ref{eq:PM_lifetimes}) and (\ref{eq:MM_lifetimes}) is shown in
Figure \ref{fi:lifet}. The main difference between these two functions
concerns the life--time of low mass stars ($ < 8\,M_\odot$). The MM
function delays the explosion of stars with mass $\magcir
1\,M_\odot$, while it anticipates the explosion of stars below
$1\,M_\odot$ with respect to PM function. Only for masses below
$1M_\odot$, the PM function predict much more long--living stars. We
have verified that, assuming a Salpeter IMF (see below), the SNIa
rate from a coeval stellar population is expected to be higher after
$\sim 3$ Gyr when the MM lifetime function is adopted. This implies
that different life--times will produce different evolution of both
absolute and relative abundances. This will be discussed in more
detail in Sect. \ref{s:res_lifet}.

We point out that the above lifetime functions are independent of
metallicity, whereas in principle this dependence can be included
in a model of chemical evolution. For instance,
\cite{1996A&A...315..105R} used the metallicity--dependent lifetimes
as obtained from the Padova evolutionary tracks
\cite{1994A&AS..106..275B}.

\subsubsection{Stellar yields}
\label{s:yields}

The stellar yields specify the quantity $p_{Z_i}(m, Z)$, which appears
in eq.\ref{eq:stev} and, therefore, the amount of different metal
species which are released during the evolution of each star
particle. In the runs presented in this work, we adopt the yields
provided by \cite{1997A&AS..123..305V} the low and intermediate mass
stars and by \cite{2003NuPhA.718..139T} for SNIa. As for SNII we adopt
the metallicity--dependent yields by \cite{1995ApJS..101..181W} in the
reference run, while we will also show the effect of using instead the
yields by \cite{2004ApJ...608..405C} (WW and CL respectively,
hereafter; see Table \ref{t:runs}).  We also assume that all the stars
having masses $>40\,M_\odot$ directly end in black holes.

Along with freshly produced elements, stars also release
non--processed elements. Sometimes, papers presenting yields table do
not explicitly take into account these non--processed yields
\citep[e.g.,][]{1997A&AS..123..305V}.  In order to account for them,
whenever necessary, we assume that the non--processed metals
are ejected along with Helium and non--processed Hydrogen. 
%{\bf Then, if
%$M_H$ is the hydrogen mass, as given by the yields table, we assume
%that it is given by $M_H=m_H + m_Z$, where $m_H$ is the actually
%ejected hydrogen mass and $m_Z$ is the ejected mass of non-processed
%elements. Since the metallicity of the ejected non--processed gas is
%expected to be equal to the initial metallicity $Z$ of the star, then
%$m_Z=m_HZ$.} 
%Hence, the mass of ejected non--processed metals can be
%written as $m_Z=M_H Z/(1+Z)$, while the mass of Hydrogen is
%$M_H/(1+Z)$. The same relations also hold for the mass of the ejected
%Helium.

%\begin{figure*}
%\centerline{
%\hbox{
%\psfig{file=Figs/integrated_yields.eps,width=8.5cm}
%\psfig{file=Figs/WW_vs_CL_O2Fe.eps,width=8.5cm}}}
%\caption{Left panel: the ratio $M_j^{WW}/M_j^{CL}$ between the mass of
%  specie $j$ produced by the SNII of a SSP when using the two sets of
%  yields by \protect\cite{1995ApJS..101..181W} and by
%  \protect\cite{2004ApJ...608..405C} for different values of the
%  initial SSP metallicity. Right panel: the ratio of the O/Fe
%  relative abundances produced by the same two sets of yields, as a
%  function of the initial SSP metallicity. Filled circles are for the
%  products of SNII only, while open squares account also for the
%  products of SNIa and AGB stars.}
%\label{fi:metyie}
%\end{figure*}

\begin{figure}
\centerline{
\includegraphics[width=8.5cm]{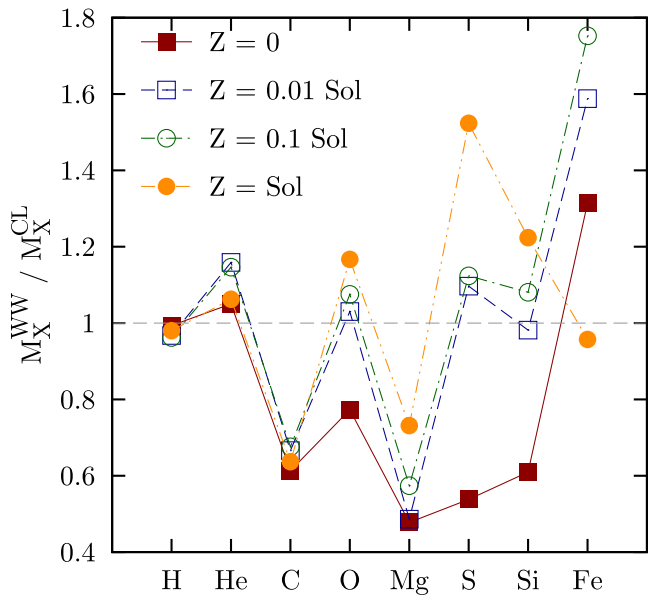}}
\caption{The ratio $M_j^{WW}/M_j^{CL}$ between the mass of species
  $j$, produced by the SNII of a SSP, when using the two sets of
  yields by \protect\cite{1995ApJS..101..181W} and by
  \protect\cite{2004ApJ...608..405C} for different values of the
  initial SSP metallicity. Different symbols are for different values
  of the initial metallicity of the SSP, as reported by the labels.}
\label{fi:metyie}
\end{figure}

Besides H and He, in the simulations presented in this paper we trace
the production of Fe, O, C, Si, Mg, S. The code can be easily modified
to include other metal species.

In Figure \ref{fi:metyie} we show the ratios between
the abundances of different elements, as expected for the WW and the
CL yields, from the SNII of a SSP. Different curves and symbols here
correspond to different values of the initial metallicity of the
SSP. Quite apparently, the two sets of yields provide significantly
different metal masses, by an amount which can sensitively change with
initial metallicity. 
%In order to highlight this dependence on the
%initial SSP metallicity, we show in the right panel of
%Fig. \ref{fi:metyie} the ratio of the relative O/Fe abundances from
%the two sets of SNII yields. Clearly, the differences become smaller
%when including also the contribution of SNIa and AGB stars (open
%squares) to the contribution of SNII only (filled circles). In
In Sect.\ref{s:yie} we will discuss the effect of changing yields on
the resulting enrichment of the ICM and of the stellar population in
simulated clusters.

\subsubsection{The initial mass function}
\label{s:imf}
The initial mass function (IMF) is one of the most important quantity
in modeling the star formation process. It directly determines the
relative ratio between SNII and SNIa and, therefore, the relative
abundance of $\alpha$--elements and Fe--peak elements.  The shape of
the IMF also determines how many long--living stars will form with
respect to massive short--living stars. In turn, this ratio affects
the amount of energy released by SNe and the present luminosity of
galaxies, which is dominated by low mass stars, and the (metal)
mass--locking in the stellar phase.

As of today, no general consensus has been reached on whether the IMF
at a given time is universal or strongly dependent on the environment,
or wheter it is time--dependent, i.e. whether local variations of the
values of temperature, pressure and metallicity in star--forming
regions affect the mass distribution of stars.

Nowadays, there are growing evidences that the IMF in the local
universe, expressed in number of stars per unit logarithmic mass
interval, is likely to be a power--law for $m_\star > 1\,M_\odot$ with
slope $x\sim 1.35$, while it becomes flat below the $1\,M_\odot$
threshold, possibly even taking a negative slope below $\sim
0.2\,M_\odot$ \citep[e.g., ][]{2001MNRAS.322..231K}.  Theoretical
arguments \citep[e.g., ][]{1998MNRAS.301..569L} suggest that the
present--day characteristic mass scale $\sim 1\,M_\odot$ should have
been larger in the past, so that the IMF at higher redshift was
top--heavier than at present. \cite{2000ApJ...528..711C} showed that
varying the IMF by decreasing the characteristic mass with time, leads
to results at variance with observations of chemical properties of the
Galaxy. While the shape of the IMF is determined by the local
conditions of the inter--stellar medium, direct hydrodynamical
simulations of star formation in molecular clouds are only now
approaching the required resolution and sophistication level to make
credible predictions on the IMF \citep[e.g.,][]{2005MNRAS.356.1201B}.

In order to explore how the IMF changes the pattern of metal
enrichment, we implement it in the code in a very general way, so that
we can easily use both single-slope and multi--slope IMFs, as well as
time--evolving IMFs. In this work we use single-slope IMFs defined as
\be \phi(m)\,=\,\hder{N}{\log m} \propto m^{-x} \ee using $x=1.35$ for
the standard Salpeter IMF \citep{1955ApJ...121..161S} in our reference
run. In the above equation, $N$ is the number of stars per unit
logarithmic mass interval. We will explore the effect of changing the
IMF by also using a top--heavier IMF with $x=0.95$ \citep[][AY
hereafter]{1987A&A...173...23A}, as well as the multi--slope IMF by
\cite{2001MNRAS.322..231K}, which is defined as
\be
\phi(m)\, \propto \,\left\{\begin{array}{ll}
m^{-1.7} & m \ge 1\,{\rm M}_\odot\\ 
\vspace{-0.2cm}&\\
m^{-1.2} & 0.5 \le m < 1\,{\rm M}_\odot\\ 
\vspace{-0.2cm}&\\
m^{-0.3} & m \le 0.5\,{\rm M}_\odot\\ 
\end{array} \right.
\label{Kroupa}
\ee

\begin{figure}
\centerline{
\includegraphics[width=8.5cm]{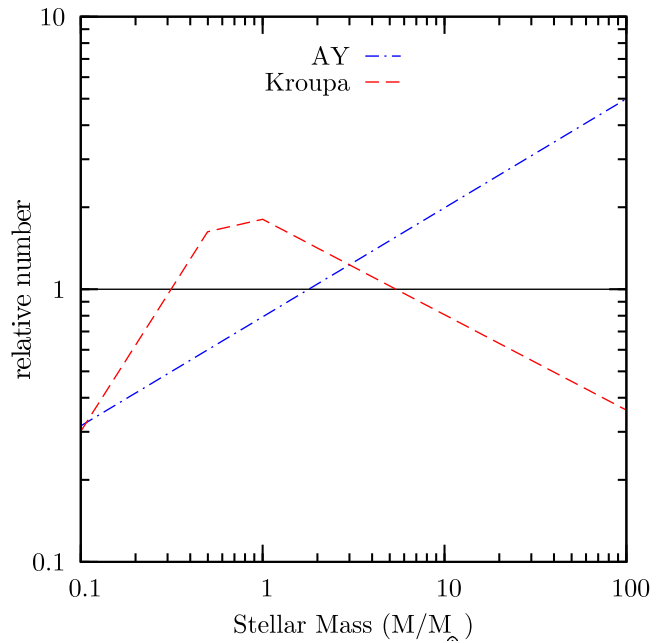} }
\caption{The dependence of the shape of the IMF on the stellar mass,
  relative to the Salpeter IMF
  \protect\citep{1955ApJ...121..161S}. Dashed and dot-dashed curves
  are for the IMF by \protect\cite{1987A&A...173...23A} and by
  \protect\cite{2001MNRAS.322..231K}, respectively. The horizontal
  solid line indicates for reference the Salpeter IMF.}
\label{fi:imfs}
\end{figure}

We show in Figure \ref{fi:imfs} the number of stars, as a function of
their mass, predicted by the AY and Kroupa IMFs, relative to those
predicted by the Salpeter IMF. As expected, the AY IMF predicts a
larger number of high--mass stars and, correspondingly, a smaller
number of low--mass stars, the crossover taking place at $\simeq 2
M_\odot$. As a result, we expect that the enrichment pattern of the AY
IMF will be characterized by a higher abundance of those elements,
like Oxygen, which are mostly produced by SNII. On the other hand,
the Kroupa IMF is characterized by a relative deficit of high--mass
stars and, correspondingly, a relatively lower enrichment in Oxygen is
expected. 

Since clusters of galaxies basically behave like ``closed boxes'', the
overall level of enrichment and the relative abundances should
directly reflect the condition of star formation. While a number of
studies have been performed so far to infer the IMF shape from the
enrichment pattern of the ICM, no general consensus has been
reached. For instance, \cite{1997ApJ...488...35R},
\cite{2002NewA....7..227P} and \cite{2004cgpc.symp..260R} argued that
both the global level of ICM enrichment and the $\alpha$/Fe relative
abundance can be reproduced by assuming a Salpeter IMF, as long as
this relative abundance is subsolar. Indeed, a different conclusion
has been reached by other authors \citep[e.g.,
][]{1996ApJ...466..695L} in the attempt of explaining $[\alpha/{\rm
Fe}]>0$ in the ICM. For instance, \cite{2004ApJ...604..579P} used a
phenomenological model to argue that a standard Salpeter IMF can not
account for the observed $\alpha$/Fe ratio in the ICM. A similar
conclusion was also reached by \cite{2005MNRAS.358.1247N}, who used
semi--analytical models of galaxy formation to trace the production of
heavy elements, and by \cite{2006MNRAS.371..548R}, who used
hydrodynamical simulations including chemical enrichment.
\cite{2006MNRAS.373..397S} analysed the galaxy population from
simulations similar to those presented here. This analysis led us to
conclude that runs with a Salpeter IMF produce a color--magnitude
relation that, with the exception of the BCGs, is in reasonable
agreement with observations. On the contrary, the stronger enrichment
provided by a top--heavier IMF turns into too red galaxy colors.

\vspace{0.3truecm} In summary, our implementation of chemical
evolution is quite similar to that adopted by
\cite{2003MNRAS.340..908K} and by \cite{2004MNRAS.347..740K}, while
differences exist with respect to other implementations. For instance,
\cite{1996A&A...315..105R} and \cite{2003MNRAS.339.1117V} also used a
scheme similar to ours, but neglected the contribution from low- and
intermediate-mass stars. \cite{2002MNRAS.330..821L} adopted a coarse
stochastic description of the ejecta from star particles: differently
from our continuous description, in which enriched gas is continuously
released by star particles, they assumed that each star particle is
assigned a given probability to be entirely converted into an enriched
gas particle. Finally, \cite{2001MNRAS.325...34M} and
\cite{2005MNRAS.364..552S} neglected delay times for SNII, assumed a
fixed delay time for SNIa and neglected the contribution to enrichment
from low- and intermediate-mass stars.

\subsection{Feedback through galactic winds}
SH03 discussed the need to introduce an efficient mechanism to
thermalize the SNe energy feedback, in order to regulate star
formation, thereby preventing overcooling. To this purpose, they
introduced a phenomenological description for galactic winds, which
are triggered by the SNII energy release. We provide here a basic
description of this scheme for galactic winds, while we refer to the
SH03 paper for a more detailed description. The momentum and the
kinetic energy carried by these winds are regulated by two
parameters. The first one specifies the wind mass loading according to
the relation, $\dot M_W=\eta \dot M_*$, where $\dot M_*$ is the star
formation rate. Following SH03, we assume in the following
$\eta=3$. The second parameter determines the fraction of SNe energy
that powers the winds, ${1\over 2}\dot M_W v_W^2= \chi
\epsilon_{SNe}\dot M_*$, where $\epsilon_{SNe}$ is the energy feedback
provided by the SNe under the IRA for each $M_\odot$ of stars formed.
In the framework of the SH03 effective model for star formation,
winds are uploaded with gas particles which are stochastically
selected among the multiphase particles, with a probability
proportional to their local star formation rate. As such, these
particles come from star--forming regions and, therefore, are heavely
metal enriched.  Furthermore, SH03 treated SPH particles that become
part of the wind as temporarily decoupled from hydrodynamical
interactions, in order to allow the wind particles to leave the dense
interstellar medium without disrupting it. This decoupling is
regulated by two parameters. The first parameter, $\rho_{\rm dec}$,
defines the minimum density the wind particles can reach before being
coupled again. Following SH03, we assumed this density to be 0.5 times
the threshold gas density for the onset of star formation. The second
parameter, $l_{\rm dec}$, provides the maximum length that a wind
particle can travel freely before becoming hydrodynamically coupled
again. If this time has elapsed, the particle is coupled again, even
if it has not yet reached $\rho_{\rm dec}$. We assumed $l_{\rm
dec}=10\,h^{-1}$kpc.

While we leave the scheme of kinetic feedback associated to wind
galactic unchanged, we decide to use instead the value of the wind
velocity, $v_w$, as a parameter to be fixed. For the reference run, we
assume $v_w=500\vel$. With the above choice for the wind mass loading
and assuming that each SN provides an energy of $10^{51}$ ergs, this
value of $v_w$ corresponds to $\epsilon_{SNe}\simeq 1$ for a Salpeter
IMF. We will also explore which is the effect of assuming instead a
stronger feedback, with $v_w=1000\vel$, on the pattern of chemical
enrichment.

% results
%\input{sect.3}
\section{Results}
\label{s:results}
In this Section we will discuss the results of a set of
simulations of one single cluster. The cluster that we
have simulated has been extracted from a low--resolution cosmological
box for a flat $\Lambda$CDM cosmology with $\Omega_m=0.3$ for the
matter density parameter, $h=0.7$ for the Hubble parameter,
$\Omega_b=0.039$ for the baryon density parameter and $\sigma_8=0.8$
for the normalization of the power spectrum of density
perturbations. Using the Zoomed Initial Condition (ZIC) technique
\citep{TO97.2}, mass resolution is increased in the Lagrangian
region surrounding the cluster, while also high--frequency modes of
the primordial perturbation spectrum are added. The main
characteristics of this cluster (Cl1 hereafter) are summarized in
Table \ref{t:clus}, along with the mass resolution and the
Plummer--equivalent softening parameter used in the runs (note that
the softenings are set fixed in physical units from $z=2$ to $z=0$,
while they are fixed in comoving units at higher redshift).

The reference run of Cl1 is performed under the following assumptions:
{\em (a)} metals produced by a star particle are distributed to
surrounding gas particles by using a SPH spline kernel with density
weighting over 64 neighbors for the distribution of metals around star
forming particles (see Sect. \ref{s:metspr}); {\em (b)} Salpeter IMF;
{\em (c)} stellar yields from \cite{2003NuPhA.718..139T} for SNIa,
\cite{1997A&AS..123..305V} for the low and intermediate mass stars and
\cite{1995ApJS..101..181W} for SNII; {\em (d)} Life--time function
from \cite{1993ApJ...416...26P}; {\em (e)} $v_w=500\vel$ for the
velocity of winds. In the following we will show the effect of
changing each one of these assumptions. Therefore, we will explore
neither the cluster-by-cluster variation in the enrichment pattern of
the ICM, nor any dependence on the cluster mass. We perform instead a
detailed study of the effect of changing a number of parameters which
specify both the numerical details of the implementation of the model
of chemical evolution model. We defer to a forthcoming paper the
properties of the ICM enrichment for a statistical ensemble of
simulated galaxy clusters. We summarize in Table \ref{t:runs} the
aspects in which the various runs of Cl1 differ from the reference
one.

In addition, we also performed simulations of another cluster (Cl2 in
Table \ref{t:clus}), whose initial conditions are generated at three
different resolutions, for the same cosmological model, by spanning a
factor 15 in mass resolution (see also
\citealt{2006MNRAS.367.1641B}). The lowest resolution at which the Cl2
cluster is simulated is comparable to that of the Cl1 runs.  As
such, this series of runs, that is performed for the same setting of
the Cl1 reference (R) run, allows us to check in detail the effect of
resolution on the ICM chemical enrichment.

Besides global properties, we will describe the details of the ICM
enrichment by showing radial profiles of the Iron abundance and of the
relative abundances of [O/Fe] and [Si/Fe]\footnote{We follow the
  standard bracket notation for the relative abundance of elements $A$
  and $B$: $[A/B]=\log(Z_A/Z_B)-\log(Z_{A,\odot}/Z_{B,\odot})$.}, the
histograms of the fraction of gas and stars having a given level of
enrichment, and the time evolution of the metal production and
enrichment. A comparison with observational data will only be
performed for the abundance profiles of Iron, which is the element
whose distribution in the ICM is best mapped. As a term of comparison,
we will use a subset of 5 clusters, out of the 16, observed with
Chandra and analysed by \cite{2005ApJ...628..655V}. These clusters are
those having temperature in the range 2--4 keV, comparable to that of
the simulated Cl1 and Cl2 clusters. Since the profiles of Fe abundance
are compared with observations, we plot them out to
$R_{500}$\footnote{In the following we will indicate with $R_\Delta$
  the radius encompassing an overdensity of $\Delta \rho_{cr}$, where
  $\rho_{cr}$ is the critical cosmic density. In this way, $M_\Delta$
  will be defined as the mass contained within $R_\Delta$.
  Furthermore, we will define \rvir to be the radius encompassing the
  virial overdensity $\Delta_{\rm vir}$, as predicted by the spherical
  top--hat collapse model. For the cosmology assumed in our
  simulations, it is $\Delta_{\rm vir}\simeq 100$.} which is the
maximum scale sampled by observations. We plot instead profiles of
[Si/Fe] and [O/Fe] out to 2\rvir in order to show the pattern of
relative enrichment over a large enough range of scales, where the
different star formation histories generates different contributions
from stars of different mass. Here and in the following we assume that
a SPH particle belongs to the hot diffuse gas of the ICM when it meets
the following three conditions: {\bf (i)} whenever the particle is
tagged as multiphase, its hot component should include at least 90 per
cent of its mass; {\bf (ii)} temperature and density must not be
  at the same time below $3\times 10^4$K and above $500\bar\rho_{bar}$
  respectively, where $\bar\rho_{bar}$ is the mean cosmic baryon
density. While observational data on the Iron abundance profiles are
used to ``guide the eye'', we emphasize that the primary aim of this
paper is not that of a detailed comparison with observations. This
would require a statistically significant set of simulated clusters,
sampling a wider range of temperatures, and is deferred to a
forthcoming paper.

\begin{table}
\centerline{
\begin{tabular}{lcccc}
\hline Cluster & $M_{\rm 200}$ & $R_{\rm 200}$ & $m_{gas}$ & $\epsilon_{Pl}$ \\
               &$10^{14}\msun$& $\hm$        &$10^8\msun$&$h^{-1}$kpc     \\
\hline
Cl1 & 2.2 &  0.98 & 5.7 & 5.0       \\
Cl2 & 1.4 &  0.85 &     &           \\
\multicolumn{3}{c}{R1} & 2.31 & 5.2 \\
\multicolumn{3}{c}{R2} & 0.69 & 3.5 \\
\multicolumn{3}{c}{R3} & 0.15 & 2.1 \\
\hline
\end{tabular}
}
\caption{Basic properties of the two simulated clusters Column 1:
  cluster name; Column 2: mass within the radius, $R_{200}$,
  encompassing an overdensity of 200 times the critical density,
  $\rho_c$; Column 3: value of $R_{200}$; Column 4: initial mass of
  the gas particles; Column 5: Plummer--equivalent softening of the 
  gravitational force at $z=0$.}
\label{t:clus}
\end{table}

\begin{table}
%\centerline{
\begin{tabular}{ll}
\hline 
R & Reference run\\
N16/N128 & B-spline kernel for metal distribution with 16/128 neighbors\\ 
%Ng12 & 12 stellar generations\\
%Ng12L & 12 stellar generation only in low--metallicity regions\\
MW & B-spline kernel for metal distribution using mass weighting\\ 
TH & Top--hat kernel for metal distribution\\
AY & Top--heavier IMF (Arimoto \& Yoshii 1987)\\
Kr & Kroupa IMF \protect\citep{2001MNRAS.322..231K}\\
CL & Yields for SNII from \protect\cite{2004ApJ...608..405C} \\
%NH02 & Threshold for star formation set to $n_H=0.2$ cm$^{-3}$\\
%ZC & Set to zero the metallicity for the cooling function\\
NW & No feedback associated to winds\\
SW & Strong winds with $v_w=1000\vel$\\
MM & Life--time function by Maeder \& Meynet (1989)\\
\hline
\end{tabular}
%}
\caption{Characteristics of the different Cl1 runs with respect to the
  reference run (see text). Column 1: Name of the run; Column 2:
  characteristic of the run.}
\label{t:runs}
\end{table}

\begin{figure*}
\centerline{
\hbox{
\includegraphics[width=5.8cm]{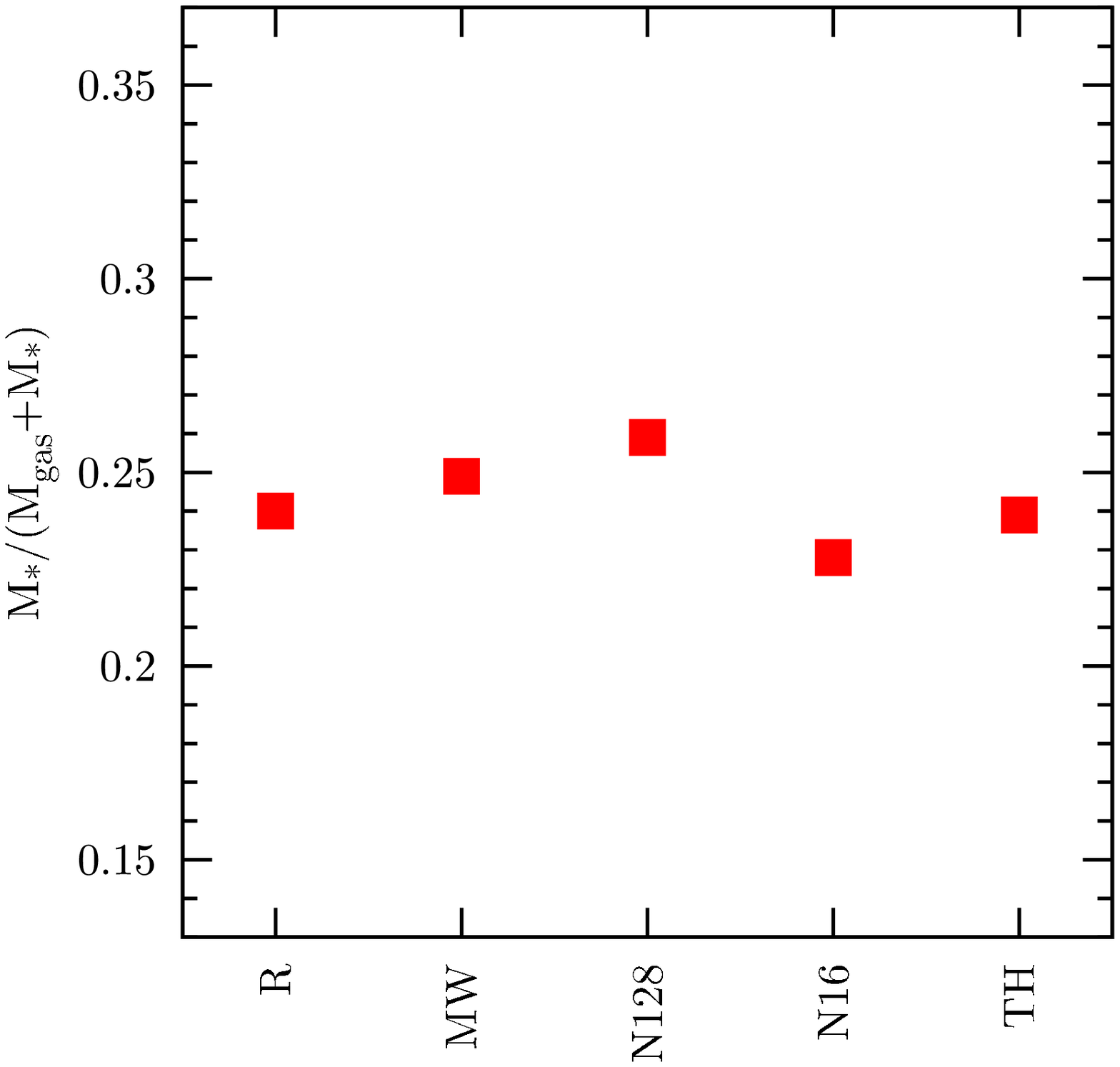} 
\includegraphics[width=5.8cm]{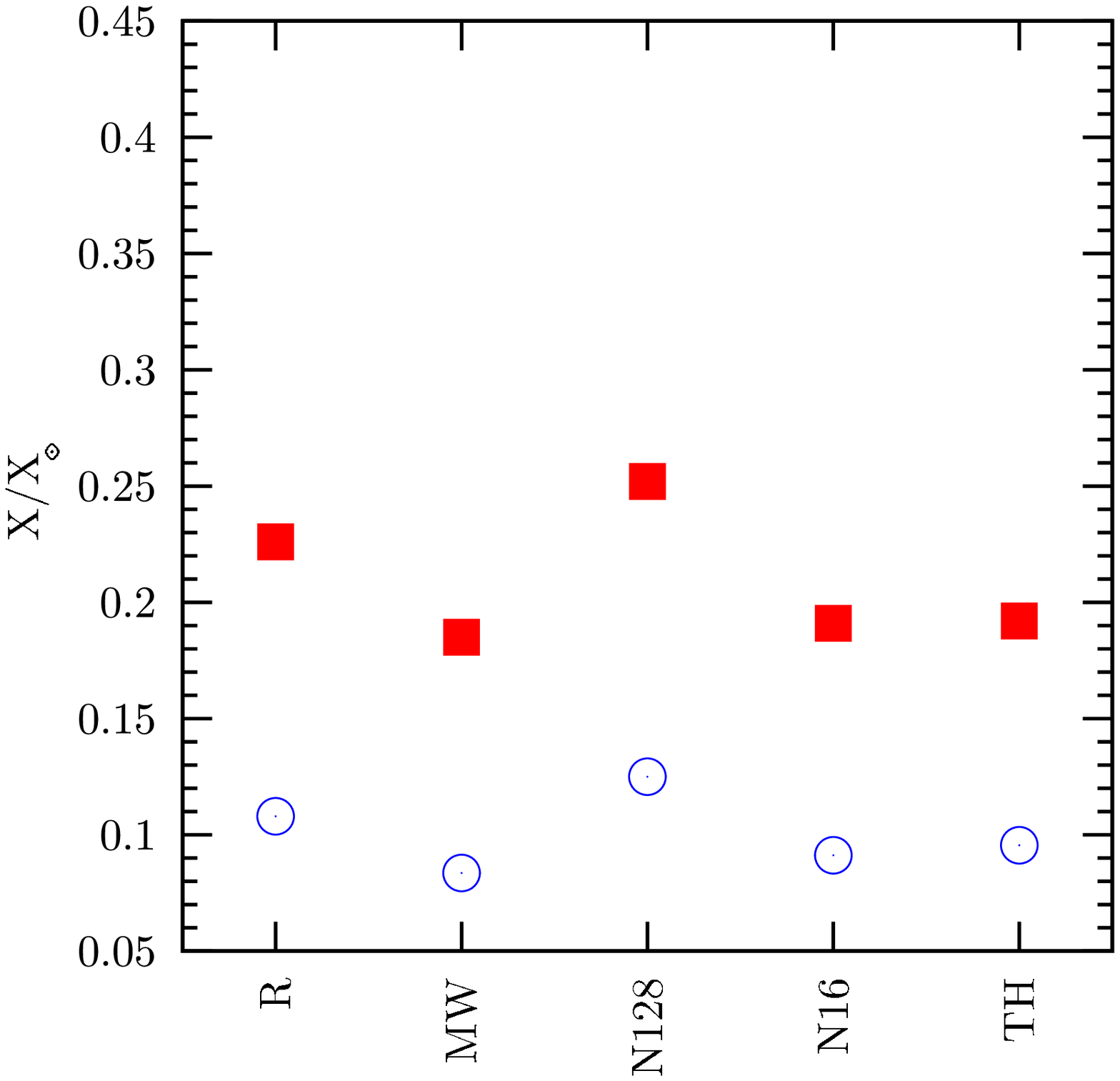} 
\includegraphics[width=5.8cm]{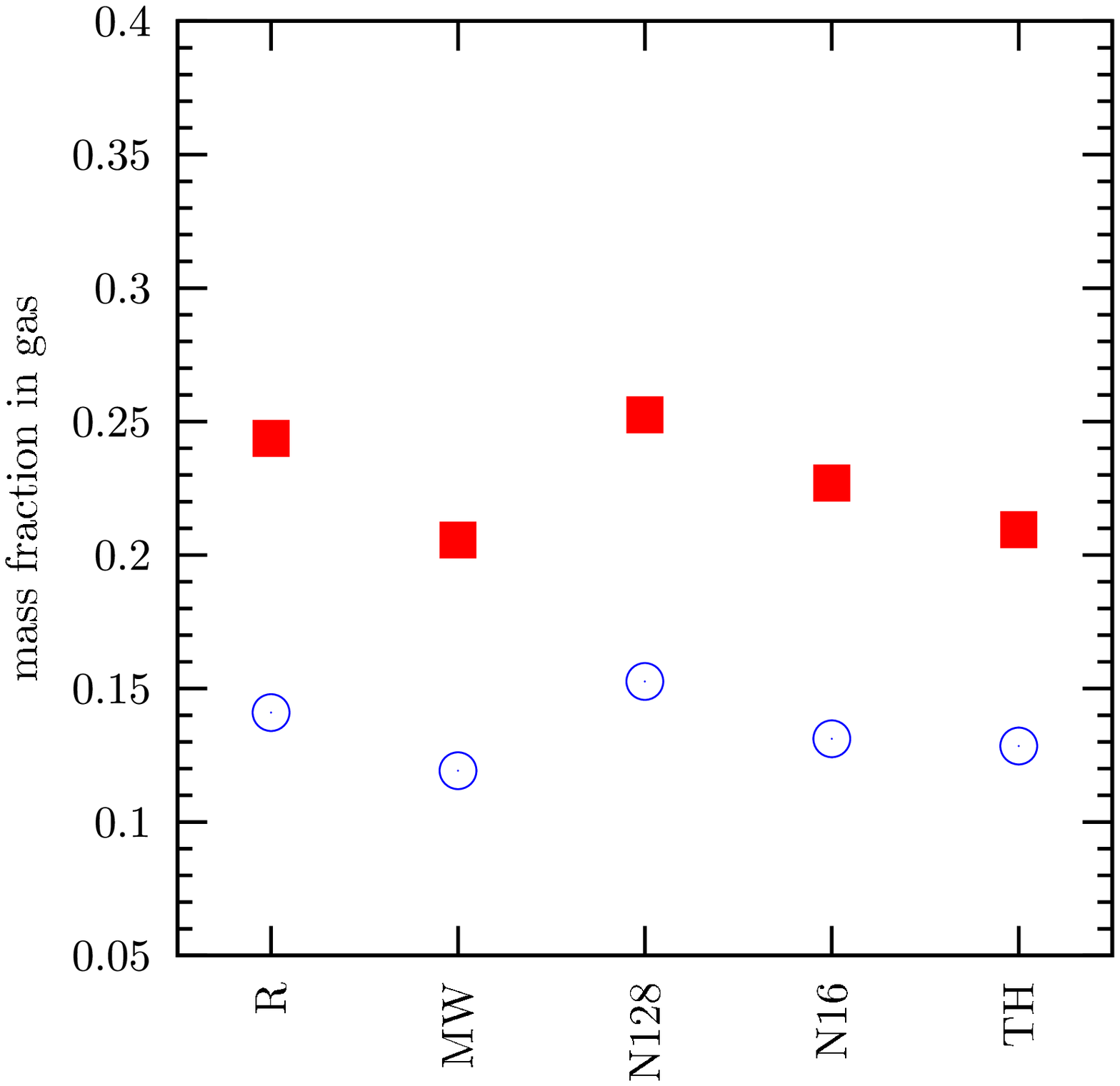} }}
\caption{The effect of the numerical parameters, which define the
  distribution of metals, on global quantities computed within the
  virial radius of the Cl1 cluster: total mass fraction of baryons in
  stars (left panel), mass--weighted Iron abundance (central panel)
  and fraction of metal mass in the diffuse gas (right panel). In the
  central and in the right panels, filled squares refer to Iron while
  open circles refer to Oxygen. The labels indicating the different
  runs are as reported in Table \ref{t:runs}.}
\label{fi:glob_num} 
\end{figure*}

\subsection{The effect of metal spreading}
\label{s:metspr}
A crucial issue in the numerical modeling of ICM enrichment concerns
how metals are distributed around the star particles, where they
are produced. In principle, the diffusion of metals around star
forming regions should be fully determined by local physical
processes, such as thermal motions of ions \citep[e.g.,][]{SA88.1},
turbulence \citep[][]{2005MNRAS.359.1041R}, ram--pressure stripping
\citep[e.g., ][]{2006A&A...452..795D}, galactic ejecta
\citep[e.g.,][]{2006astro.ph.11859S}. However, for these mechanisms to
be correctly operating in simulations, it requires having high enough
resolution for them to be reliably described. A typical example is
represented by turbulent gas motions
\citep[e.g.,][]{2005MNRAS.359.1041R}. The description of turbulence
requires not only sampling a wide enough dynamical range, where the
energy cascade takes place, but also a good knowledge of the plasma
viscosity, which determine the scale at which turbulent energy is
dissipated \citep[e.g.,][]{2005MNRAS.364..753D}. While approaches to
include physical viscosity in the SPH scheme have been recently
pursued \citep{2006MNRAS.371.1025S}, a full accounting for its effect
requires including a self--consistent description of the magnetic
field which suppresses the mean free path of ions.

\begin{figure*}
\centerline{
\vbox{
\hbox{
\includegraphics[width=7.5cm]{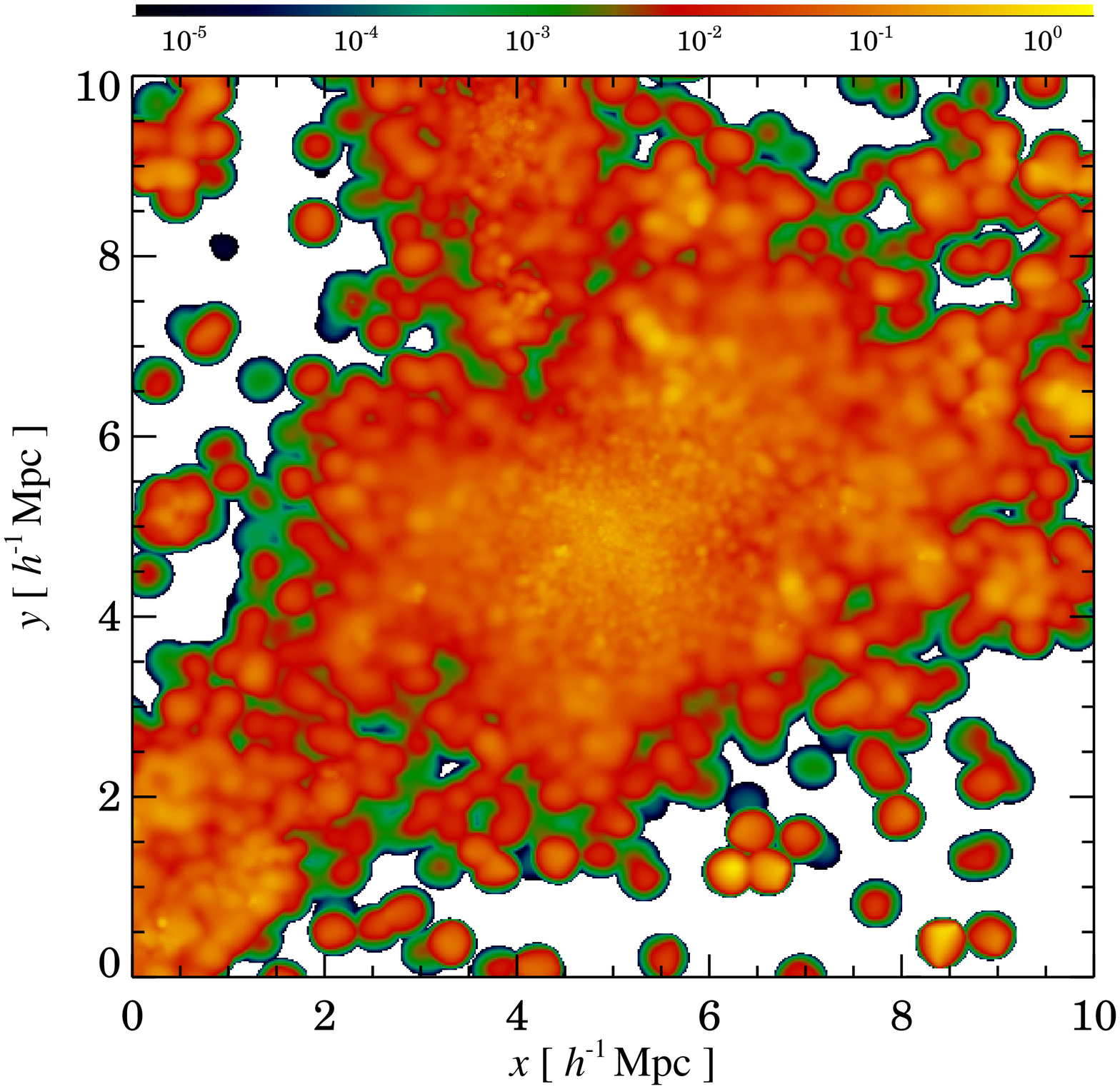} 
\includegraphics[width=7.5cm]{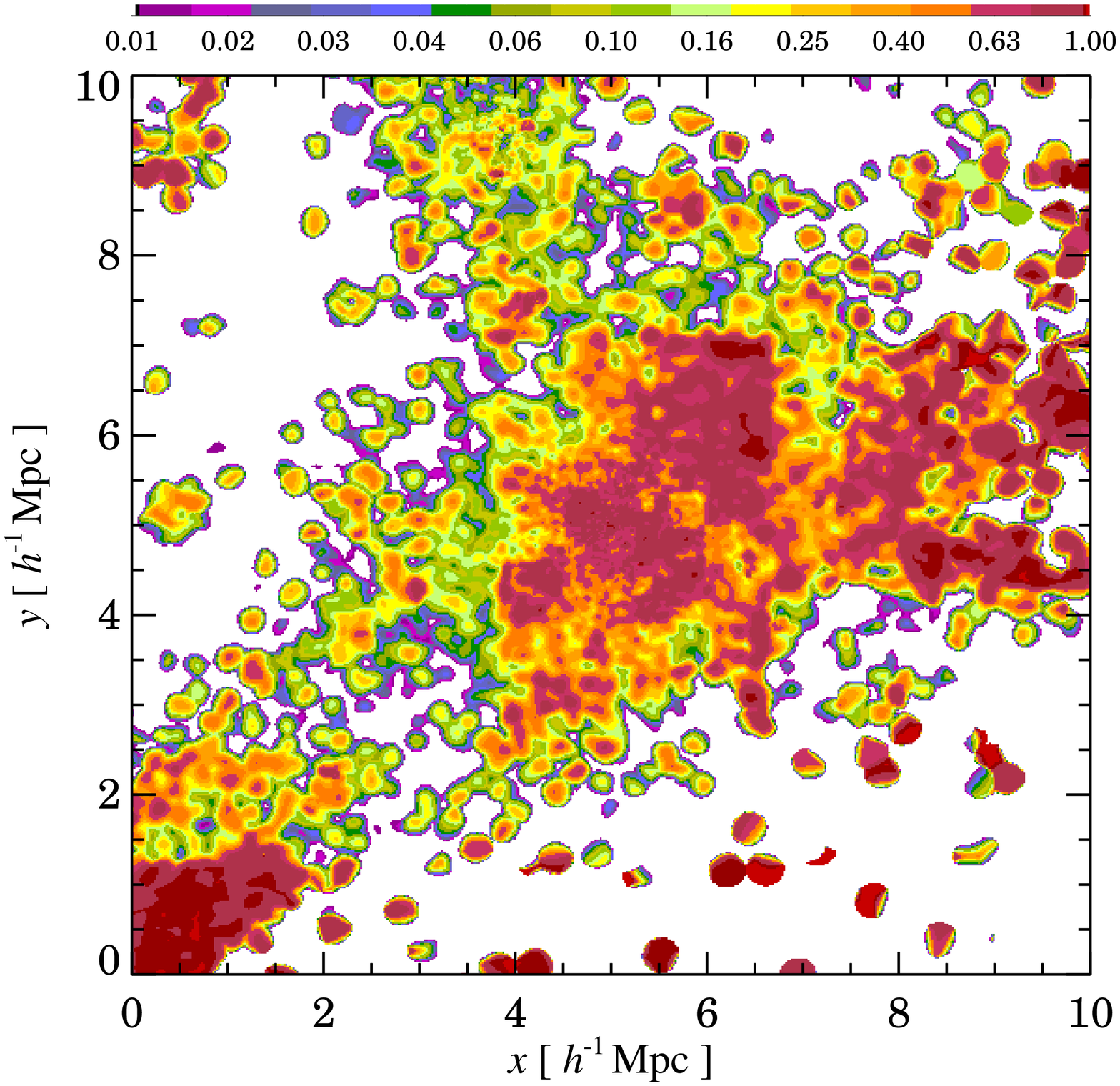} 
}
\hbox{
\includegraphics[width=7.5cm]{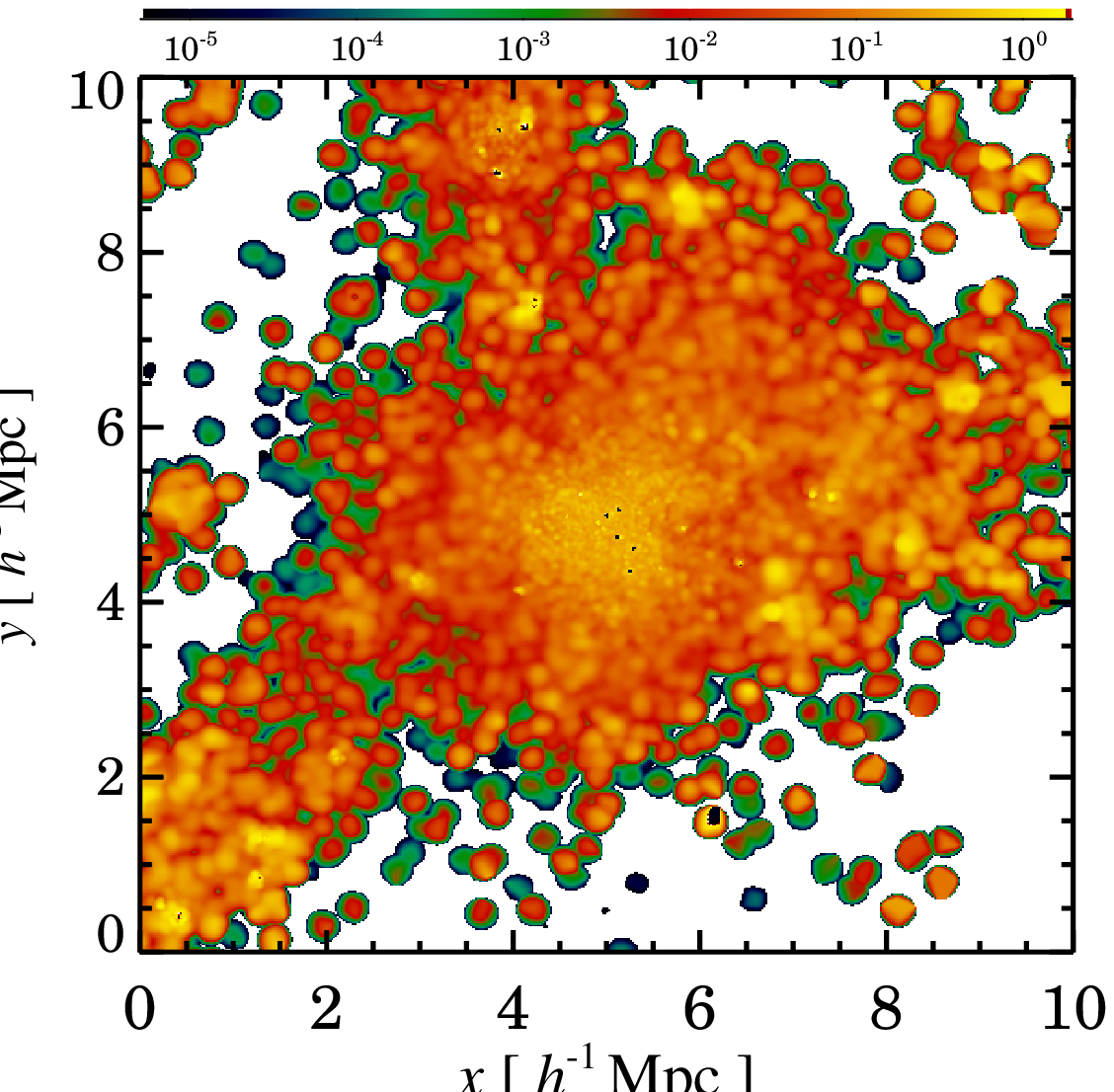} 
\includegraphics[width=7.5cm]{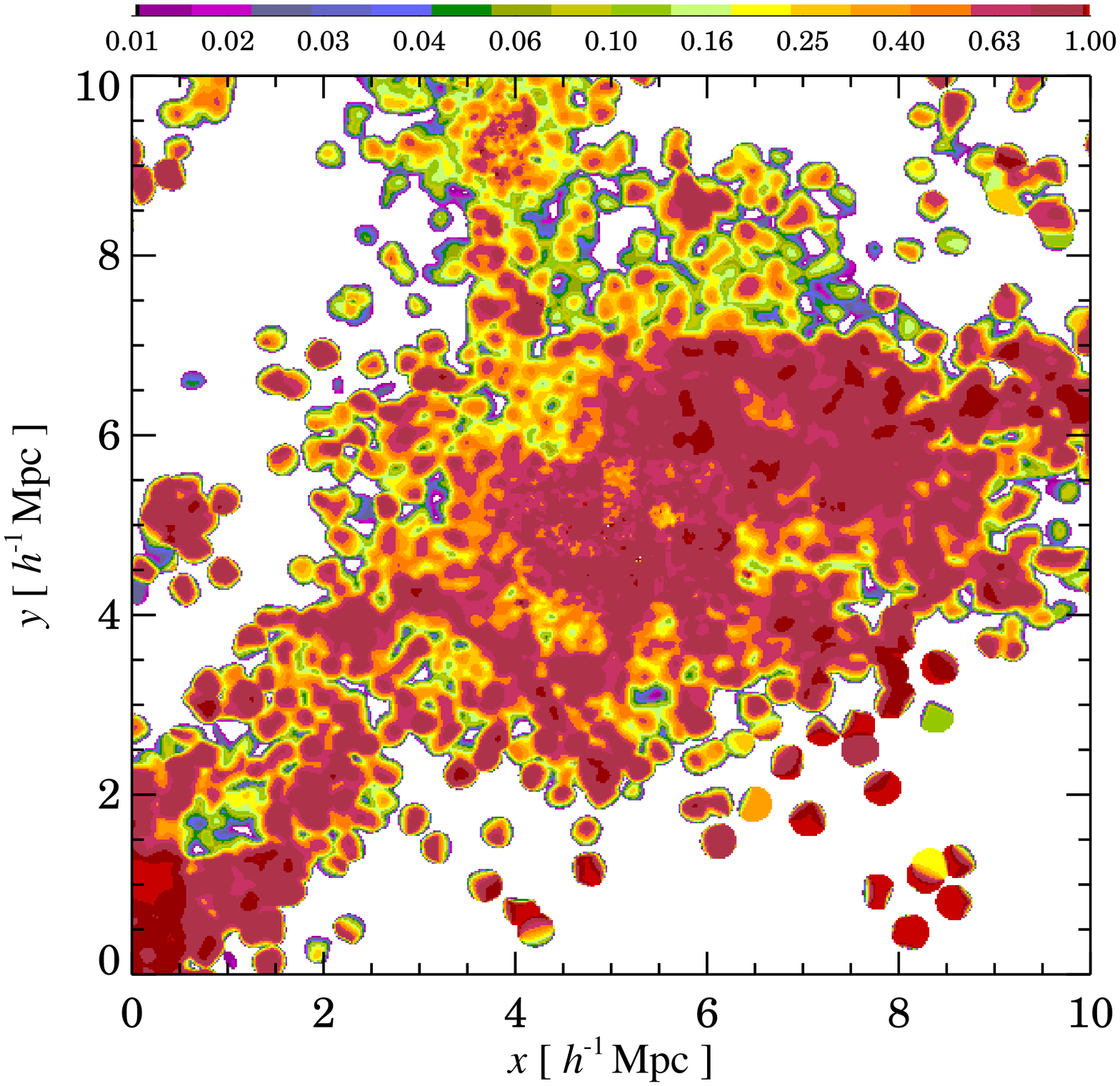} 
}
\hbox{
\includegraphics[width=7.5cm]{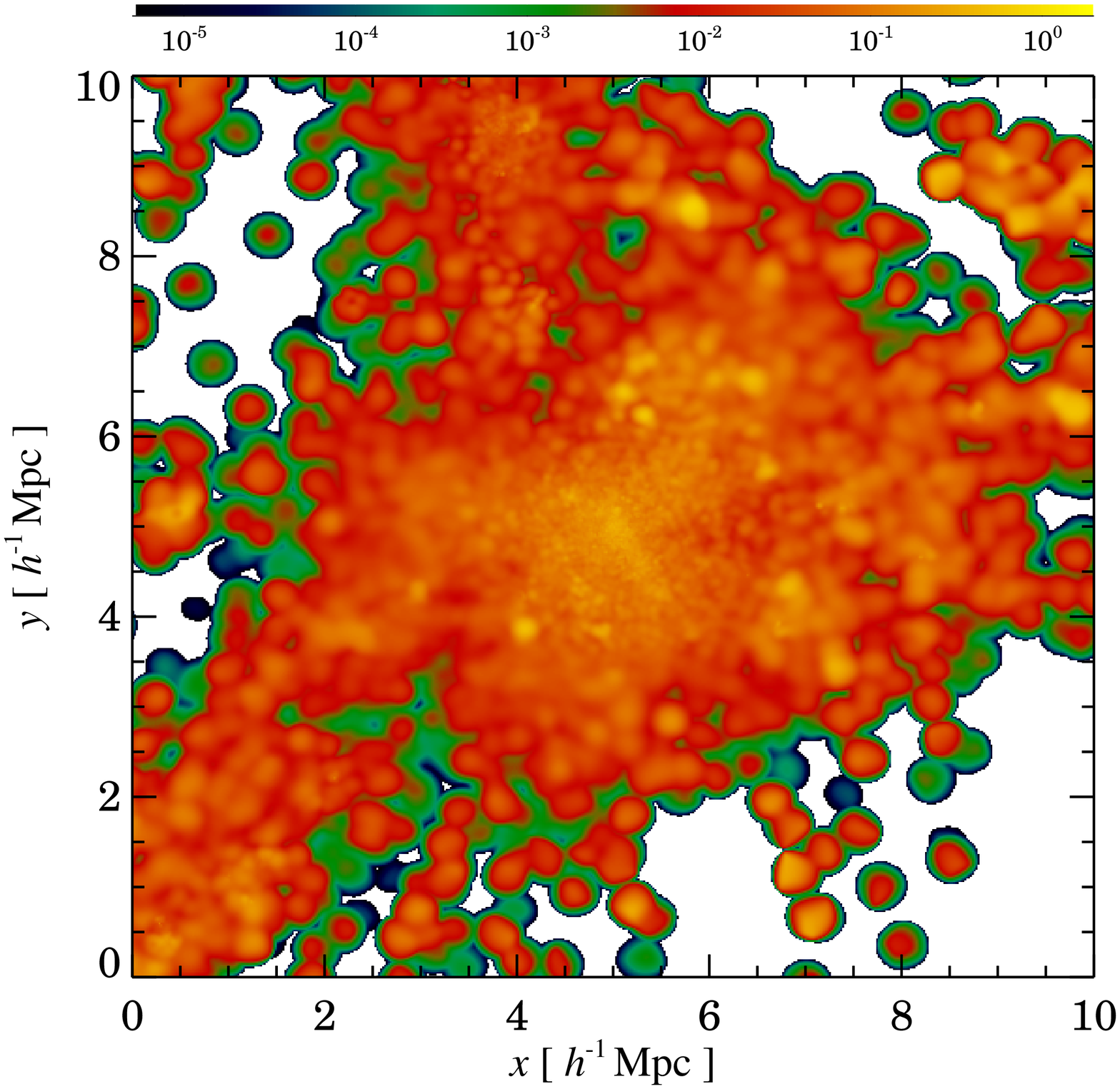} 
\includegraphics[width=7.5cm]{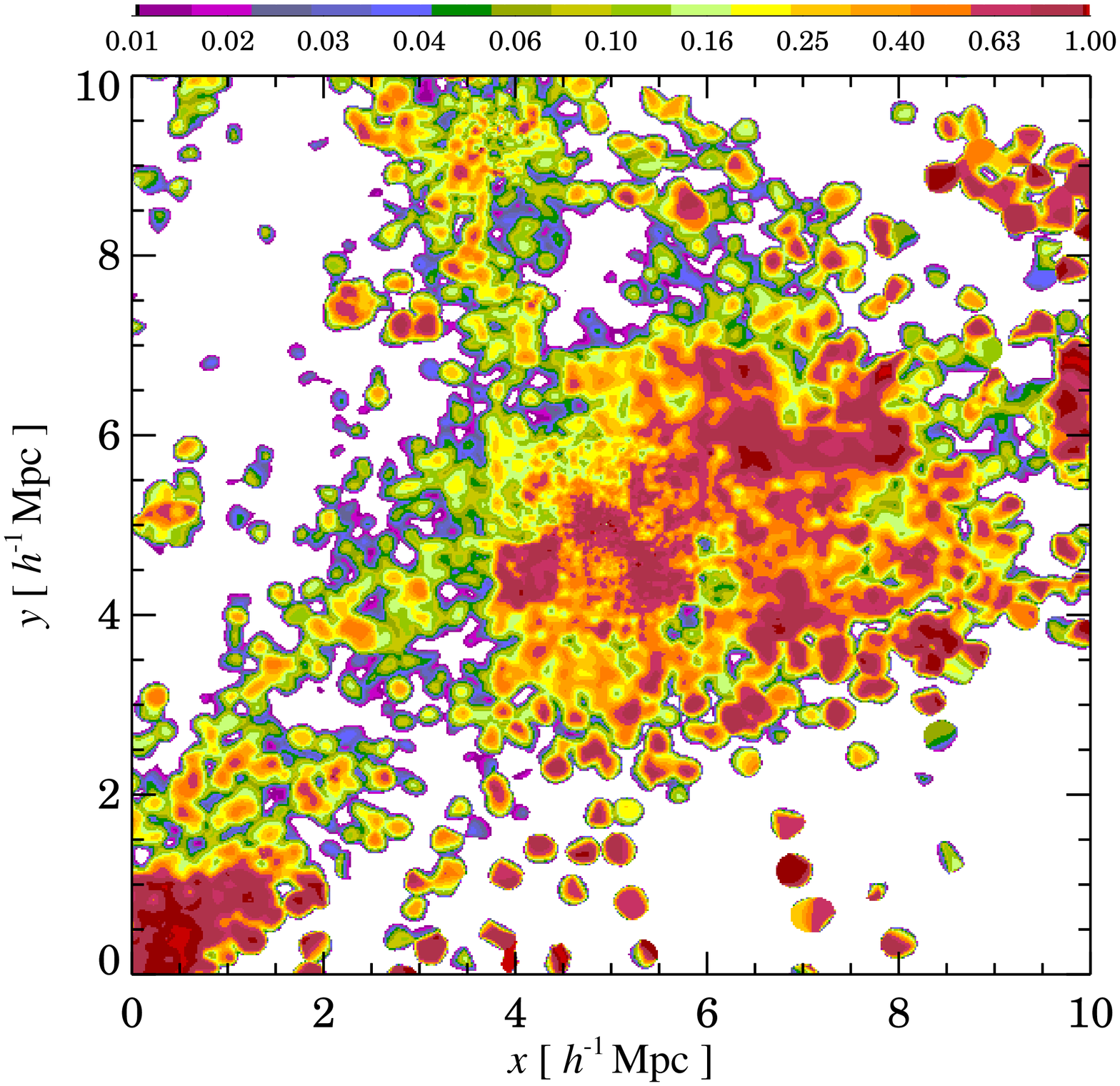}
} 
}}
\caption{Maps of gas metallicity for the three different IMFs. Top,
  central and bottom panels refer to the
  \protect\cite{1955ApJ...121..161S} IMF, to the top--heavy
  \protect\citep{1987A&A...173...23A} IMF and to the
  \protect\cite{2001MNRAS.322..231K} IMF, respectively. The left
  panels are for the Iron abundance, while the right panels show the
  fractional contribution of SNII to the global metallicity. For
  reference, the virial region of the cluster is marked with the white
  square. The maps have been done by smoothing the contribution of
  each gas particle with a SPH kernel using the corresponding
  smoothing length. The projection has been done through a slice of
  $1\hm$ thickness, around the centre of the cluster.}
\label{fi:maps_imf} 
\end{figure*}

Owing to the present limitations in a fully self--consistent numerical
description of such processes, we prefer to adopt here a simplified
numerical scheme to describe the diffusion of metals away from star
forming regions. We will then discuss the effect of modifying this
scheme, so as to judge the robustness of the results against the
uncertainties in the modeling of the metal transport and diffusion.

Every time a star particle evolves, the stars of its population, the
ejected heavy elements and energy must be distributed over the
surrounding gas. The code accomplishes this task by searching for a
number $N_g$ of gas neighbours and then using a kernel to spread
metals and energy according to the relative weights that the kernel
evaluation assigns to each gas particle. In this way, the fraction of
metals assigned to the $i$--th neighbour gas particle can be written as
\be
w_i\, =\, \frac{W_{i}}{\sum_{j=1}^{N_{nb}} W_j}\,.
\ee
where $W_i$ is kernel value at the position of the $i$-th particle
$N_{nb}$ is the number of neighbors over which metals are distributed
and the denominator enforce conservation of the mass in metals.
\begin{figure*}
\centerline{
\hbox{
\includegraphics[width=8.5cm]{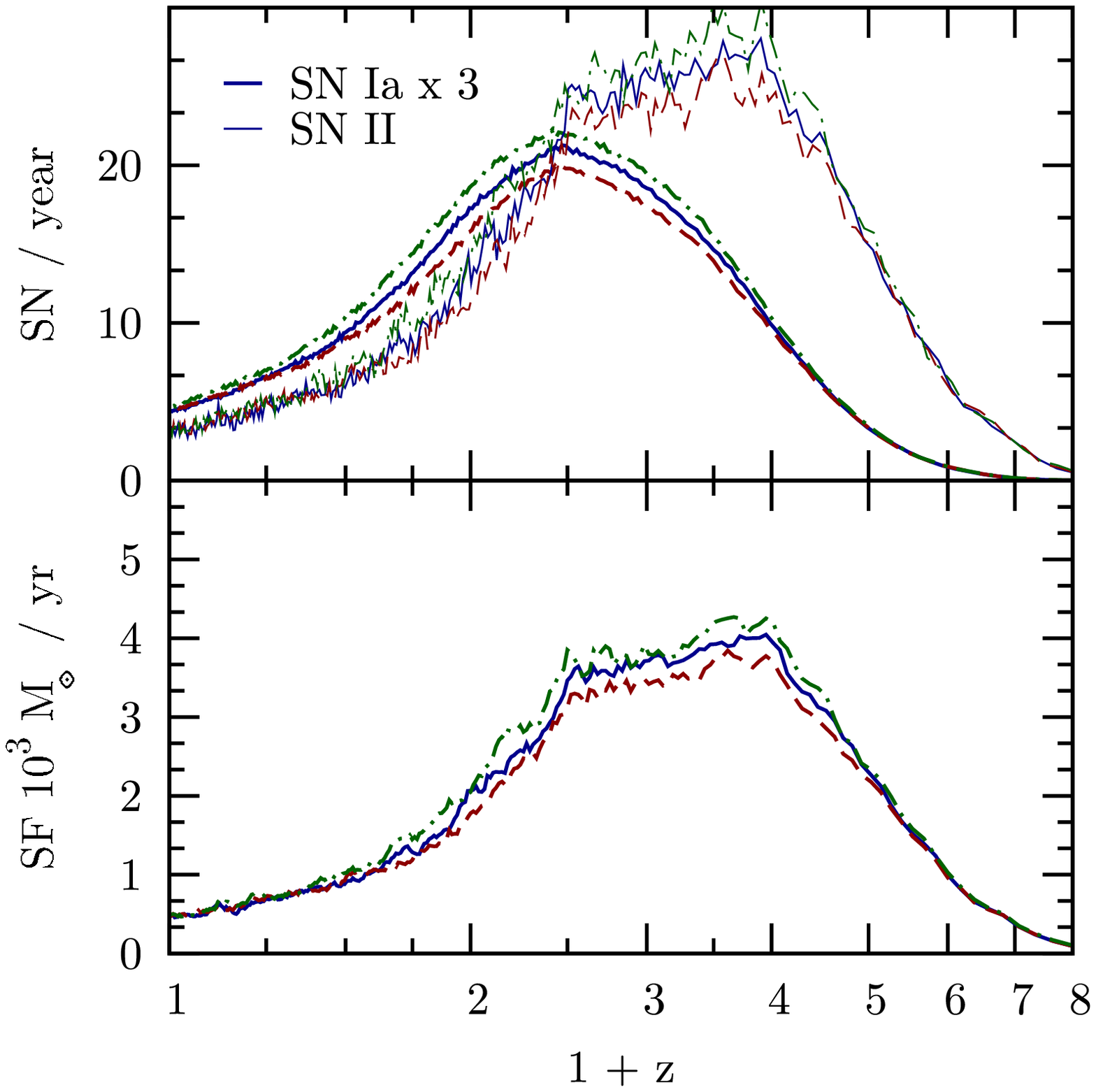}
\includegraphics[width=8.5cm]{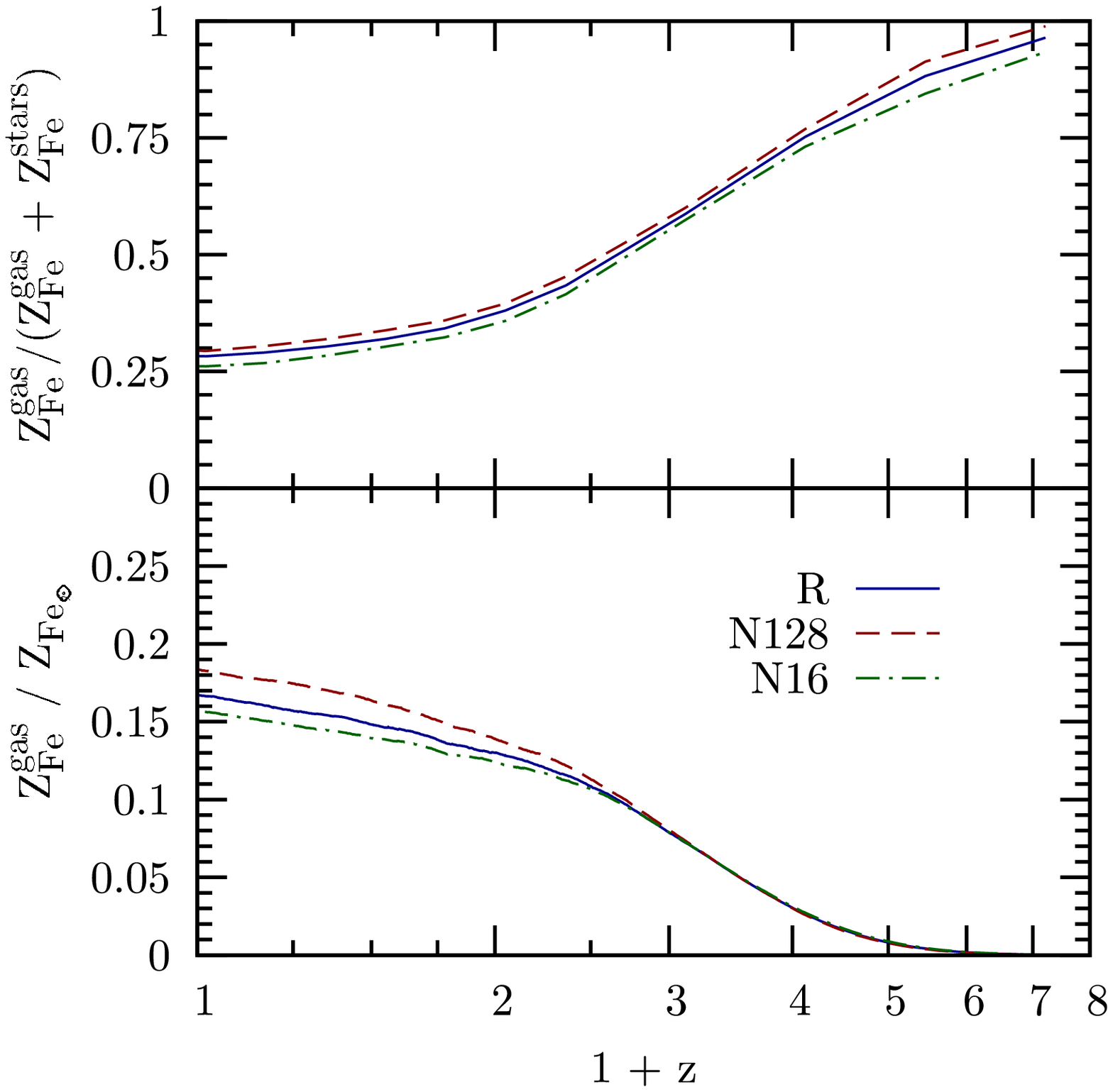} }}
\caption{Left panel: the effect of changing the number of neighbours
  for metal spreading on the star formation rate (bottom) and on the
  SN rates (top; heavy and solid lines are for SNIa and SNII,
  respectively). Right panel: the enrichment history of Iron. The
  bottom panel shows the evolution of the mean Iron abundance in the
  gas; the top panel is for the evolution of the fraction of the total
  mass of Iron which is in the diffuse gas, the rest being locked in
  stars.  In all panels, solid, dashed and dot-dashed curves are for
  the reference run (R), for the runs using 128 (N128) and 16 (N16)
  neighbors for metal spreading, respectively, when using the B-spline
  kernel with density weighting.}
\label{fi:enr_ng}
\end{figure*}
As for the weighting function $W$, the most natural choice, which
is the one usually adopted in the SPH chemo--dynamical models so far
presented in the literature, is to use the same B-spline kernel used
for the computation of the hydrodynamical quantities, also using the
same number of neighbors \citep[e.g., ][]{2001MNRAS.325...34M}. Since
it is a peaked function that rapidly declines to zero, it seems
suitable to mimic what would be the expected behaviour of the metal
deposition process. Nevertheless, it may be argued that the stochastic
nature of the star formation algorithm, that samples the underlying
``real'' star formation on a discrete volume, should require weighting
equally all the volume surrounding each stars particle. In order to
judge the sensitivity of the resulting enrichment pattern on the
choice of the weighting kernel, we will discuss in the following the
effects of using instead a top--hat filter.

Once the functional form of $W$ is defined, one should choose the
physical quantity with respect to which the weight is assigned. A
first possibility is to weight according to the mass $m_g$ of each gas
particle, in such a way that more massive particles receive a
relatively larger amount of metals. An alternative choice is to weight
instead according to the SPH volume carried by each particle,
i.e. using the quantity $m_g/\rho_g$, where $\rho_g$ is the gas density
associated to a particle. With this choice one gives more weight to
those particles which sample a larger volume and, therefore, collect a
larger amount of metals, under the assumption that they are released
in a isotropic way by each star particle. Therefore, one expects that
weighting according to the volume assigns more metals to
gas particles which are relatively more distant from the star forming
regions and, therefore, at lower density.

In the following, we will assume in our reference run that the
spreading of metals is performed with the spline kernel, by weighting
over 64 neighbors according to the volume of each particle. In order
to check the stability of the results, we also modified the scheme for
metal and energy spreading in the following ways: {\bf (i)} use the
same kernel and density weighting, but with 16 and 128 neighbours
(N16 and N128 runs, respectively); {\bf (ii)} weight according to the
mass of the gas particle, instead of according to its volume, using
$N_{nb}=64$ (MW run); {\bf (iii)} use a top--hat window encompassing 64
neighbors, weighting by mass (TH run).

Figure \ref{fi:glob_num} shows the global properties of the simulated
cluster at $z=0$, in terms of amount of stars produced, mass--weighted
ICM metallicity and fraction of metals in the gas, for the different
numerical schemes used to distribute metals. In general, we note that
changing the details of the metal spreading has only a modest effect
on the overall stellar population and level of enrichment. We find
that the amount of stars within the cluster virial region ranges
between 23 and 26 per cent of the total baryon budget. As for the
global ICM enrichment, it is $Z_{Fe}\simeq (0.20-0.25)Z_{Fe,\odot}$
and $Z_{O}\simeq 0.15 Z_{O,\odot}$ for the mass--weighted Iron and
Oxygen abundance, respectively. Quite interestingly, only about a
quarter of the total produced Iron is in the diffuse hot gas, while
this fraction decreases to about 15 per cent for Oxygen. This
different behaviour of Oxygen and Iron can be explained on the ground
of the different life--times of the stars which provide the dominant
contribution to the production of these elements. Since Oxygen is
produced by short--living stars, it is generally released in star
forming regions and therefore likely to be distributed among
star--forming gas particles. For this reason, Oxygen has a larger
probability to be promptly locked back into stars. On the other hand,
Iron is largely contributed by long--living stars. Therefore, by the
time it is released, the condition of star formation around the parent
star particle may be significantly changed. This may happen both as a
consequence of the suppression of star formation inside galaxies or
because the star particle may have later taken part to the diffuse
stellar component \citep[e.g., ][]{2004ApJ...607L..83M}. In both
cases, Iron is more likely to be distributed among non star--forming
particles and, therefore, contributes to the ICM enrichment, instead
of being locked back in stars.

This effect is clearly visible in the right panels of Figure
\ref{fi:maps_imf}, where we show the map of the fractional
contribution of SNII to the global metal enrichment. Clearly, SNII
provide a major contribution (magenta--red color) in high--density
patchy regions, within and around the star forming sites. On the
contrary, the contribution from SNIa and AGB dominates in the diffuse
medium. This is a typical example of how the pattern of the ICM
enrichment is determined by the competing effects of the chemical
evolution model and of the complex dynamics taking place in the dense
environment of galaxy clusters. It confirms that a detailed study of
the chemical enrichment of the diffuse gas indeed requires a correct
accounting of such dynamical effects. We find that the contribution in
Oxygen within the cluster virial radius from SNII, from SNIa and from
low-- and intermediate mass stars is of about 70, 5 and 25 per cent
respectively, while that in Iron is 25, 70 and 5 per cent. This
demonstrates that none of these three sources of metals can be
neglected in a detailed modelling of the chemical enrichment of the
ICM. As shown in the left panels of Fig. \ref{fi:maps_imf}, the
distribution of Iron generally follows the global large--scale
structure of gas inside and around the cluster, with an excess of
enrichment inside the virial region and along the filaments from which
the cluster accrete pre--enriched gas. We will comment in Sect.
\ref{s:res_imf} the dependence of this enrichment pattern on the IMF.

Quite interestingly, \cite{2003AJ....125.1087G} discovered two SNIa
not associated to cluster galaxies in Virgo and argued that up to
about 30 per cent of the SNIa parent stellar population is
intergalactic \citep[see also ][]{2005ApJ...632..847M}. This is
exactly the SNIa population that in our simulations is responsible for
the diffuse enrichment in Iron. As the statistics of the observed
population of intergalactic SNIa population improves, it will be
interesting to compare it with the predictions of our simulations and
to better quantify their contribution to the ICM enrichment.

\begin{figure}
\centerline{
\includegraphics[width=8.5cm]{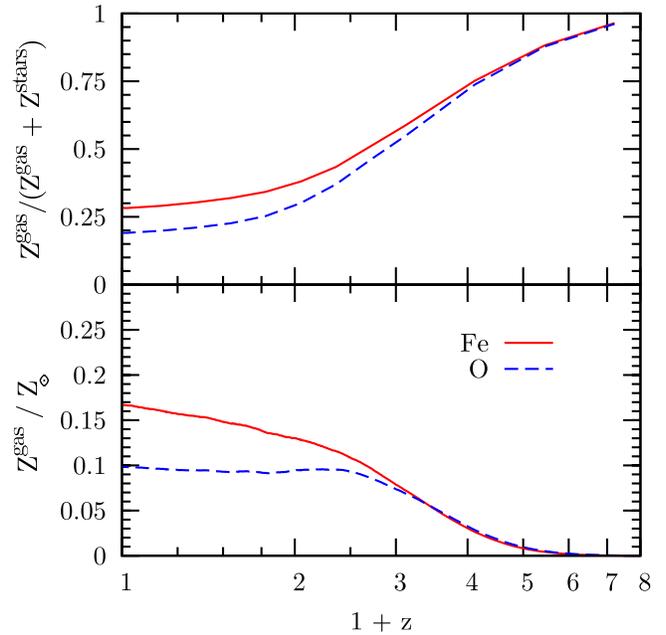} }
\caption{The evolution of enrichment for the reference (R) run in
  both Iron (solid curves) and Oxygen (dashed curves).  The bottom
  panel shows the evolution of the mean abundances in the gas, while
  the top panel is for the evolution of the fraction of the total
  metal mass which is in the diffuse gas, the rest being locked in
  stars.}
\label{fi:enr_ref}
\end{figure}

Figure \ref{fi:enr_ng} shows how the history of enrichment and of star
formation changes by changing the scheme for distributing metals. In
the left panel, the star--formation rate (SFR) is compared to the rate
of SN explosions. As expected, the SNII rate follows quite closely the
SFR (see left panel), and peaks at $z\simeq 2$--3. On the other hand,
the rate of SNIa is substantially delayed, since they arise from stars
with longer life--times. Therefore, this rate peaks at $z\simeq 1.5$
with a slower decline at low redshift than for SNII. Since the rate of
SNIa is given by the combined action of SFR, life--time function and
IMF, computing it by adding a constant time--delay to the SFR is
generally too crude and inaccurate an approximation. As shown in
Figure \ref{fi:enr_ref}, the different redshift dependence of the SNIa
and SNII rates is reflected by the different histories of enrichment
in Iron and Oxygen. Since Oxygen is only contributed by short--living
stars, the corresponding gas enrichment stops at $z\mincir 1$, when
the SFR also declines. On the contrary, the gas enrichment in Iron
keeps increasing until $z=0$, as a consequence of the significant SNIa
rate at low redshift. An interesting implication of
Fig.\ref{fi:enr_ref} is that the relative abundance of Oxygen and Iron
in the ICM can be different from that implied by the stellar yields.
Therefore, the commonly adopted procedure to compare X--ray
measurements of relative ICM abundances to stellar yields \citep[e.g.,
][]{1999MNRAS.305..325F,2005ApJ...620..680B} may lead to biased
estimates of the relative contributions from SNIa and SNII
\cite[e.g.,][]{2005PASA...22..49}. We finally note that observations
of SNIa rates in clusters indicates fairly low rates, for both nearby
and distant objects \citep[e.g., ][]{2002MNRAS.332...37G}, also
consistent with the rates observed in the field. We postpone to a
future analysis a detailed comparison between the observed and the
simulated SNIa rates, and the implications on the parameters defining
the model of chemical enrichment.

\begin{figure*}
\centerline{
\hbox{
\includegraphics[width=8.5cm]{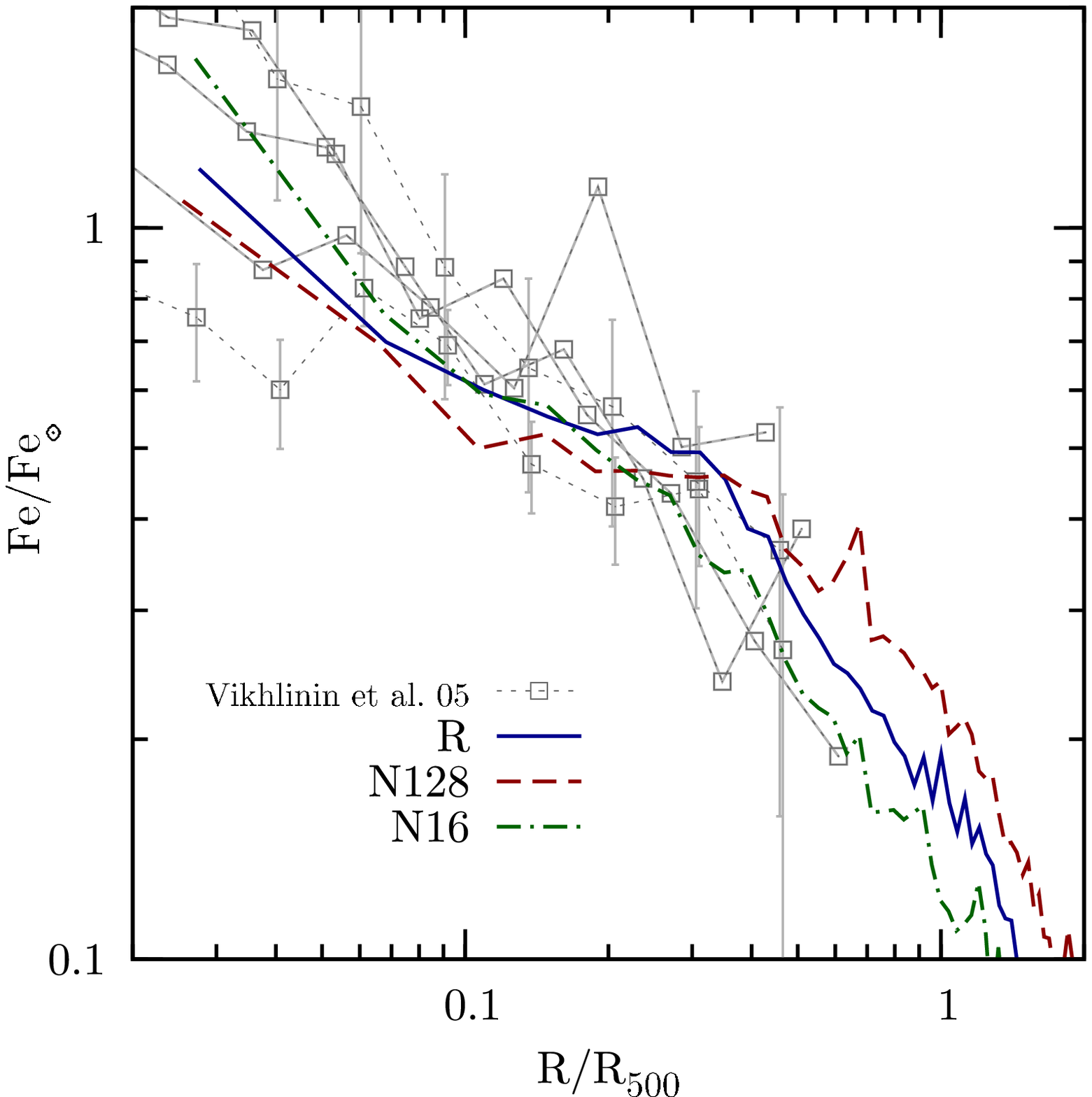} 
\includegraphics[width=8.5cm]{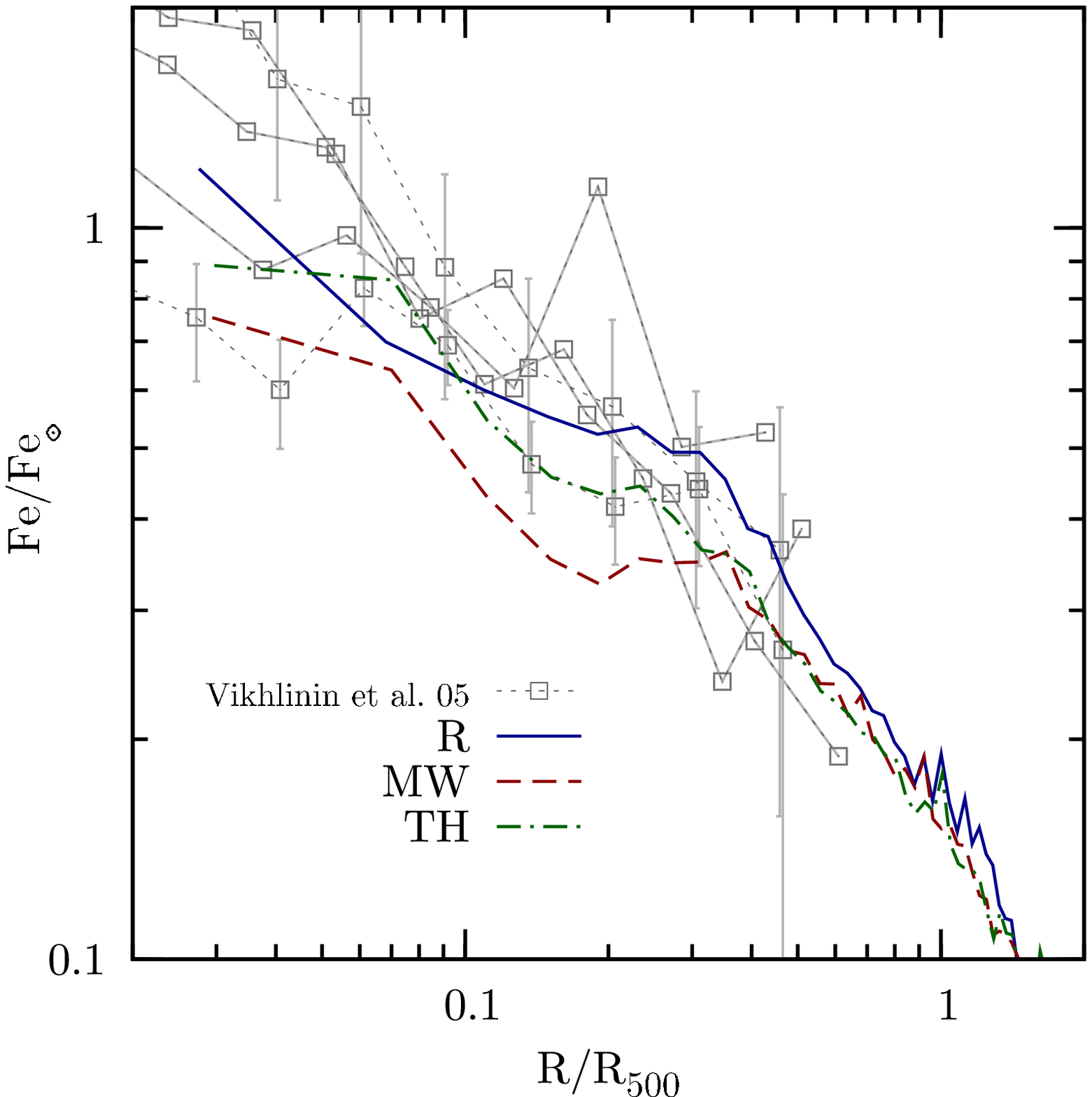} 
}}
\caption{The effect of changing the metal spreading on the Iron
  abundance profiles. Left panel: comparison between the reference run
  (solid curve) and runs done by changing the number of neighbours to
  16 (N16; short--dashed line) and to 128 (N128; long--dashed line).
  Right panel: comparison between the reference run (solid curve) 
  and the runs in which metal spreading is
  performed using mass--weighting with the SPH kernel (MW; short--dashed
  line) and the top--hat kernel (TH; long--dashed line). 
%comparison between reference run (solid curve) and runs
%  done by increasing the number of stellar generations to 12 (Ng12;
%  short--dashed line). 
The data points refer to a subset of four
  clusters, out of 16 observed with Chandra ana analysed by
  \protect\cite{2005ApJ...628..655V}, which have temperature in the range 2--4
  keV. For reasons of clarity, errorbars are not reported. }
\label{fi:prof_ng}
\end{figure*}

In general, using a different number of neighbours to spread metals
has only a minor impact on the history of star formation and
enrichment (see Fig. \ref{fi:enr_ng}). We only note that increasing
the number of neighbors turns into a slight increase of the SFR at all
redshifts, with a corresponding slight increase of the metallicity. In
fact, increasing the number of neighbours has the effect of
distributing metals more uniformly among gas particles. This causes a
larger number of particles to have a more efficient cooling and,
therefore, to become eligible for star formation.

%\begin{figure}
%\centerline{
%%\hbox{
%\psfig{file=Figs/Profiles/Metals/z_0.00_prof_Fe_numerics.w.eps,width=8.5cm} 
%%\psfig{file=Figs/Profiles/Metals/z_0.00_prof_O-Si2Fe_numerics.w.eps,width=8.5cm}}
%} 
%\caption{The effect of changing the weighting scheme in the metal
%  spreading on the profile of the
%  mass--weighted Iron abundance, out to $R_{500}$, for the reference
%  run (R; solid line) and for the runs in which metal spreading is
%  done using mass--weighting with the SPH kernel (MW; short--dashed
%  line) and using the top--hat kernel (TH; long--dashed line). The
%  observational data points in the left panel are the same as in
%  Figure \ref{fi:prof_ng}.}
%\label{fi:prof_w}
%\end{figure}

In the left panel of Figure \ref{fi:prof_ng} we show the effect of
changing the number of neighbors over which to distribute metals on
the profile of the Iron abundance. As expected, increasing the number
of neighbors corresponds to an increasing efficiency in distributing
metals outside star forming regions. As a result, metallicity profiles
becomes progressively shallower, with a decrease in the central
regions and an increase in the outer regions. Although this effect is
not large, it confirms the relevance of understanding the details of
the mechanisms which determine the transport and diffusion of the
metals.

As for the comparison with data, we note that the differences between
the different weighting schemes are generally smaller than the
cluster-by-cluster variations of the observed abundance gradients. In
general, the simulated profiles are in reasonable agreement with the
observed ones. 
%
%Still, simulated profiles tend to steepen at $R\magcir 0.3R_{500}$,
%while at these scales the profiles measured by
%\cite{2005ApJ...628..655V} tend to flatten, consistent also with the
%Beppo--SAX results by \cite{2004A&A...419....7D}.
%
A more detailed comparison with observed abundance profiles will be
performed in a forthcoming paper, based on a larger set of simulated
clusters (Fabjan et al. in preparation).
 
%In the right panel of Fig. \ref{fi:prof_ng} we show the effect of
%increasing from 3 to 12 the number of generations of stars that a
%star--forming gas particle can spawn. As discussed in
%Sect. \ref{s:NumericalMethod}, increasing this number turns into a
%more accurate description of the process of star formation and,
%therefore, to a more accurate representation of the process of
%chemical enrichment. Increasing the number of stellar generations
%from 3 to 12 causes a significant overhead in the computational cost of the
%simulation, which increases by about a factor two. While the overhead
%associated to the computation of gravity is negligible, most of it is
%due to the much larger number of stars for which the equations of
%chemical evolution are solved. As for the profile of Iron abundance,
%the effect is rather small at all scales, with a marginal tendency to
%produce a flatter profile only at the smallest resolved radii. 
%%
%%This result demonstrates that the overall enrichment pattern is left
%%largely unaffected by the accuracy of the star formation description.
%%
%This conclusion is also supported by the stability of the
%global star fraction and enrichment pattern, as shown in
%Fig. \ref{fi:glob_num}.

Finally, we show in the right panel of Figure \ref{fi:prof_ng} the
variation of the Iron abundance profile when changing the weighting
scheme for the distribution of metals, while keeping fixed to 64 the
number of neighbors. As for the Iron profile, using volume, instead of
mass, in the SPH kernel has a rather small effect. Only in the
innermost bin, the Iron abundance increases when weighting according
to the mass as a result of the less effective spreading to less dense
gas particles. As for the top--hat kernel, its Iron profile lies below
the other ones at all radii, although by a rather small amount.

\subsection{The effect of resolution}
\label{s:res}
In this Section we present the results of the simulations of the Cl2
cluster, done at three different resolutions (see Table
\ref{t:clus}). 

\begin{figure*}
\centerline{
\hbox{
\includegraphics[width=5.8cm]{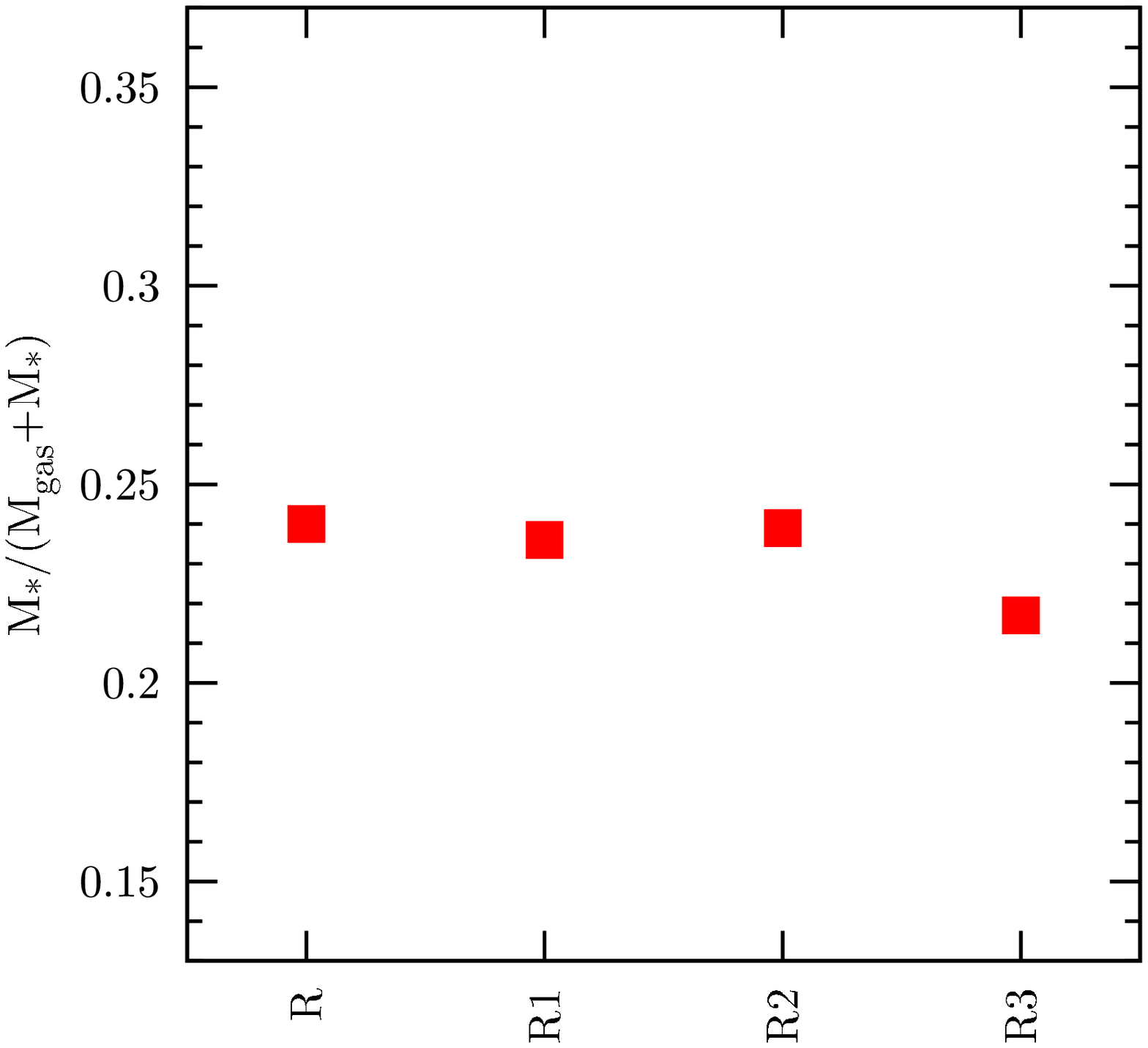} 
\includegraphics[width=5.8cm]{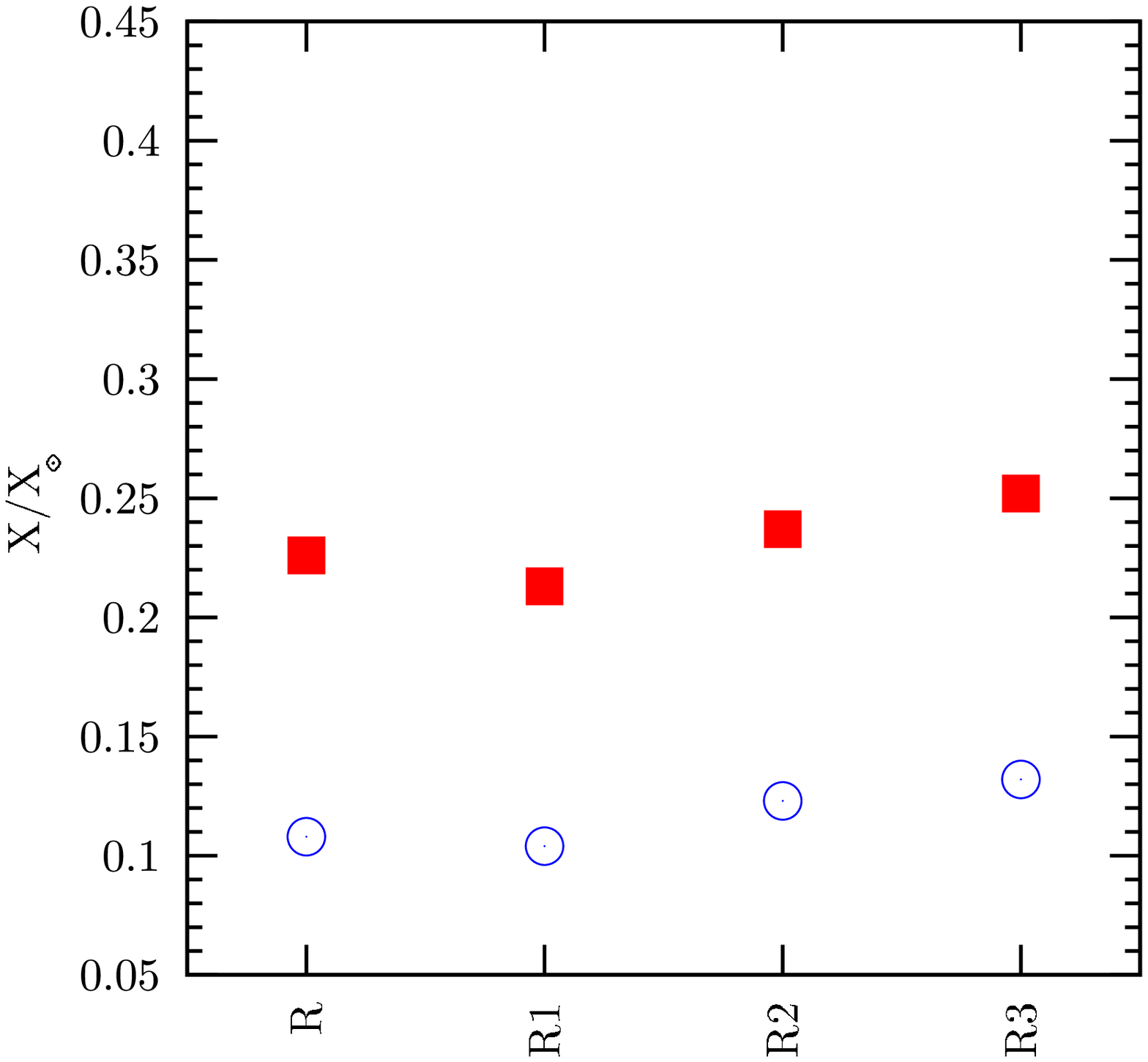} 
\includegraphics[width=5.8cm]{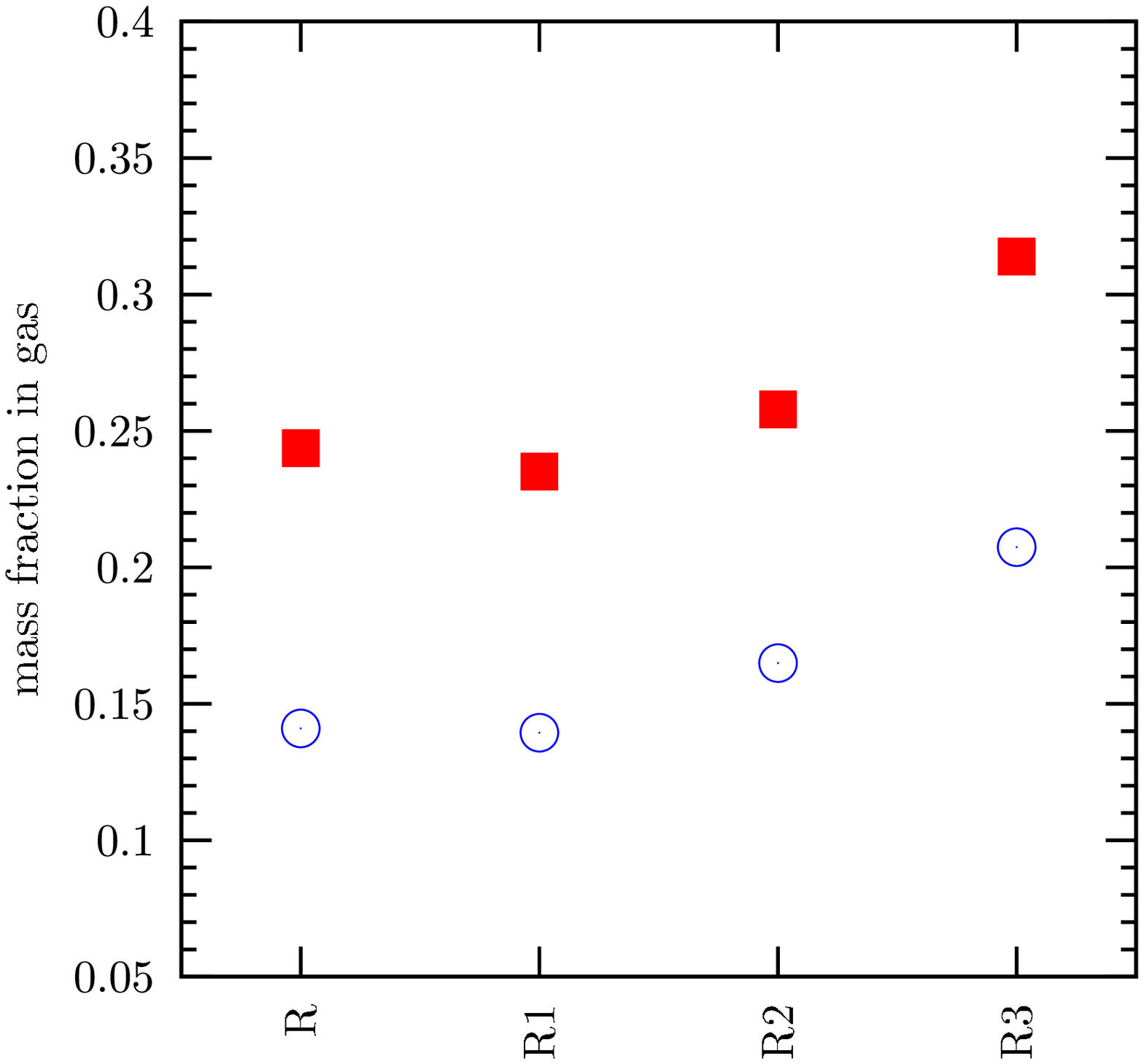} }}
\caption{The same as in Figure \ref{fi:glob_num}, but for the runs of
  the Cl2 cluster at three different resolutions.
  The labels indicating the different runs are as reported
  in Table \ref{t:runs}.}
\label{fi:glob_res} 
\end{figure*}

\begin{figure*}
\centerline{
\hbox{
\includegraphics[width=8.5cm]{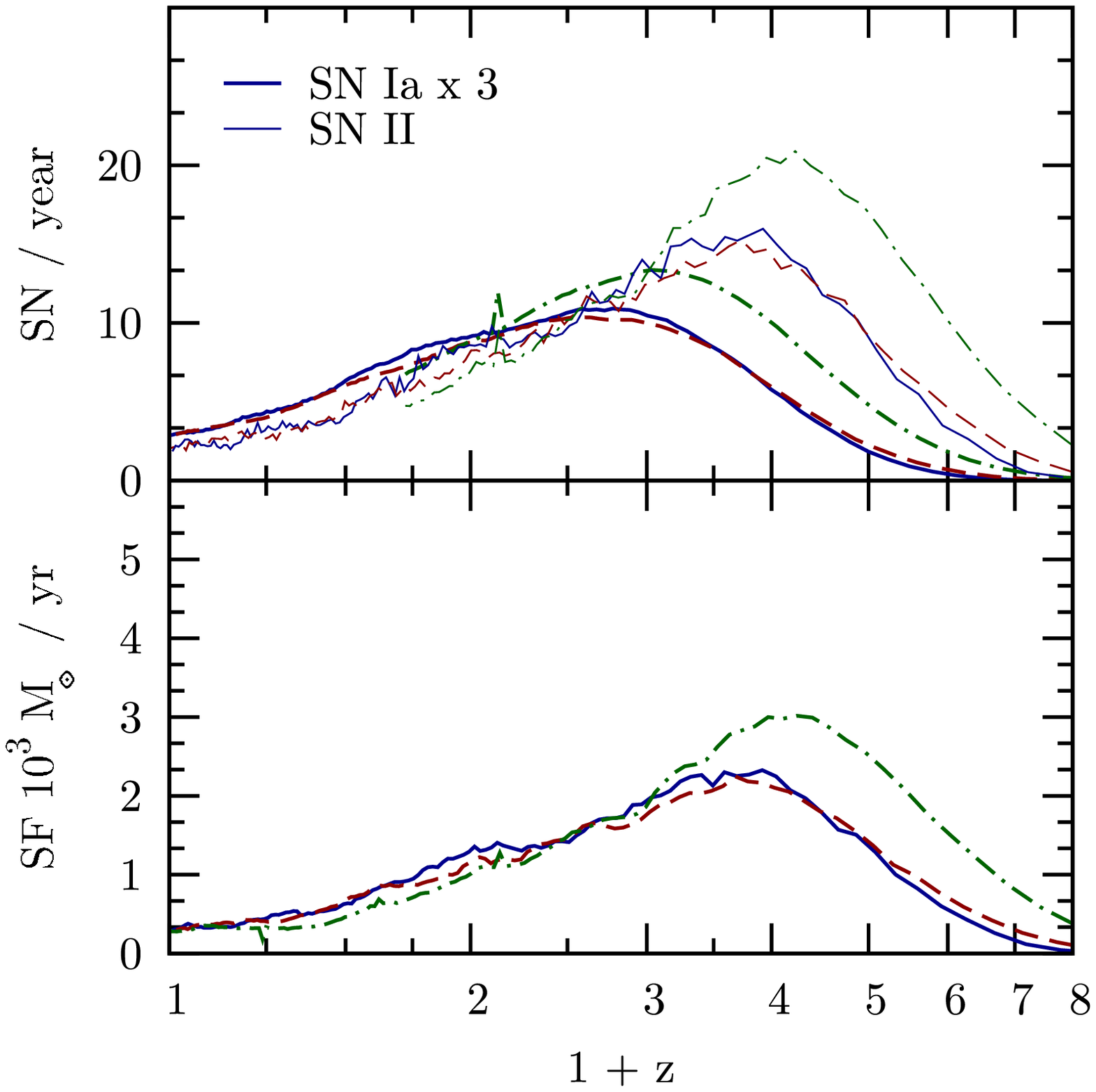} 
\includegraphics[width=8.5cm]{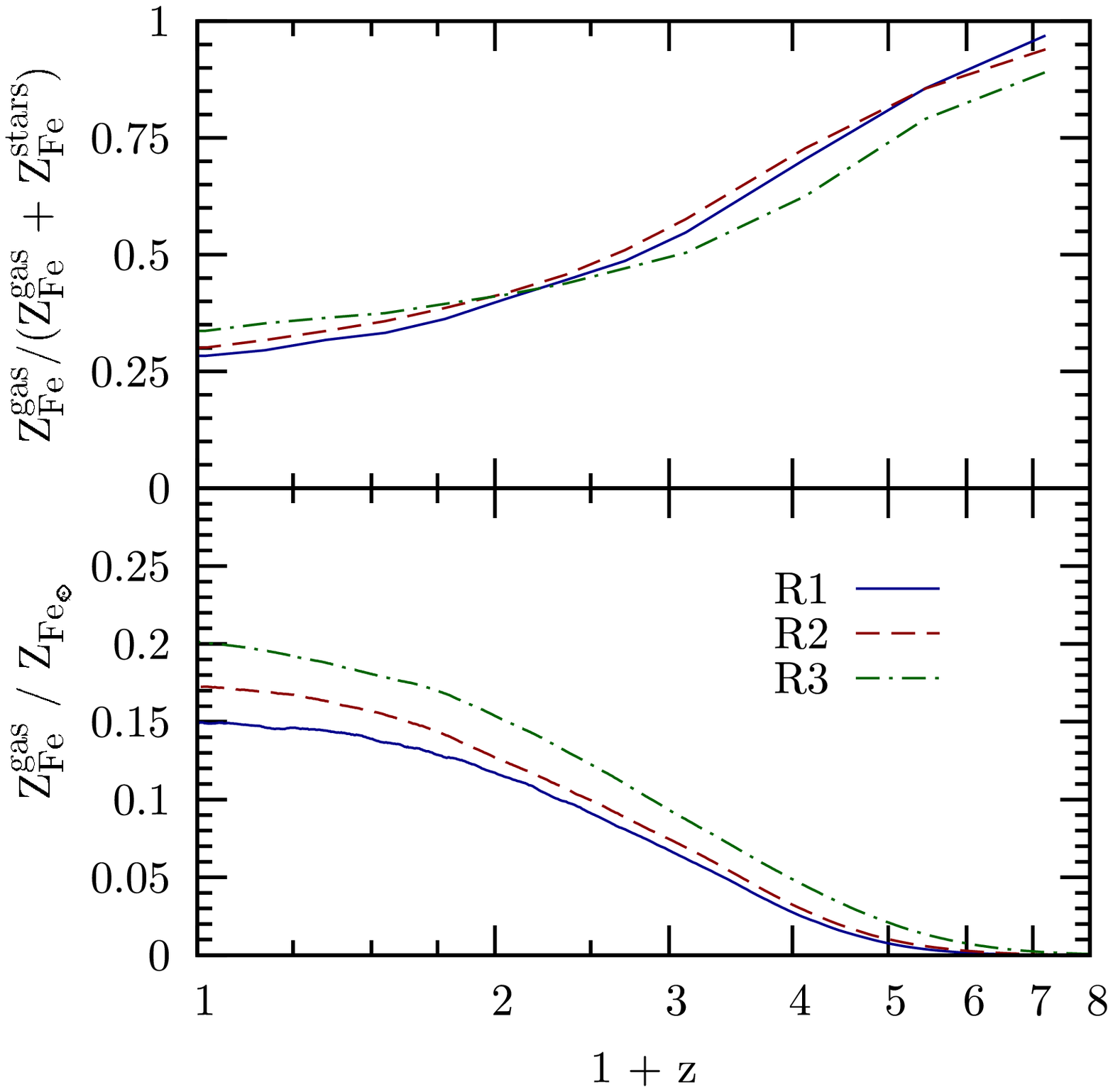} 
}}
\caption{Left panel: the effect of resolution on the star formation
  rate (bottom) and supernova rates (top; tick lines: SNIa; thin
  lines: SNII). Right panel: the effect of resolution on the effect on
  the enrichment history. The upper panel shows the evolution of the
  fraction of Iron in contained in gas, while the lower panel shows
  the evolution of the Iron abundance in the gas. Solid, dashed and
  dot--dashed lines are for the R1, R2 and R3 runs, respectively.}
\label{fi:enr_res} 
\end{figure*}

Figure \ref{fi:enr_res} shows the effect of resolution on the rates of
star formation and SN explosions (left panel) and on the history of
chemical enrichment (right panel). As expected, increasing resolution
enhances the high--redshift tail of star formation, as a consequence
of the larger number of resolved small halos which collapse first and
within which gas cooling occurs efficiently. Quite interestingly, the
increase with resolution of the high--$z$ star formation rate is
compensated by a corresponding decrease at lower, $z\mincir 1$,
redshift. As a net result, the total amount of stars formed within the
cluster virial region by $z=0$ (left panel of Figure
\ref{fi:glob_res}) turns out to be almost independent of
resolution. This results is in line with that already presented by
\cite{2006MNRAS.367.1641B} for a similar set of simulations, but not
including the chemical enrichment. On the one hand, increasing
resolution increase the cooling consumption of gas at high redshift,
thereby leaving a smaller amount of gas for subsequent low--$z$ star
formation. On the other hand, smaller halos, forming at higher
redshift when increasing resolution, generate winds which can more
easily escape the shallow potential wells. As a result, the gas is
pre--heated more efficiently, so as to reduce the later star
formation.

\begin{figure*}
\centerline{
\hbox{
\includegraphics[width=8.5cm]{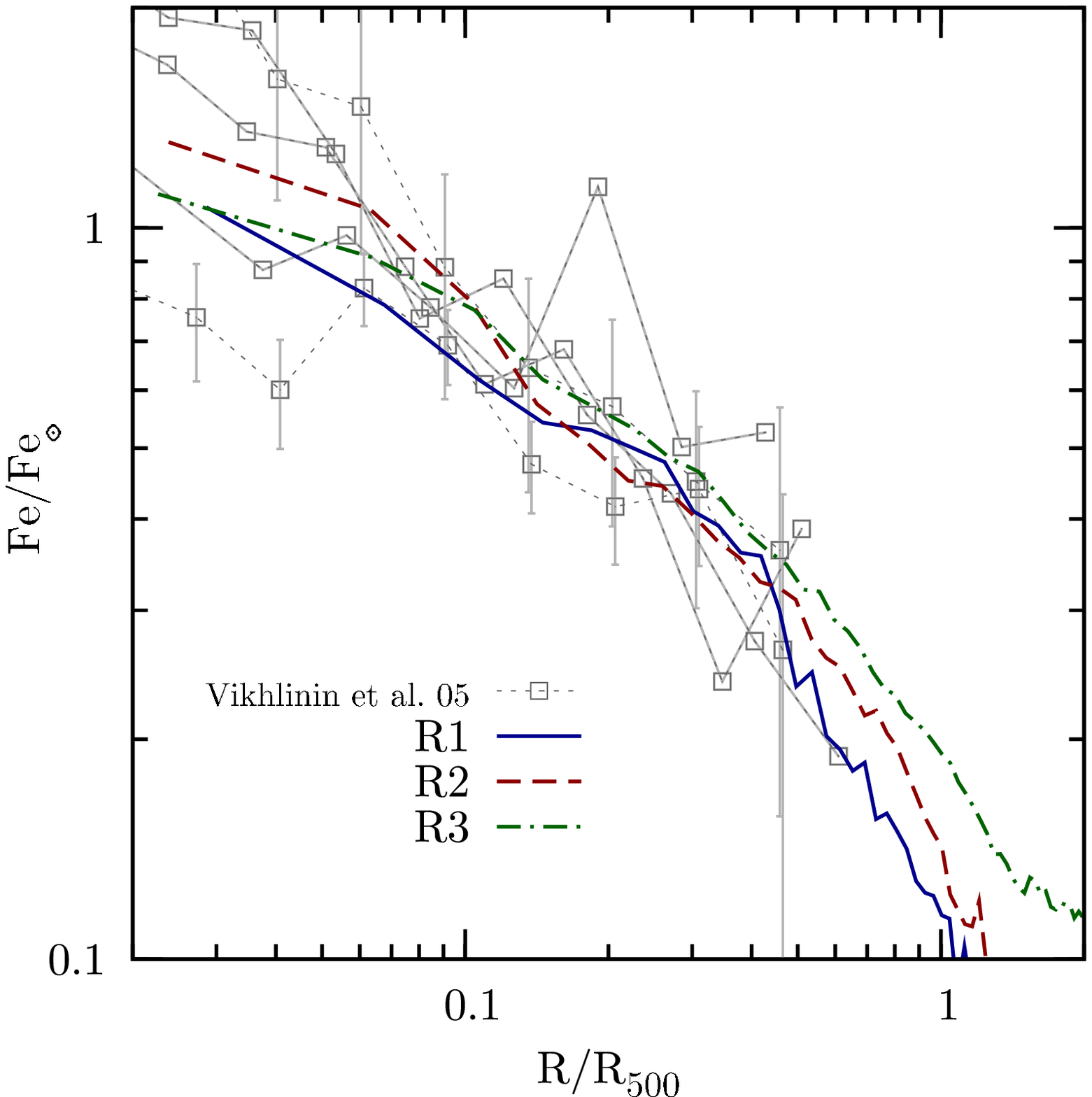} 
\includegraphics[width=8.8cm]{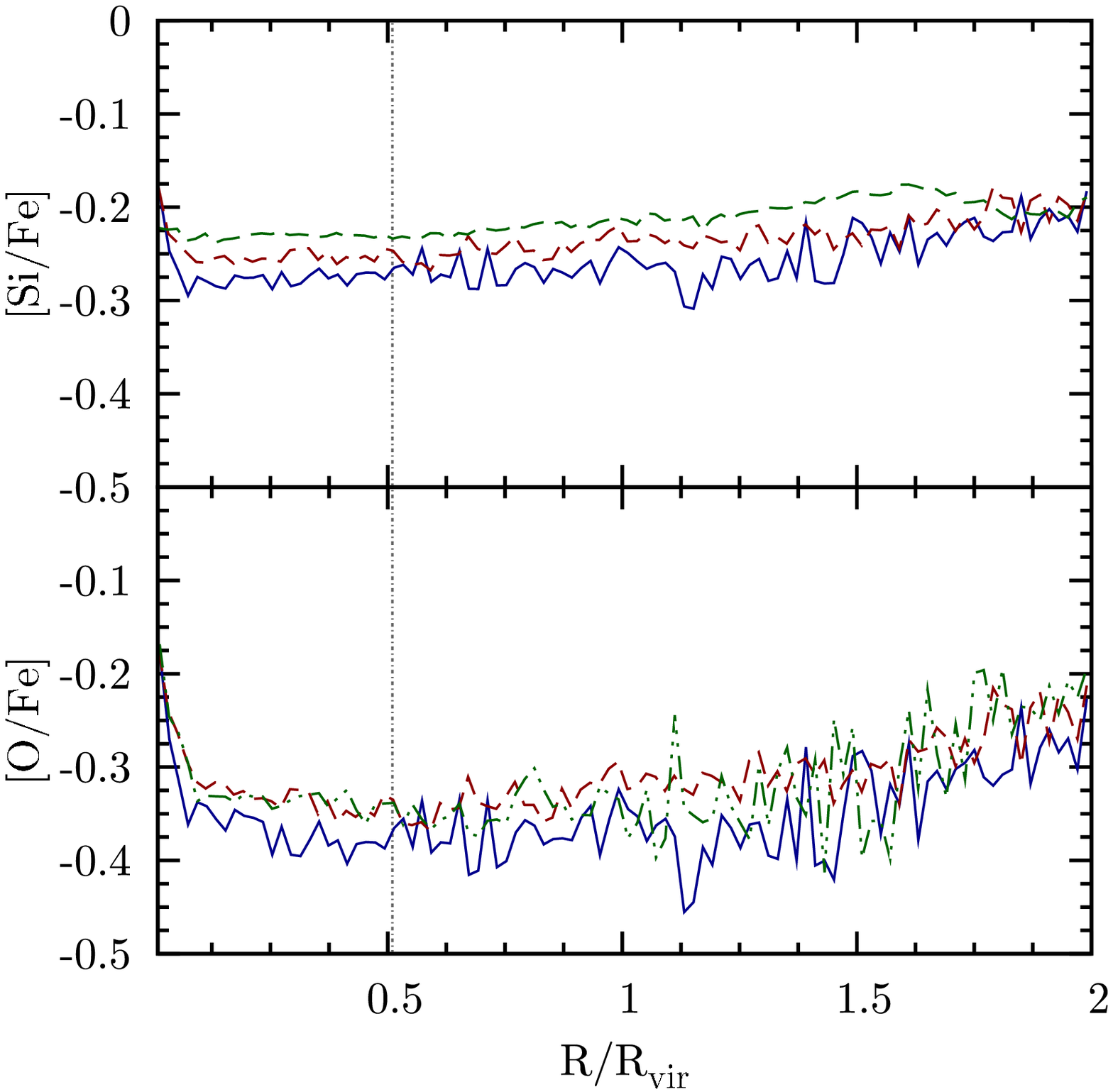} 
}}
\caption{The effect of resolution on the abundance profiles of the Cl2
  cluster. Left panel: profile of the mass--weighted Iron abundance,
  out to $R_{500}$.  The observational data points in the left panel
  are the same as in Figure \ref{fi:prof_ng}. Right panel: the
  profiles of the relative abundance of Silicon (top) and Oxygen
  (bottom) with respect to Iron, plotted out to 2\rvir. The
  meaning of the different line types is the same as in the left panel
  of Figure \ref{fi:prof_res}. The dotted vertical line indicates
  the value of $R_{500}$.}
\label{fi:prof_res} 
\end{figure*}

In spite of the stable star fraction, the level of ICM enrichment,
both in Iron and in Oxygen (central panel of Fig. \ref{fi:glob_res})
increases with resolution. The reason for this is the larger fraction
of metals which are found in the diffuse gas at increasing resolution
(left panel of Fig.\ref{fi:glob_res}). The fact that increasing
resolution corresponds to a more efficient distribution of metals is
also confirmed by the behaviour of the Iron abundance profiles
(left panel of Figure \ref{fi:prof_res}), which become systematically
shallower at large radii, $R\magcir 0.5R_{500}$. The reason for this
more efficient spreading of metals from star--forming regions has two 
different origins.
First of all, by increasing resolution one better resolves processes, like
``turbulent'' gas motions and ram--pressure stripping, which are
effective in diffusing metals away from the star forming regions, thus
preventing them to be locked back in stars. Furthermore, the
better--resolved and more wide--spread star formation at high redshift
releases a larger fraction of metals in relatively shallower potential
wells. Therefore, galactic winds are more efficient in distributing
metals in the IGM.
% In order to decide which one of these two mechanisms is the leading
% one, one can verify whether the extra amount of metals produced at
% high redshift when resolution is increased is enough to account for
% the increasing level of metallicity at large cluster-centric
% distances.

% We note from the left panel of Figure \ref{fi:enr_res} that both the
% SFR and the Sn rate at high redshift increase with resolution. This
% is expected, since the improved resolution allows to resolved
% smaller halos at higher redshift, where cooling takes place with
% increased efficiency. At the same time, galactic winds associated
% with this pristine star formation can easely escate the potential
% wells of the small halos, thereby providing an efficient mean to
% heat the IGM. In turn, this heating prevents the low--redshift star
% formation.  This is the reason why the low--$z$ star formation
% slighly decreases with resolution.
As for the history of enrichment (right panel of Figure
\ref{fi:enr_res}), we note that increasing resolution has the effect
of progressively increasing the gas Iron abundance at all
redshifts. While the overall effect is of about $\simeq 30$ per cent
at $z=0$ it is by a factor of 2 or more at $z\magcir 4$. This
illustrates that, while resolution has a modest, though sizable,
effect at low redshift, it must be increased by a substantial factor
to follow the enrichment process of the high--redshift IGM. As for the
evolution of the fraction of Iron in gas, we note that it decreases
with resolution at high redshift, while increasing at low
redshift. The high--$z$ behaviour is consistent with the increasing
star--formation efficiency with resolution, which locks back to stars
a larger fraction of metals. On the other hand, at low redshift this
trend is inverted (see also the right panel of
Fig.\ref{fi:glob_res}). This transition takes place at about the same
redshift, $z\sim 2.5$, at which the star formation rate of the R3 run
drops below that of the R1 runs, thus confirming the link between
star--formation efficiency and locking of metals in stars.

\begin{figure}
\centerline{
\includegraphics[width=8.5cm]{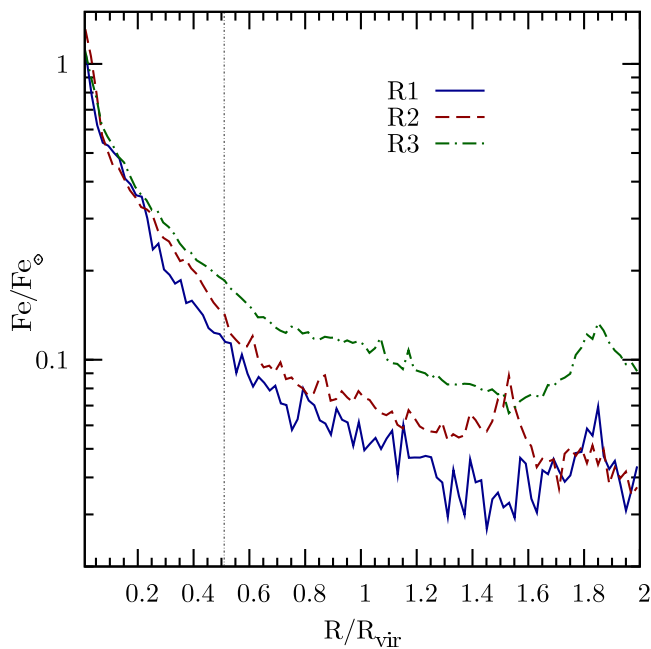} 
}
\caption{The effect of resolution on the profiles or Iron abundance,
  out to 2\rvir. Linear scales are used for the radius, to
  emphasize the behaviour of the profiles in the cluster outskirts.}
\label{fi:proflin_res} 
\end{figure}

An increased efficiency with resolution in distributing metals in the
diffuse medium is also confirmed by the the Iron abundance profile
(left panel of Figure \ref{fi:prof_res}). While we do not detect any
obvious trend at small radii, $R\mincir 0.2R_{500}$, there is a clear
trend for profiles to be become shallower at larger radii as
resolution increases. In order to better show this effect, we plot in
Figure \ref{fi:proflin_res} the Iron abundance profile out to 2\rvir,
by using linear scales for the cluster--centric distance. This allows
us to emphasize the regime where the transition from the ICM to the
high--density Warm-Hot Intergalactic Medium (WHIM) takes place
\citep[e.g., ][]{2006ApJ...650..560C}. Quite interestingly, the effect
of resolution becomes more apparent in the outskirts of clusters.
Behind the virial radius, the abundance of Iron increases by 50 per
cent from the low--resolution (LR) to the high--resolution (HR) runs.
In these regions the effect of ram--pressure stripping is expected to
be less important, owing to the lower pressure of the hot gas. This
demonstrates that the increasing of the ICM metallicity with
resolution is mainly driven by a more efficient ubiquitous
high--redshift enrichment, rather than by ram-pressure stripping of
enriched gas from merging galaxies.

As for the profiles of the relative abundance (right panel of
Fig.  \ref{fi:prof_res}), they are rather flat out to \rvir, with a
relatively higher abundance for Silicon. In the innermost regions, the
abundance ratios increase, with a more pronounced trend for
[O/Fe]. The reason for this increase is an excess of recent star
formation taking place at the cluster centre. As a consequence,
elements mainly produced by short living stars, such as Oxygen, are
released in excess with those, like Iron, which are mostly contributed
by long--living stars. This also explains why the central increase is
less apparent for [Si/Fe], being Silicon contributed by SNIa more than
Oxygen is. An excess of star formation in the central regions of
galaxy clusters is a well known problem of simulations, like those
discussed here, which include only stellar feedback. For instance,
\cite{2006MNRAS.373..397S} analysed a set of simulations, similar to
those presented here, to study the properties of the galaxy
population. They concluded that the brightest cluster galaxies (BCGs)
are always much bluer than observed, as a consequence of the low
efficiency of SN feedback to regulate overcooling in the most massive
galaxies.

In the outer regions, $R\magcir$\rvir, the two relative abundances
tend to increase, again more apparently for Oxygen.
%
%The reason for this increase is completely different than that for the
%increase in the central regions.
%
The outskirts of galaxy clusters have been enriched at relatively
higher redshift (see also discussion in Sect. \ref{s:res_lifet},
below), when the potential wells are shallower and the enrichment
pattern tends to be more uniformly distributed. This causes the
products of short--living stars to be more effectively distributed to
the diffuse gas than at lower redshift. In this sense, the increasing
trend of the relative abundances produced by short-- and long--living
stars represents the imprint of the different enrichment epochs.  As
for the dependence on resolution, we note a systematic trend for an
increase of Oxygen and, to a lesser extent, of Silicon, at least for
$R\magcir 0.1R_{500}$. This behaviour is consistent with the increased
star formation rate at high redshift. Since Oxygen is relatively more
contributed by short--living stars, then it is released at higher
redshift than Iron and, therefore, has a more uniform distribution, an
effect that increases with resolution. In the innermost cluster
regions, $R\mincir 0.1R_{500}$, resolution acts in the direction of
reducing the excess of Oxygen and Silicon. This can also be explained
in terms of the dependence of the SFR on resolution: since most of the
low--redshift SFR is concentrated at the cluster centre, its reduction
at low redshift also reduces in these regions the relative amount of
metals released from short--living stars.

%In order to further illustrate how resolution affects the pattern of
%metal enrichment, we plot in Figure \ref{fi:mz_res} the histogram of
%the mass fraction of stars and gas having a given level of
%enrichment. The first clear outcome of this plot is that for stars
%this distribution is more skewed toward a higher enrichment than for
%the gas. As resolution, increases the mass fraction associated to the
%highest metallicity bin decreases, and so it does also the mass
%fraction in the low metallicity tail.  This is consistent with the
%expectation that an improved resolution helps in better resolving
%those dynamical processes which mix and transport highly enriched gas
%away from the high--density, possibly star--forming, regions. However,
%we note that the dependence on resolution is generally not
%large. Quite interestingly, a significant population of gas and star
%particles, characterized by pristine abundance, are always found,
%almost independent of resolution.

%\begin{figure}
%\centerline{
%\psfig{file=Figs/Enrich/z_0.00_mz.res.eps,width=8.5cm}}
%\caption{The histograms showing the effect of resolution on the
%  fraction of gas (thick lines) and stars (think lines) having a given
%  level of metal enrichment, for the Cl2 cluster. The top and bottom
%  panels are for Iron and Oxygen, respectively. The solid, dashed and
%  dot-dashed lines are runs at increasing resolution, R1, R2, and R3,
%  respectively.}
%\label{fi:mz_res}
%\end{figure}

In conclusion, our resolution study demonstrates that the general
pattern of the ICM chemical enrichment is rather stable in the central
regions of clusters, $R\mincir 0.3R_{500}$. However, the situation is
quite different in the cluster outskirts, where both absolute and
relative abundances significantly change with resolution, as a
consequence of the different efficiency with which high--redshift star
formation is described. On the one hand, this lack of numerical
convergence becomes quite apparent on scales $\magcir R_{500}$, which
can be hardly probed by the present generation of X--ray
telescopes. On the other hand, resolution clearly becomes important
in the regime which is relevant for the study of the WHIM, which is
one of the main scientific drives for X--ray telescopes of the next
generation \citep[e.g., ][]{2004PASJ...56..939Y}.

\subsection{Changing the model of chemical evolution}

\begin{figure*}
\centerline{
\hbox{
\includegraphics[width=5.8cm]{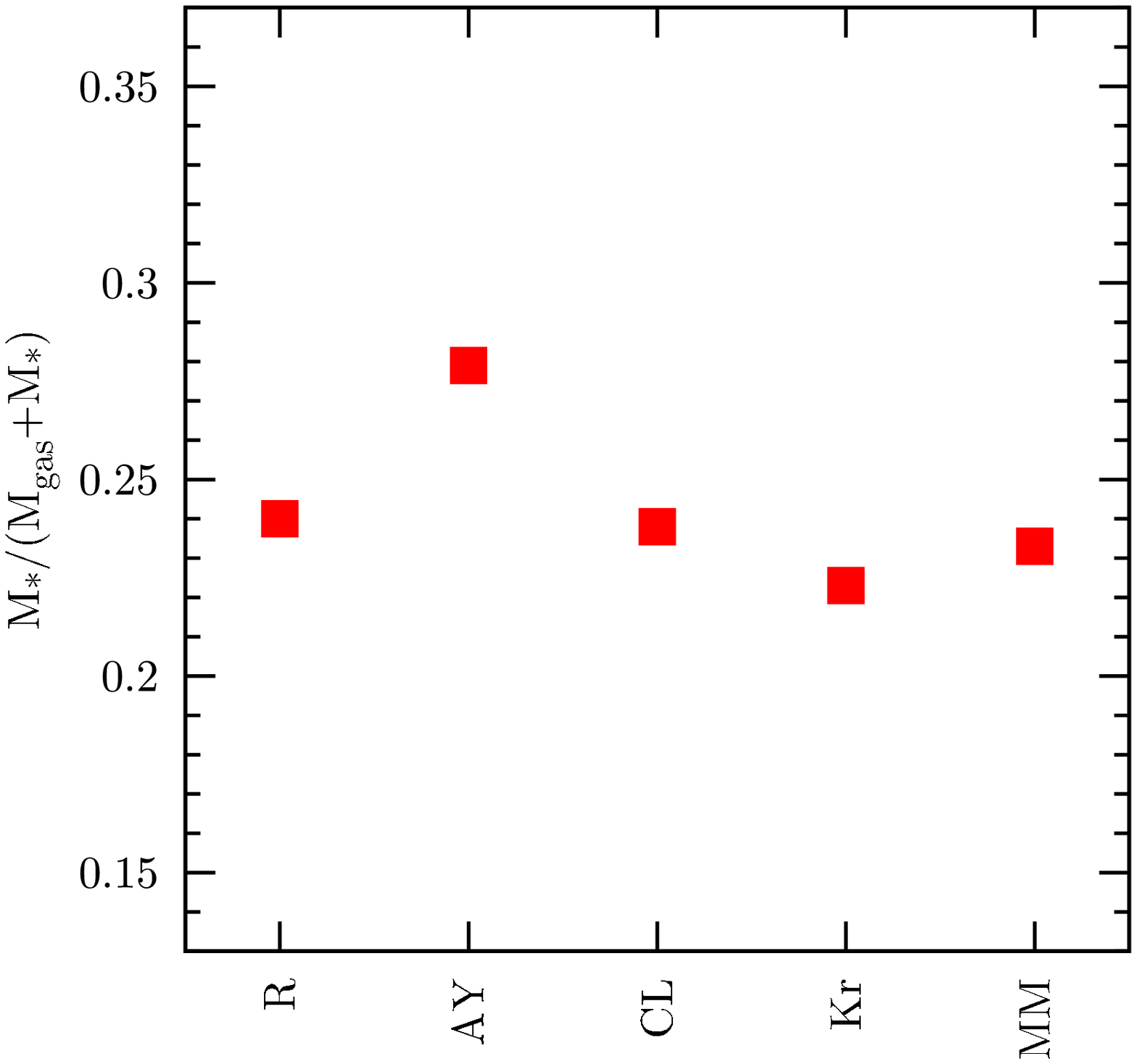} 
\includegraphics[width=5.8cm]{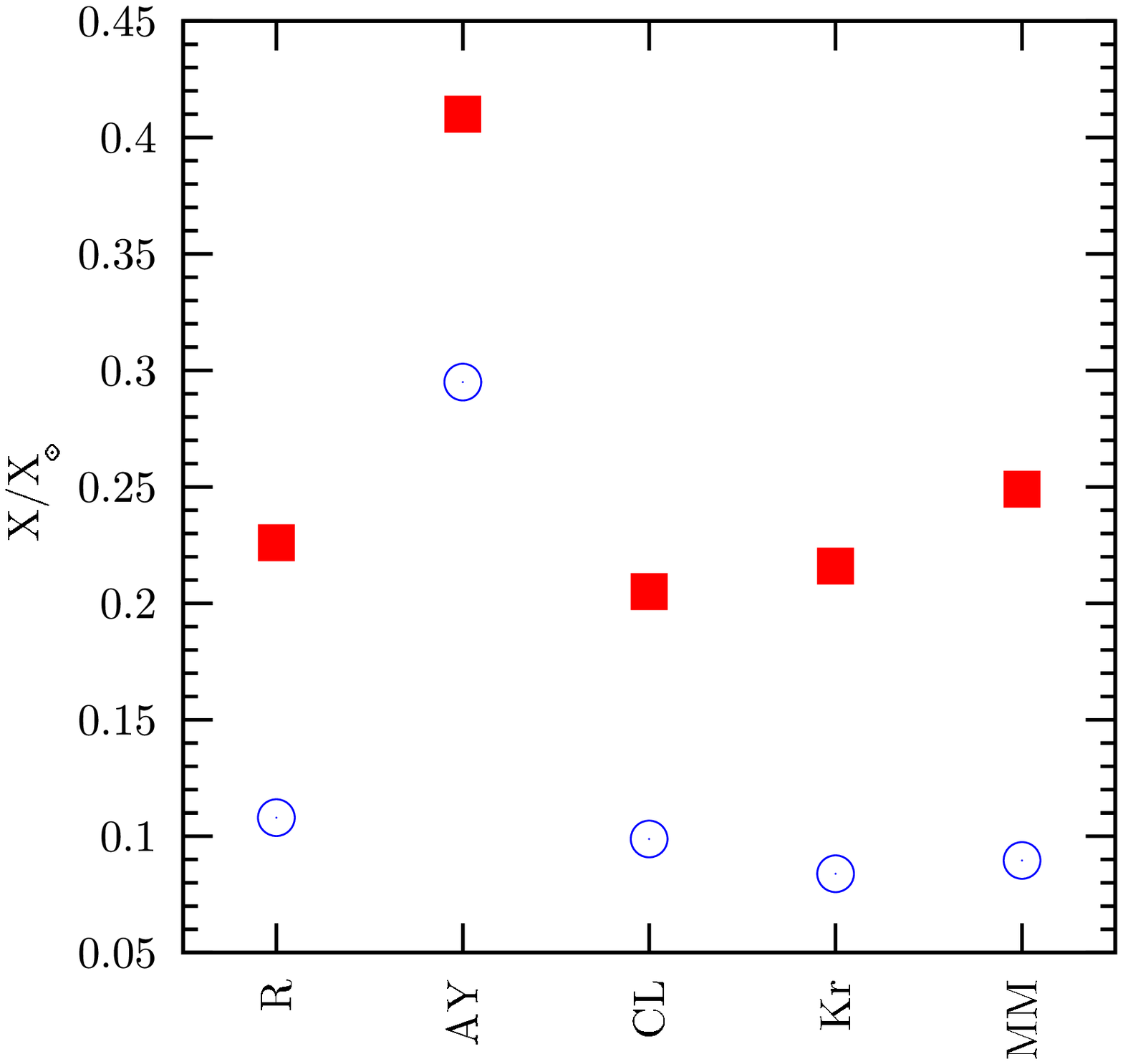} 
\includegraphics[width=5.8cm]{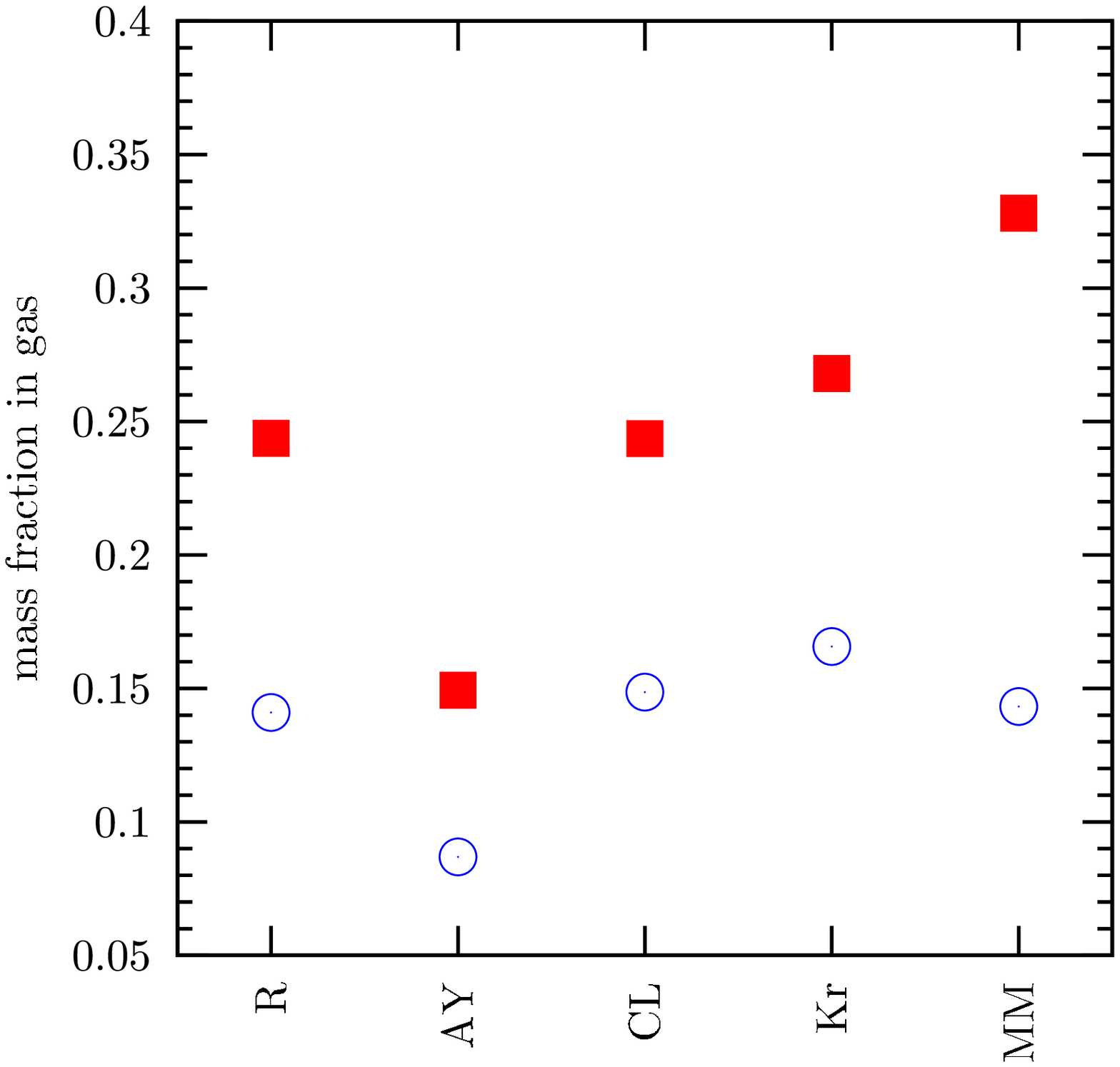} }}
\caption{The same as in Figure \ref{fi:glob_num}, but changing
  parameters related to the model of chemical evolution and to the
  feedback. The labels indicating the different runs are as reported
  in Table \ref{t:runs}.}
\label{fi:glob_ph} 
\end{figure*}

\subsubsection{The effect of the IMF}
\label{s:res_imf}
As already discussed in \ref{s:imf}, a vivid debate exists
in the literature as to whether the level of the ICM enrichment can be
accounted for by a standard, Salpeter--like, IMF or rather requires a
top--heavier shape. The absolute level of enrichment in one element,
e.g. Iron, does not necessarily represent an unambiguous imprint of
the IMF in our simulations. 
%Indeed, although the total amount of Iron
%depends on the IMF, an incorrect description of the total amount of
%stars produced in the simulation could led to incorrect
%conclusions. 
For instance, an exceedingly top--light IMF could still produce an
acceptable Iron abundance in the presence of an excess of star
formation in simulations. For this reason, it is generally believed
that a more reliable signature of the IMF is provided by the relative
abundances of elements which are produced by SNIa and SNII.

\begin{figure*}
\centerline{ \hbox{ \includegraphics[width=8.5cm]{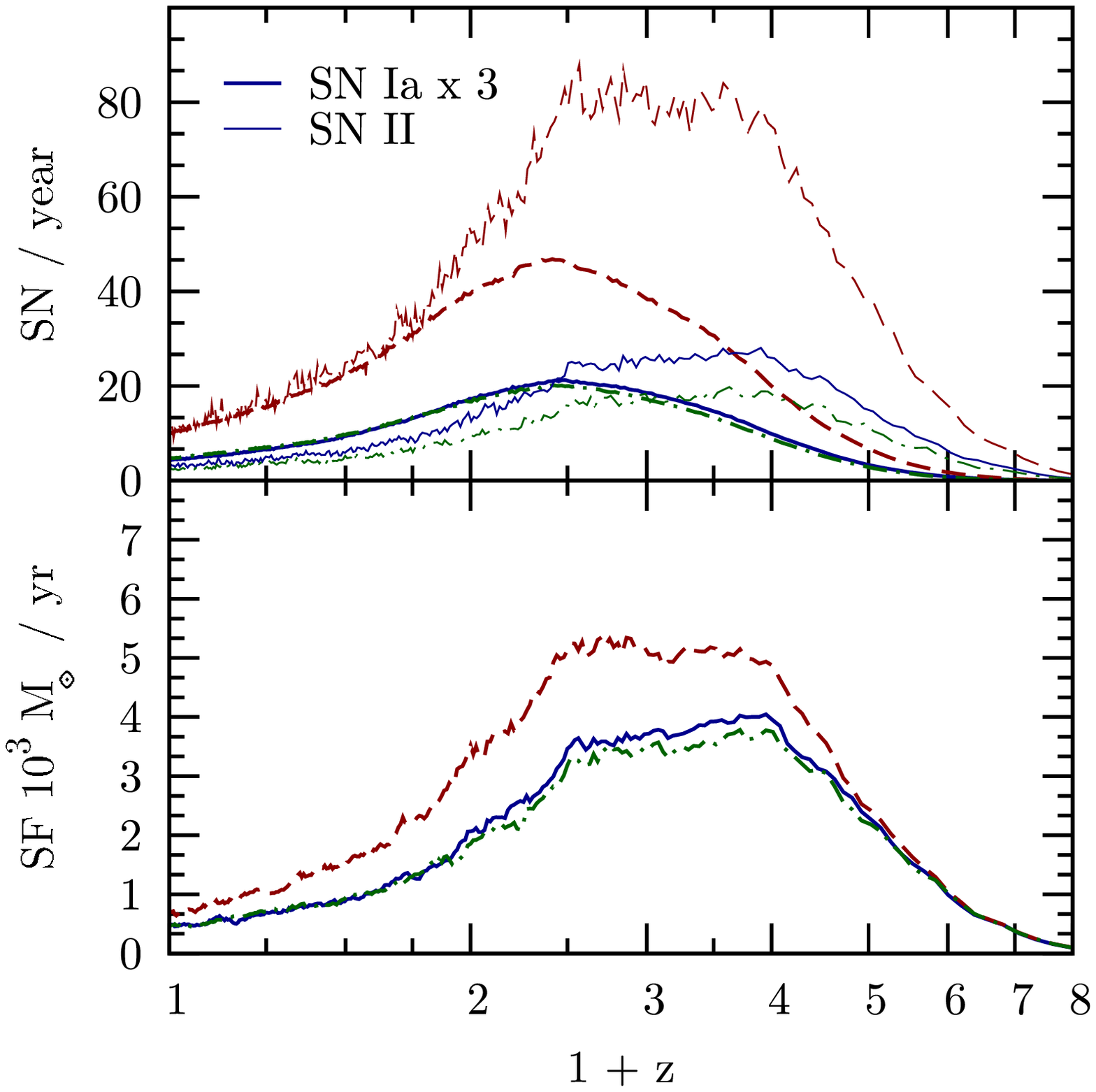}
\includegraphics[width=8.5cm]{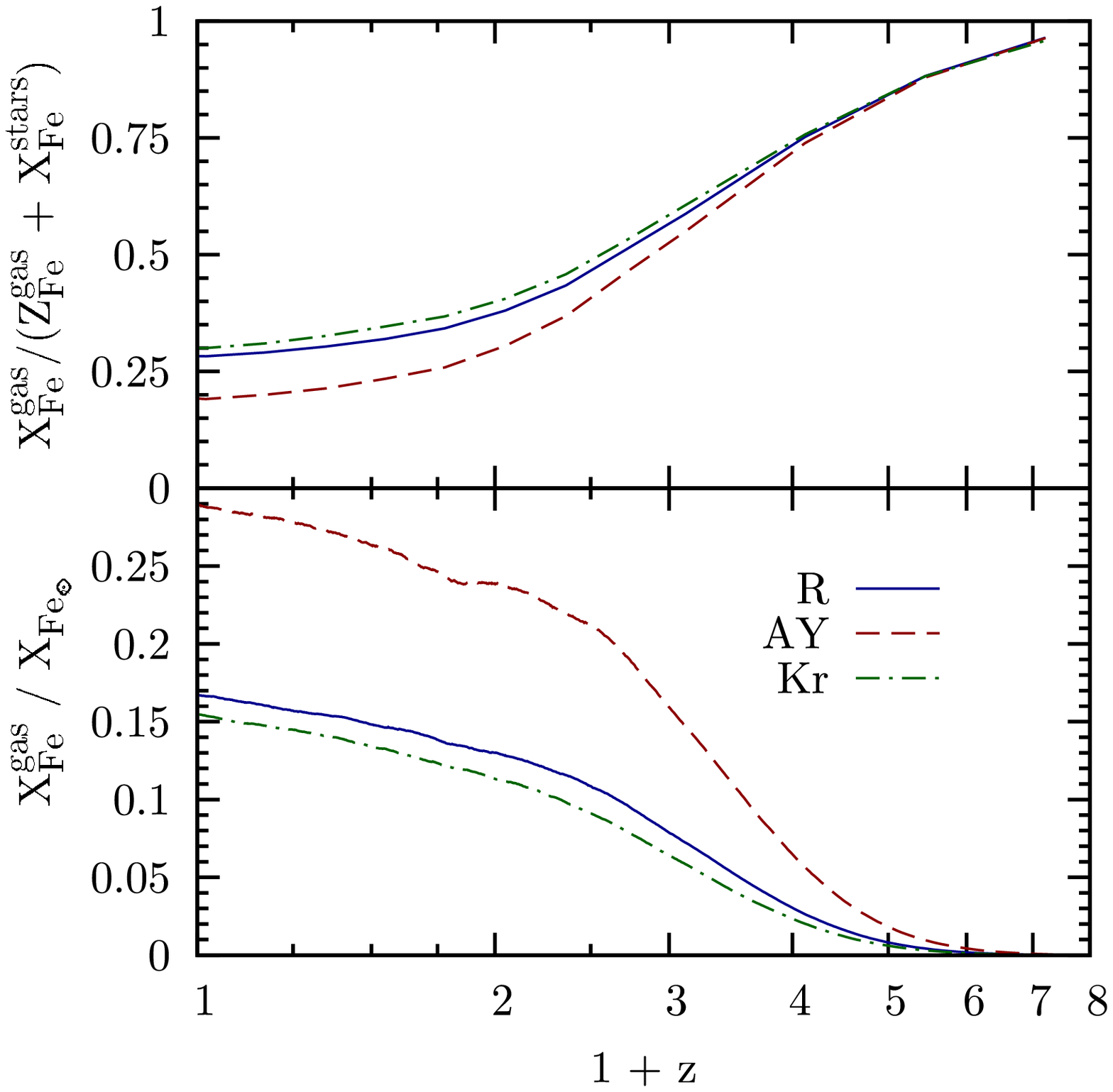} }}
\caption{The same as Figure \ref{fi:enr_ng}, but changing the IMF. The
  solid lines correspond to the reference run, while the short--dashed
  and the long--dashed lines are for the IMFs by
  \protect\cite{1987A&A...173...23A} and by
  \protect\cite{2001MNRAS.322..231K}, respectively.}
\label{fi:enr_imf}
\end{figure*}

As shown in Figure \ref{fi:enr_imf} the effect of assuming an IMF,
which is top--heavier than the Salpeter one, is that of significantly
increasing the number of SNII and, to a lesser extent, also the number
of SNIa. This is consistent with the plot of Fig. \ref{fi:imfs},
which shows that an Arimoto--Yoshii IMF predicts more stars than the
Salpeter one already for $M_*\simeq 1.5 M_\odot$. The larger number of
SN clearly generates a higher level of enrichment at all redshifts
(bottom right panel of Fig. \ref{fi:enr_imf}. A higher level of gas
enrichment increases the cooling efficiency and, therefore, the
star--formation rate (bottom left panel of \ref{fi:enr_imf}). A higher
star formation efficiency has, in turn, the effect of increasing the
fraction of Iron which is locked in the stellar phase (top--right
panel of \ref{fi:enr_imf}).

In Figure \ref{fi:maps_imf} we show the maps of Iron abundance (left
panels) and of the fractional enrichment from SNII  for the three 
IMFs. The effect of a top--heavy IMF is confirmed to increase
the overall level of enrichment in Iron. At the same time, the
contribution of SNII becomes more important, thus consistent with the
increase of the number of massive stars.

\begin{figure*}
\centerline{
\hbox{
\includegraphics[width=8.5cm]{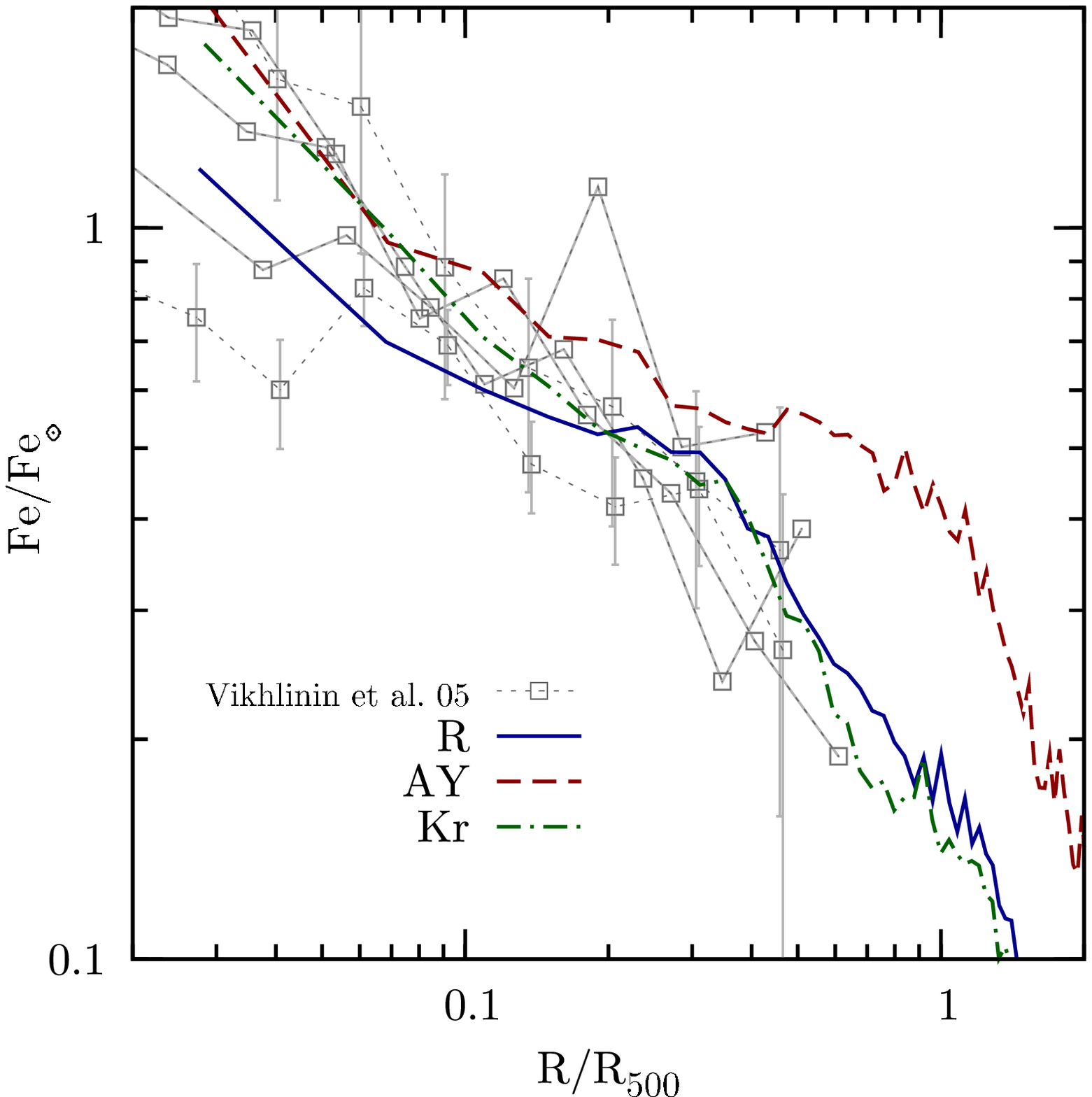} 
\includegraphics[width=8.8cm]{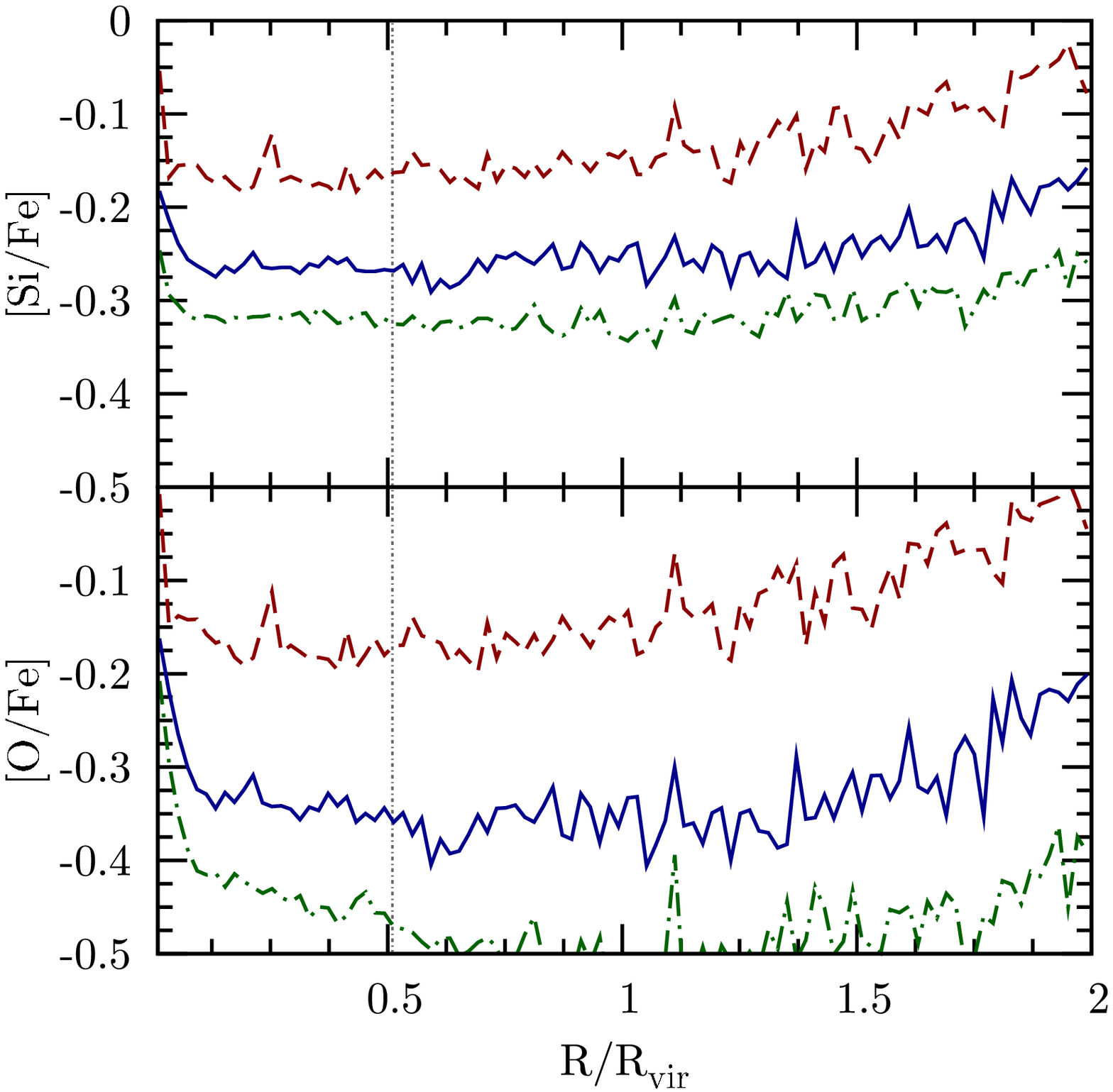}}} 
\caption{The effect of changing the IMF and the yields on the
  abundance profiles. Left panel: profiles of the mass--weighted Iron
  abundance. The data points are the same as in Figure
  \ref{fi:prof_ng}. Right panel: the relative abundance of Silicon
  (top) and Oxygen (bottom) with respect to Iron, for the same runs
  shown in the left panel. The different line--types have the same
  meaning as in Figure \ref{fi:enr_imf}.}
\label{fi:prof_imf}
\end{figure*}

The effect of assuming a top--heavier IMF is quite apparent on the
profiles of Iron abundance (left panel of Figure
\ref{fi:prof_imf}). The level of enrichment increases quite
significantly, up to factor of two or more at $R\magcir 0.5R_{500}$,
bringing it to a level in excess of the observed one. As expected, the
larger fraction of core--collapse SN also impacts on the relative
abundances (right panel of Fig. \ref{fi:prof_imf}), especially for
[O/Fe]. Since Oxygen is largely contributed by SNII,
its relative abundance to Iron increases by about 60 per cent. 

This result goes in the direction of alleviating the tension between
the largely sub--solar [O/Fe] values found for the Salpeter IMF and
the nearly solar value reported by observational analyses \citep[e.g.,
][]{2004A&A...420..135T}.  However, an overestimate of Oxygen from the
spectral fitting, used in the analysis of observational data, may
arise as a consequence of a temperature--dependent pattern of
enrichment.  Rasia et al. (in preparation) analysed mock XMM--Newton
observations of simulated clusters, including chemical enrichment with
the purpose of quantifying possible biases in the measurement of ICM
metallicity. As for the Iron abundance, they found that its
emission--weighted definition is a good proxy of the spectroscopic
value. On the contrary, the spectroscopic measurement of the Oxygen
abundance turns out to be significantly biased high with the respect
to the intrinsic abundance.  The reason for this bias is that, unlike
Iron, the Oxygen abundance is obtained from emission lines which are
in the soft part of the spectrum. On the other hand, relatively colder
structures such as filaments, seen in projection and surrounding the
ICM, give a significant contribution to the soft tail of the
spectrum. Since these structures are on average more enriched than the
hot ICM, they are over--weighted when estimating element abundances
from soft ($\mincir 1$ keV) transitions. This is the case of Oxygen,
whose abundance is generally estimated from the O-VIII line, which is
at about 0.65 keV.

Addressing the issue of observational biases in the measurement of the
ICM enrichment is outside the scope of this paper. Still, the above
example illustrates that such biases need definitely to be understood
in detail if the enrichment pattern of the ICM has to be used as a
fossil record of the past history of star formation in cluster
galaxies.

As for the results based on the IMF by \cite{2001MNRAS.322..231K}, we
note that it induces only a small difference in the SFR with respect
to the Salpeter IMF. While the SNIa rate is also left essentially
unchanged, the SNII rate is now decreased by about 50 per cent. The
reason for this is that the Kroupa IMF falls below the Salpeter one in
the high mass end, $\magcir 5M_\odot$ (see Figure \ref{fi:imfs}). This
is consistent with the maps shown in the bottom panels of
Fig. \ref{fi:maps_imf}. The global pattern of Iron distribution is
quite similar to that provided by the Salpeter IMF, while the relative
contribution from SNII is significantly reduced.  Consistent with this
picture, the profile of the Iron abundance shown in
Fig. \ref{fi:prof_imf} does not show an appreciable variation with
respect to the reference case. On the contrary, the profile of the
Oxygen abundance and, to a lesser extent, of Silicon abundance,
decreases significantly.

Our results confirm that $[\alpha/Fe]$ relative abundances are
sensitive probes of the IMF. However, we have also shown that
different elements are spread in the ICM with different
efficiencies. This is due to the fact that long--lived stars can
release metals away from star forming regions and, therefore, their
products have an enhanced probability to remain in the diffuse medium
(see discussion in Sect. \ref{s:metspr}). Therefore, both a
correct numerical modeling of such processes and an understanding of
observational biases are required for a correct interpretation of
observed metal content of the ICM.

\subsubsection{Changing the yields}
\label{s:yie}

\begin{figure*}
\centerline{
\hbox{
\includegraphics[width=8.5cm]{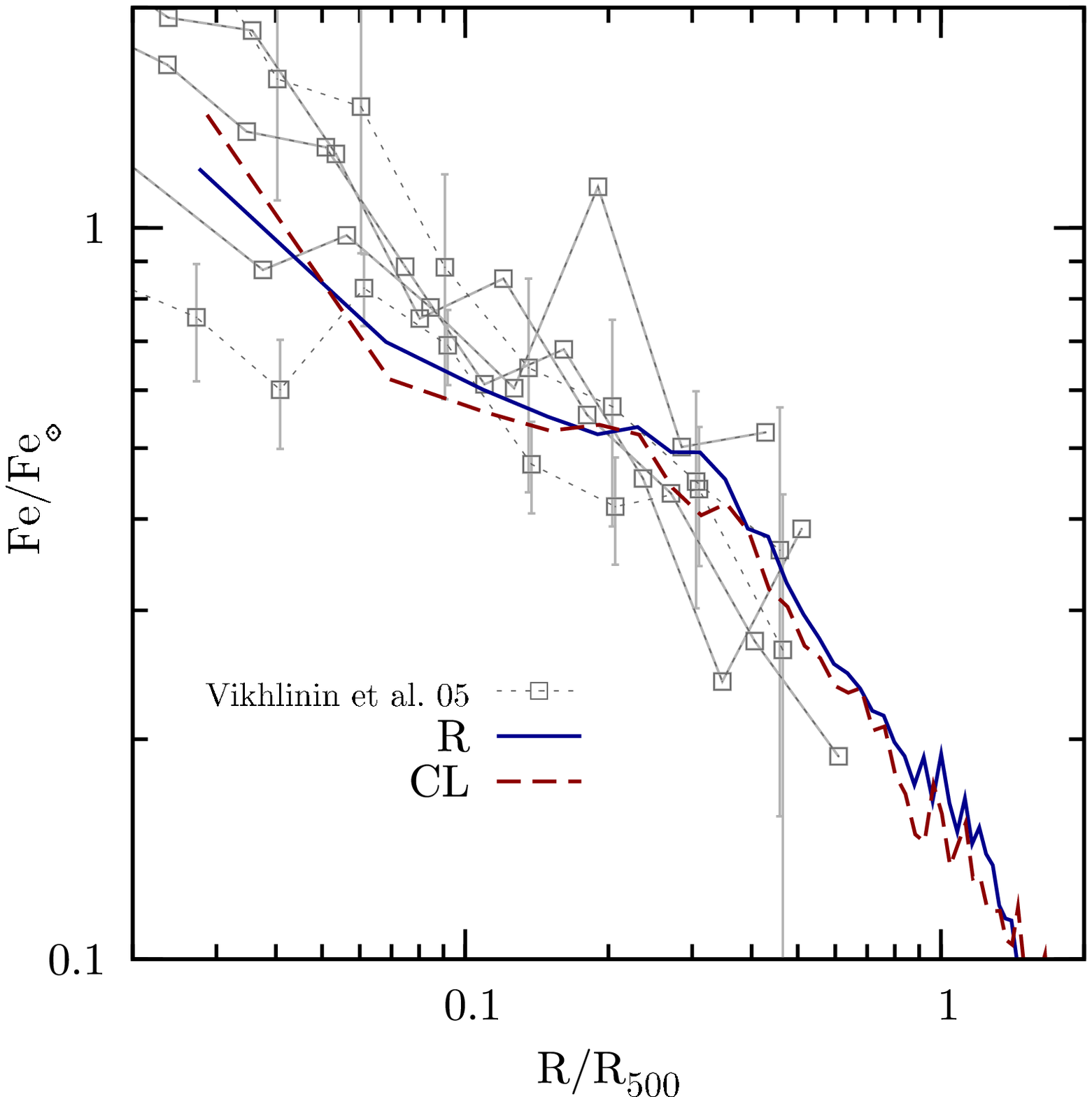}
\includegraphics[width=8.8cm]{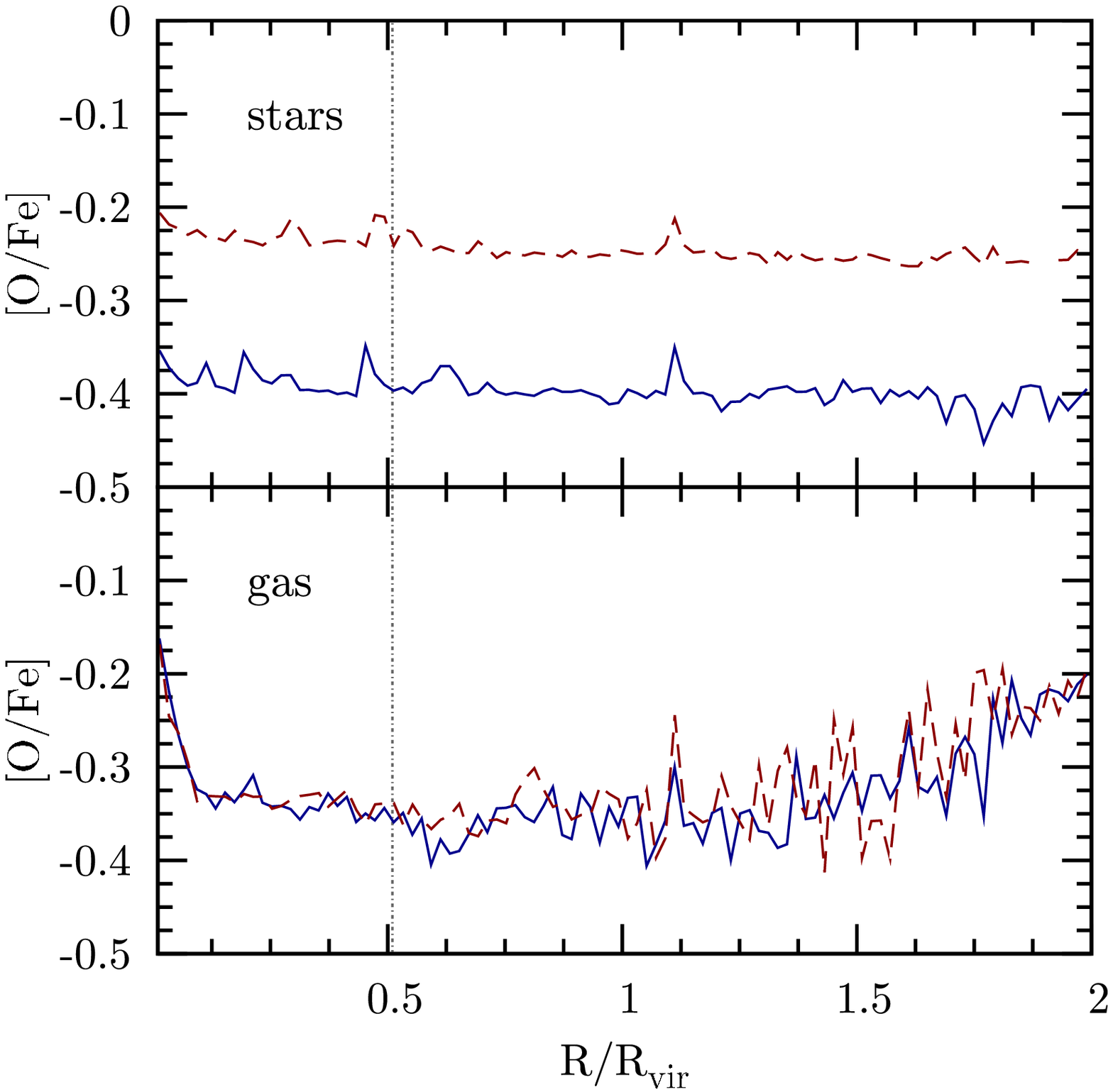} 
}}
\caption{The effect of changing the SNII yields on the abundance
  profiles. Left panel: profile of the Fe abundance in the ICM; the
  data points are the same as in Figure \ref{fi:prof_ng}. Right panel: 
  the profiles of the [O/Fe] relative abundances for the ICM (lower panel)
  and for the stars (upper panel). Solid and dashed curves refer to
  the reference (R) run, based on the yields by
  \protect\cite{1995ApJS..101..181W}, and to a run (CL) in which the
  yields by \protect\cite{2004ApJ...608..405C} are used instead.}
\label{fi:prof_yields}
\end{figure*}

In Figure \ref{fi:prof_yields} we show the effect of using the yields
by \cite{2004ApJ...608..405C} (CL) for the SNII, instead of those by
\cite{1995ApJS..101..181W} (WW) as in the reference run, on the Iron
density profile. As for the enrichment pattern of the ICM, using
either one of the two sets of yields gives quite similar results, both
for the abundance of Iron and for the [O/Fe] relative abundance. This is
apparently at variance with the results shown in Fig.\ref{fi:metyie},
where we have shown the differences in the production of different
metals from an SSP for the two sets of yields. However, we note from
that figure that such differences in the metal production have a non
trivial dependence on the SSP characteristics, being different
elements over-- or under--produced for one set of yields, depending on
the SSP initial metallicity. On the other hand, the metallicity of
each star particle in the simulations depends on the redshift at which
it is created, with the star--formation history being in turn affected
by the enrichment pattern. 

As for the enrichment in Iron (left panel of
Fig. \ref{fi:prof_yields}), we remind that this element is mostly
contributed by SNIa, while the contribution from SNII is not only
sub-dominant, but also preferentially locked back in stars. Since we
are here changing the yields of the SNII, there is not much surprise
that the effect of the profiles of $Z_{Fe}$ is marginal.  The
situation is in principle different for Oxygen (right panel of
Fig.\ref{fi:prof_yields}). However, Fig.\ref{fi:metyie} shows that the
WW yields for Oxygen are in excess or in defect with respect to the CL
ones, depending on the initial metallicity. As a result, we find that
[O/Fe] for the diffuse gas is left again substantially unchanged,
while a significant variation is found for stars, whose [O/Fe] is
about 40 per cent larger when using the CL yields table. The profiles
of [O/Fe] for stars show a mild decrease at large radii, consistent
with the fact that Oxygen is more efficiently spread around stars in
the regions where enrichment takes place at higher redshift (see
discussion in Sect. \ref{s:metspr}). Once again, this difference in
the enrichment pattern of gas and stars reflects the different
efficiency that gas dynamical processes have in transporting metals
away from star forming regions.

As a final remark, we would like to emphasize that several other
tables of stellar yields have been presented in the literature,
besides the ones by WW and CL considered here, which refer only to
massive stars, and those by \cite{1997A&AS..123..305V} and by
\cite{2003NuPhA.718..139T} that we adopted for low and intermediate
mass stars and for SNIa, respectively. Besides the yields for
intermediate--mass stars computed by \cite{1981A&A....94..175R}, other
sets of metallicity--dependent yields have been provided by
\cite{1999ApJS..125..439I} and \cite{2001A&A...370..194M} for
intermediate and low--mass stars, and by \cite{1998A&A...334..505P}
for massive stars. The differences among such sets of yields are due
to the different stellar evolutionary tracks used and/or to the
different way of describing the structure of the progenitor star. A
different approach has been followed by \cite{2004A&A...421..613F},
who inferred stellar yields by requiring that their model of chemical
evolution reproduces the observed enrichment pattern of the Milky Way.

Assessing in detail which one of them should be preferred to
trace the cosmic evolution of metal production with hydrodynamical
simulations is beyond the scopes of this paper. It is however clear
that any uncertainty in the models on which yields computations are
based can produce significant changes in the pattern the ICM
enrichment \citep[e.g.,][]{1997MNRAS.290..623G}.

\begin{figure*}
\centerline{
\hbox{
\includegraphics[width=8.8cm]{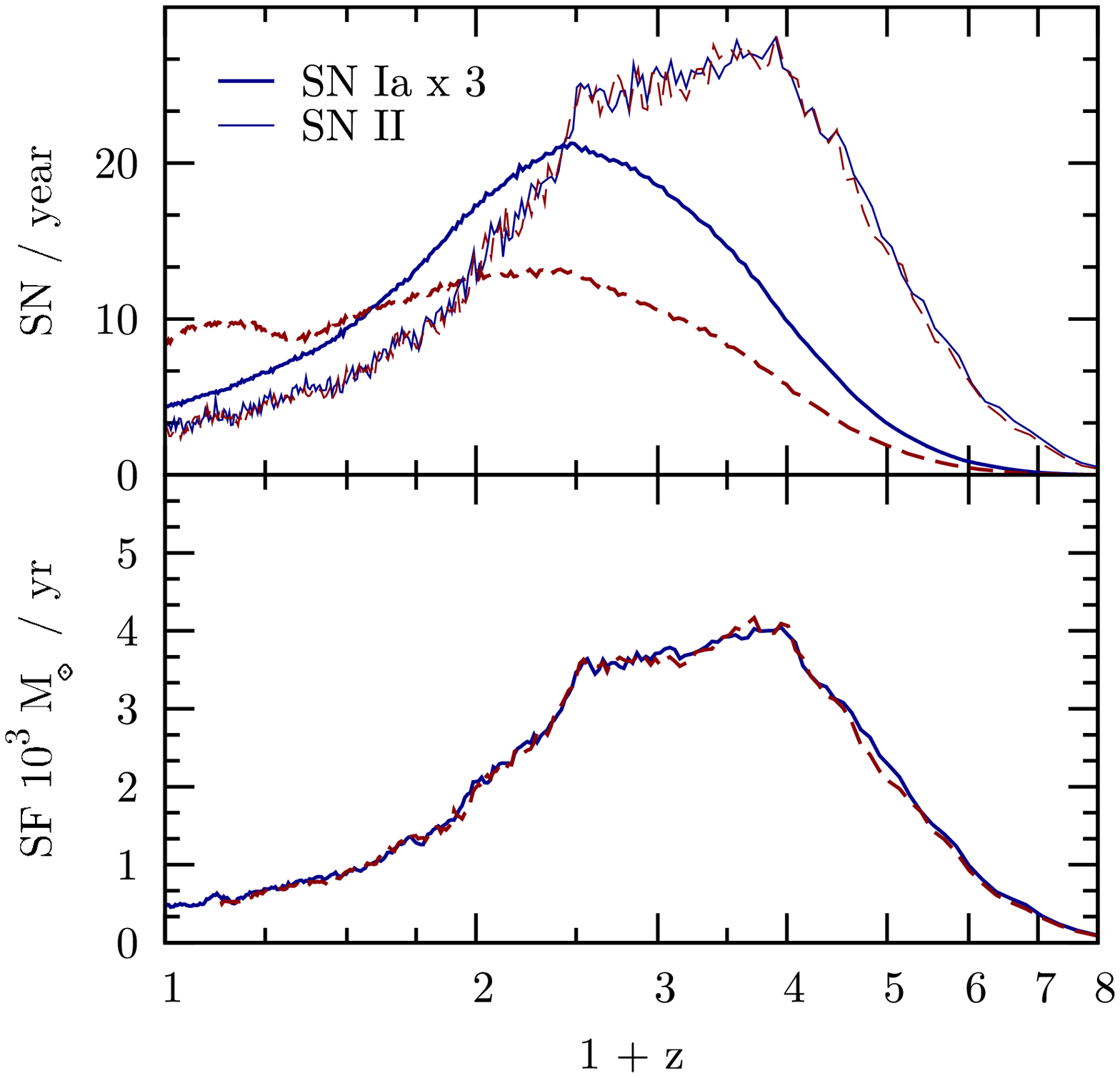}
\includegraphics[width=8.5cm]{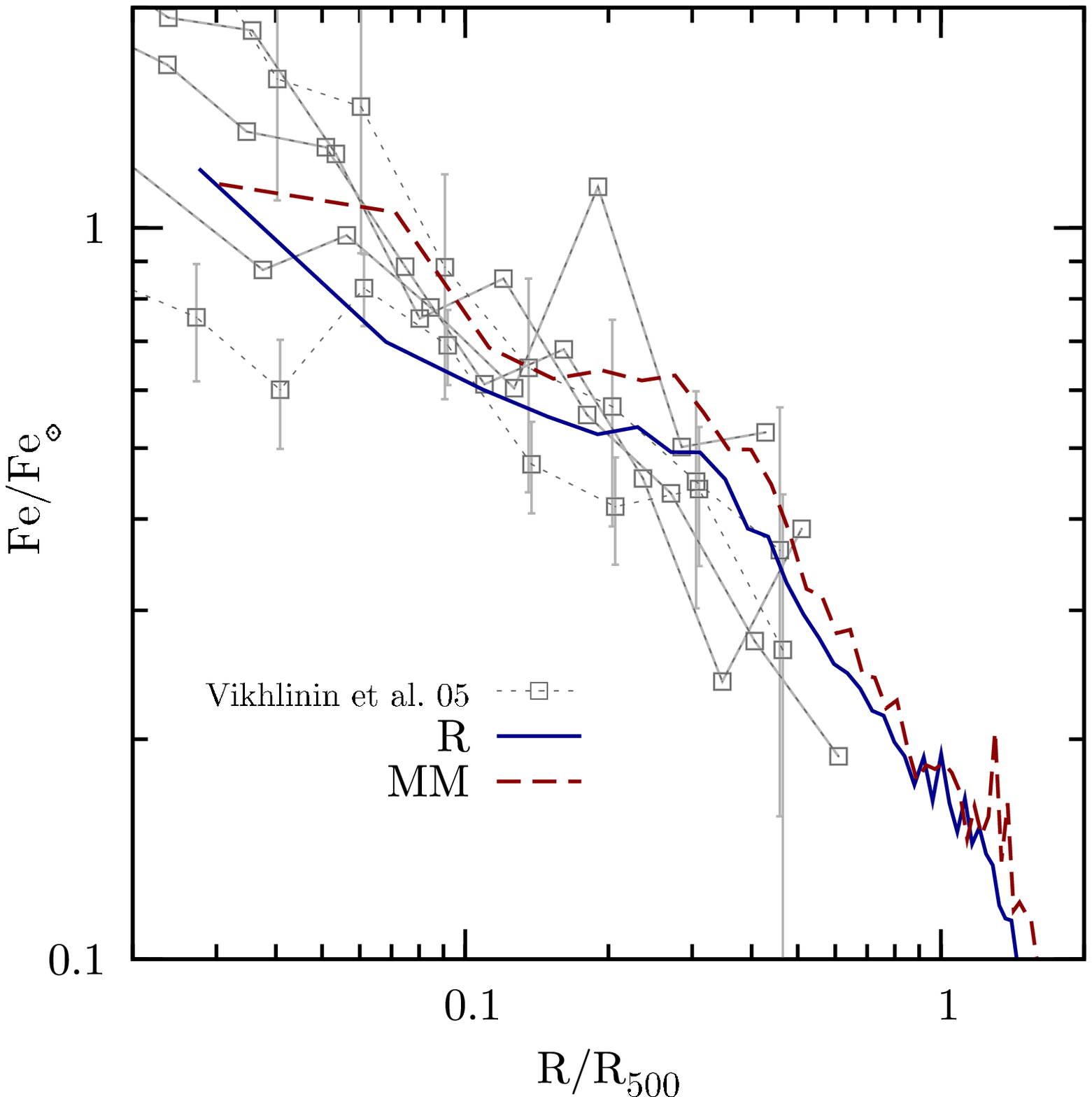}
}}
\caption{Left panel: the effect of changing the lifetime function on
  the star formation rate (bottom panel) and supernova rates (top
  panel; tick lines: SNIa; thin lines: SNII). The meaning of the
  different line types is the same as in the left panel of Figure
  \ref{fi:prof_res}. Right panel: the effect of the lifetime function
  on the profile of the Iron abundance (right panel). The data points
  are the same as in Figure \ref{fi:prof_ng}. The solid line is for the
  reference (R) run, which uses the lifetime function by
  \protect\cite{1993ApJ...416...26P}, while the dashed line is for the
  MM run which uses the lifetime function by
  \protect\cite{1989A&A...210..155M}.}
\label{fi:res_lifet}
\end{figure*}

\subsubsection{The effect of the lifetimes}
\label{s:res_lifet}
As we have shown in Figure \ref{fi:lifet}, using the lifetime function
by \cite{1989A&A...210..155M}, instead of that by
\cite{1993ApJ...416...26P}, corresponds to significantly increasing
the delay time for the release of energy and metals from low--mass
stars with $M\mincir 1 M_\odot$, while leaving substantially unchaged
that of massive stars. As a result, we expect that the SNII rate will
remain substantially unchanged, while shifting towards lower redshift
the rate of SNIa. This expectation is completely in line with the
results shown in the left panel of Figure \ref{fi:res_lifet}: changing
the lifetime function has a small enough effect on the history of
enrichment that the resulting SFR and SNII rate are left completely
unchanged; on the other hand the SNIa rate is suppressed at high
redshift, while it is substantially increased at $z\mincir 0.5$. 

\begin{figure}
\centerline{
\includegraphics[width=8.5cm]{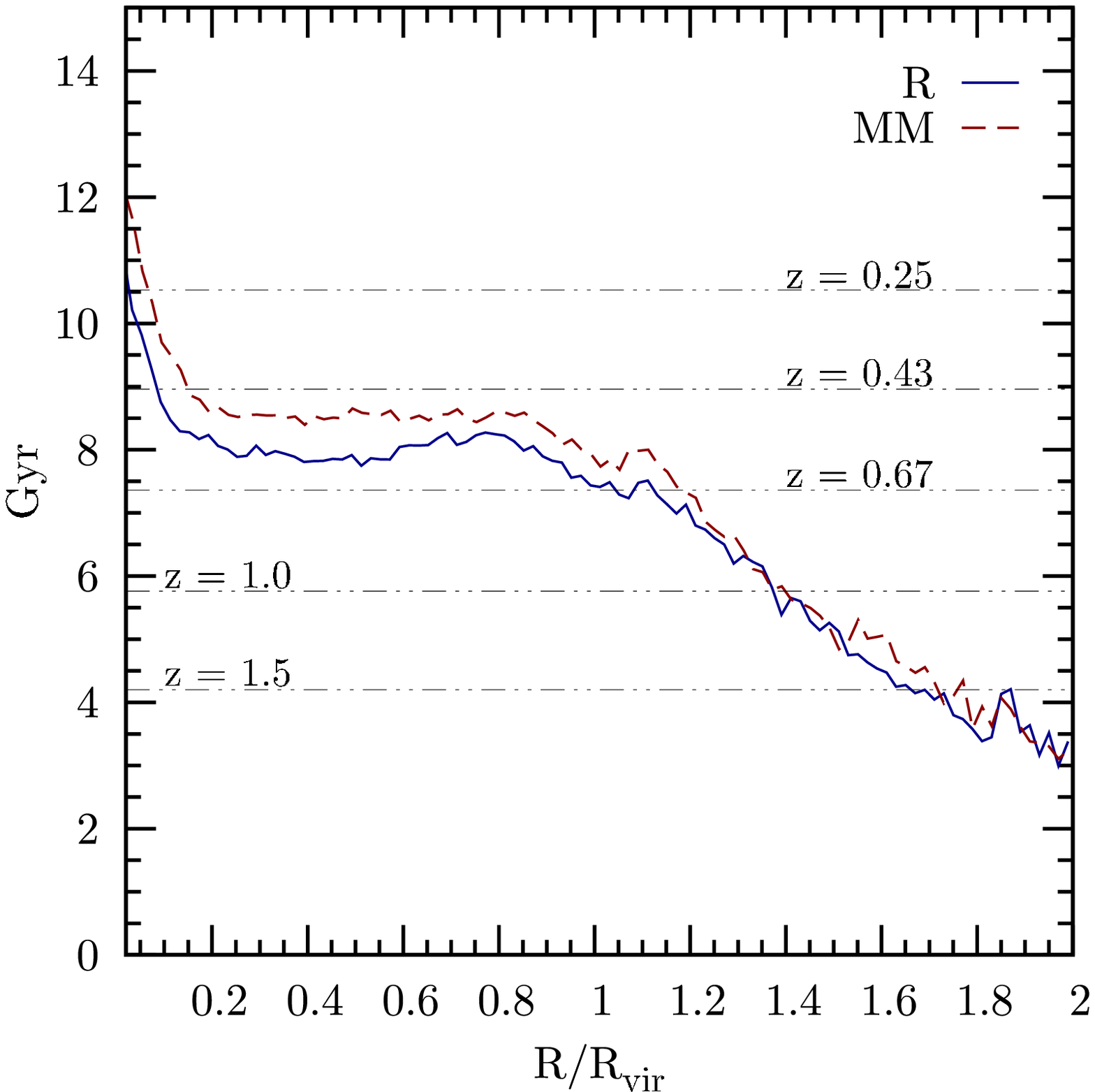} }
\caption{The radial dependence of the mean value of the age of the
  Universe at which the ICM has been enriched. The solid curve is
  for the reference (R) runs, which uses the lifetimes by
  \protect\cite{1993ApJ...416...26P}, while the dashed curve is when
  using the lifetimes by \protect\cite{1989A&A...210..155M}.
  Horizontal dashed lines indicate different values of redshift
  corresponding to difference cosmic times.}
\label{fi:z_age}
\end{figure}

The consequence of changing the lifetimes on the typical epoch of
the ICM enrichment can be can be appreciated from Fig. \ref{fi:z_age}.
To this purpose, we computed for each gas particle the mean value of
the cosmic time at which it has been enriched. This quantity is
obtained by weighting each epoch, at which an enrichment episode takes
place, by the mass of metals received within that time--step. In
Figure \ref{fi:z_age} we plot the radial profile of the mean cosmic
epoch of enrichment, which is computed by averaging over all the gas
particles falling inside different radial intervals. Outside the
virial radius, for both lifetime functions the enrichment is
progressively more pristine as we go to larger cluster-centric
distances. While at \rvir the mean redshift of enrichment is about
0.7, this increases to $\magcir 1.5$ at 2\rvir. 
%This already shows that
%an accurate description of the enrichment pattern in the outskirts of
%galaxy clusters require an accurate description of star formation at
%high redshift. We will further discuss this point in
%Sect. \ref{s:res}, here below. 
Quite interestingly, inside $\simeq 0.8$\rvir the age of enrichment
becomes nearly constant down to the central regions, $\sim
0.2$\rvir. This change of behaviour further indicates that processes
other than star formation, e.g. related to gas dynamical processes,
become important in determining the enrichment pattern. In the core
region, $\mincir 0.2$\rvir, there is a sudden decrease of the
enrichment redshift. This is due to the excess of recent star
formation, which takes place in the central cluster galaxy, and which
also causes the spike in the [O/Fe] relative abundance.

Using the lifetimes by
\cite{1989A&A...210..155M} turns into a more recent enrichment of the
ICM. We should remember here that this figure shown the mean age at
which gas has been enriched in all metals. Since the global
metallicity is dominated by Oxygen, mostly produced by SNII, we
expect a more pronunced effect on the enrichment in Iron. In
principle, a more recent enrichment due to longer delay times of SNIa
could help explaining the observational evidences for an evolution of
the ICM iron content at $z\mincir 0.5$ \citep{2007A&A...462..429B}. We
will postpone a more detailed comparison with the observed ICM
metallicity evolution to a forthcoming paper (Fabjan et al., in
preparation).

As for the profile of the Iron abundance, we note that there is a
small but sizeable increase, especially at small radii. This increase
is due to a more efficient unlocking of Iron from star forming
regions. Since the delay time is enhanced for SNIa, they have an
larger probability to release metals outside star forming regions. On
the other hand, we verified that the enrichment is Oxygen is left
unchanged, as expected from the stability of the SNII rates. As a
result, we find that [O/Fe] decreases by about 0.1 at $R_{500}$ and by
0.2 in the innermost regions.

\begin{figure*}
\centerline{
\hbox{
\includegraphics[width=5.8cm]{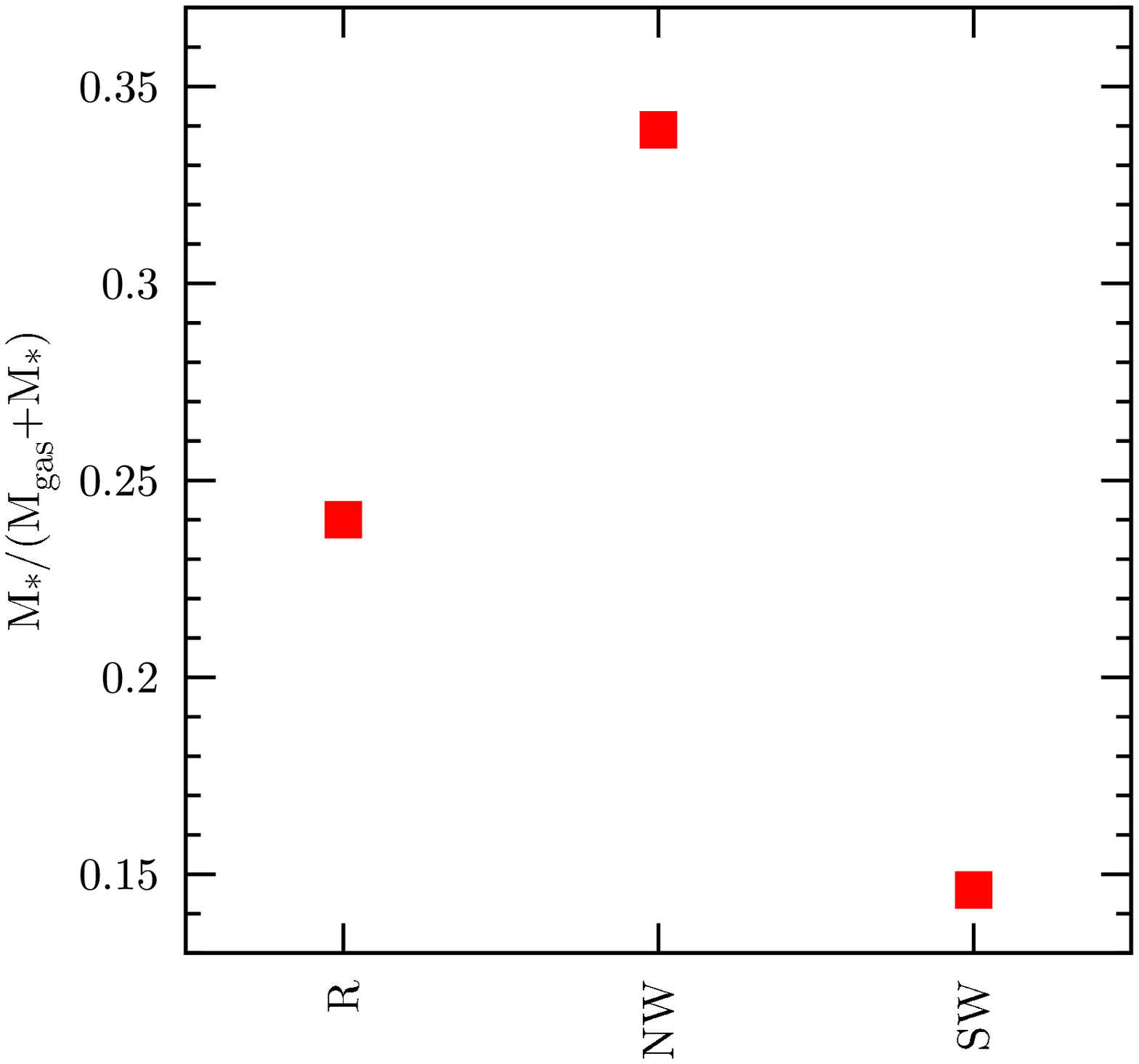} 
\includegraphics[width=5.8cm]{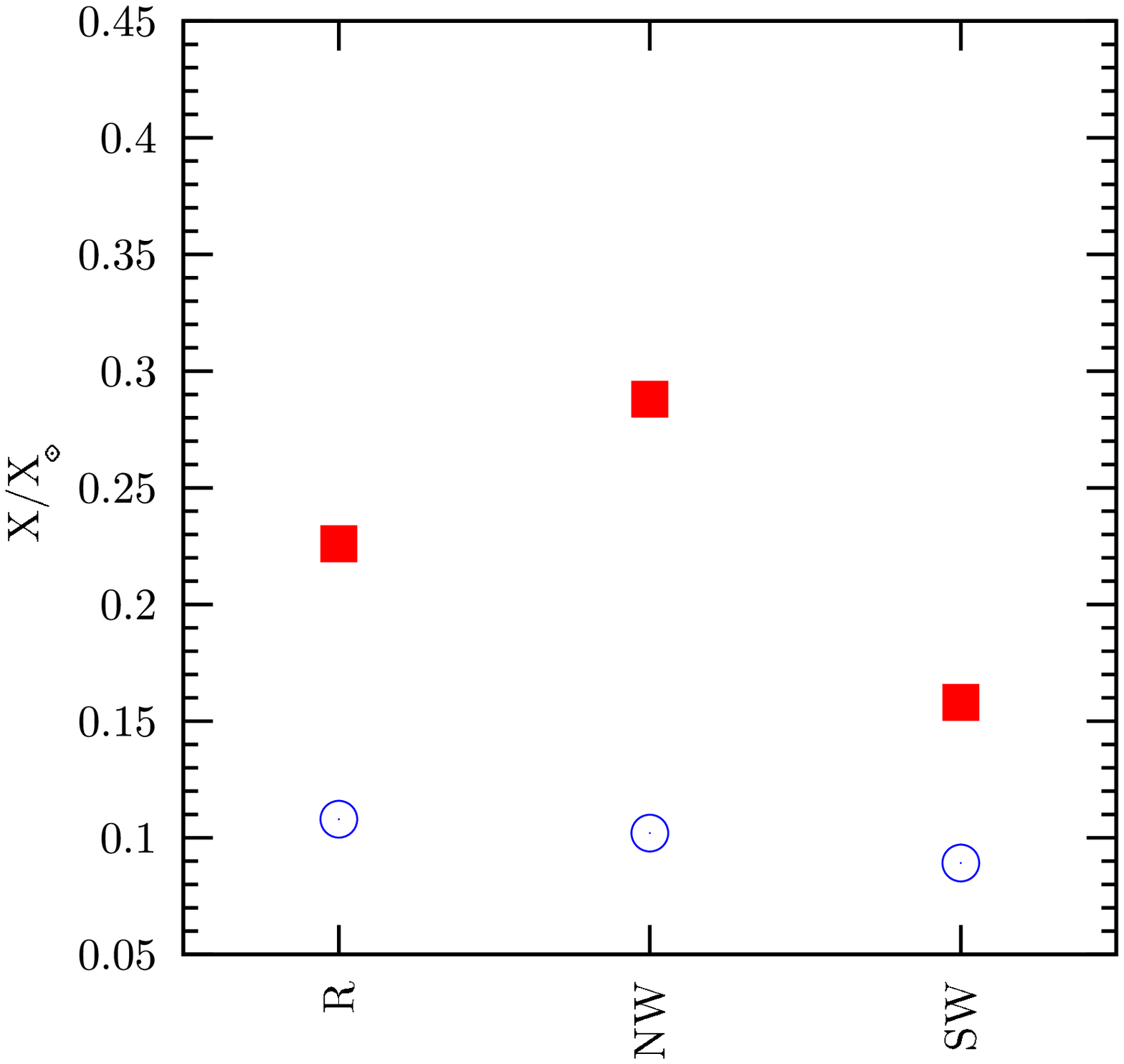} 
\includegraphics[width=5.8cm]{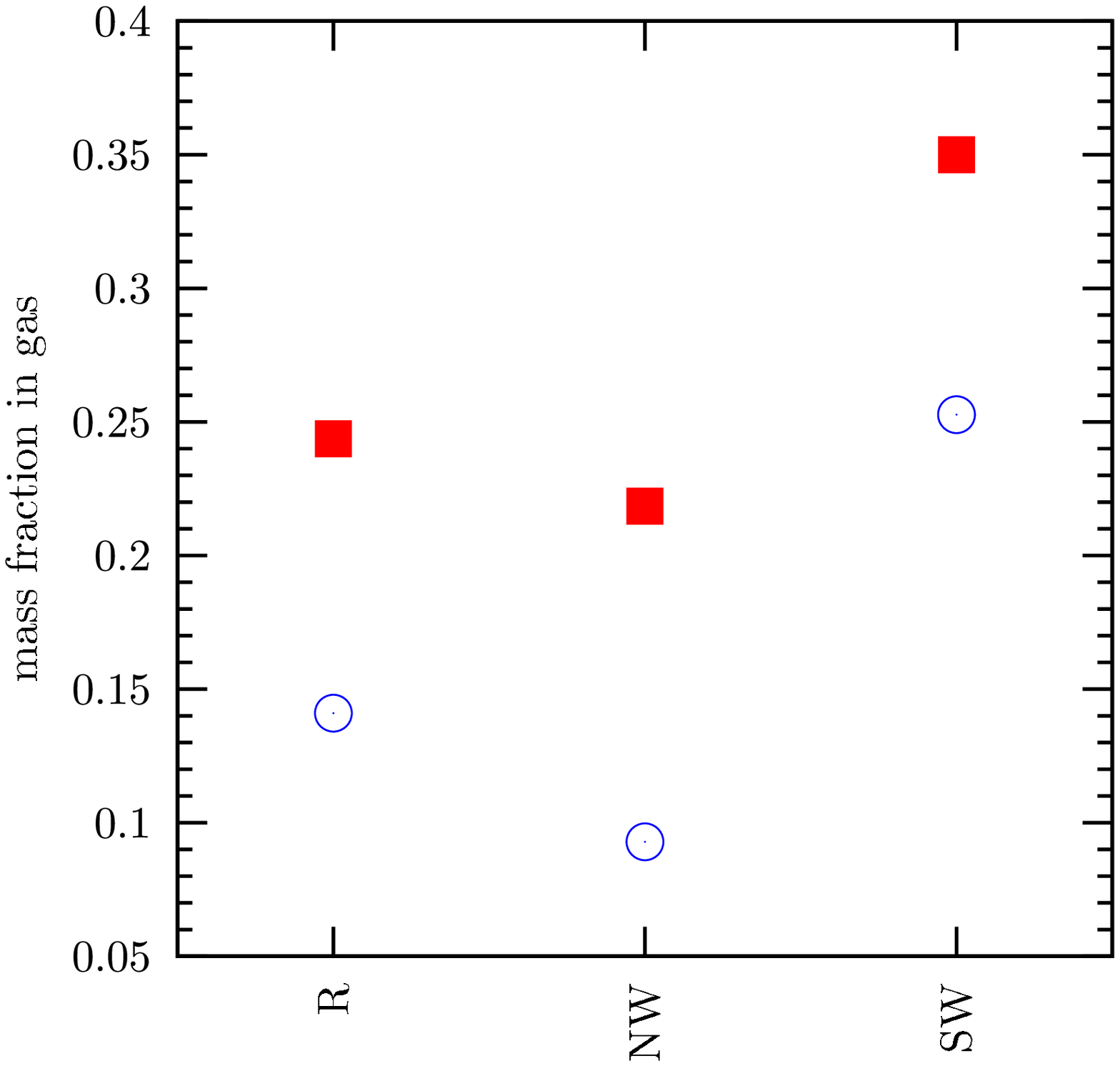} }}
\caption{The same as in Figure \ref{fi:glob_num}, but changing the
  velocity of the galactic winds. The labels indicating the different
  runs are as reported in Table \ref{t:runs}.}
\label{fi:glob_fdbk} 
\end{figure*}

\begin{figure*}
\centerline{ \hbox{ \includegraphics[width=8.5cm]{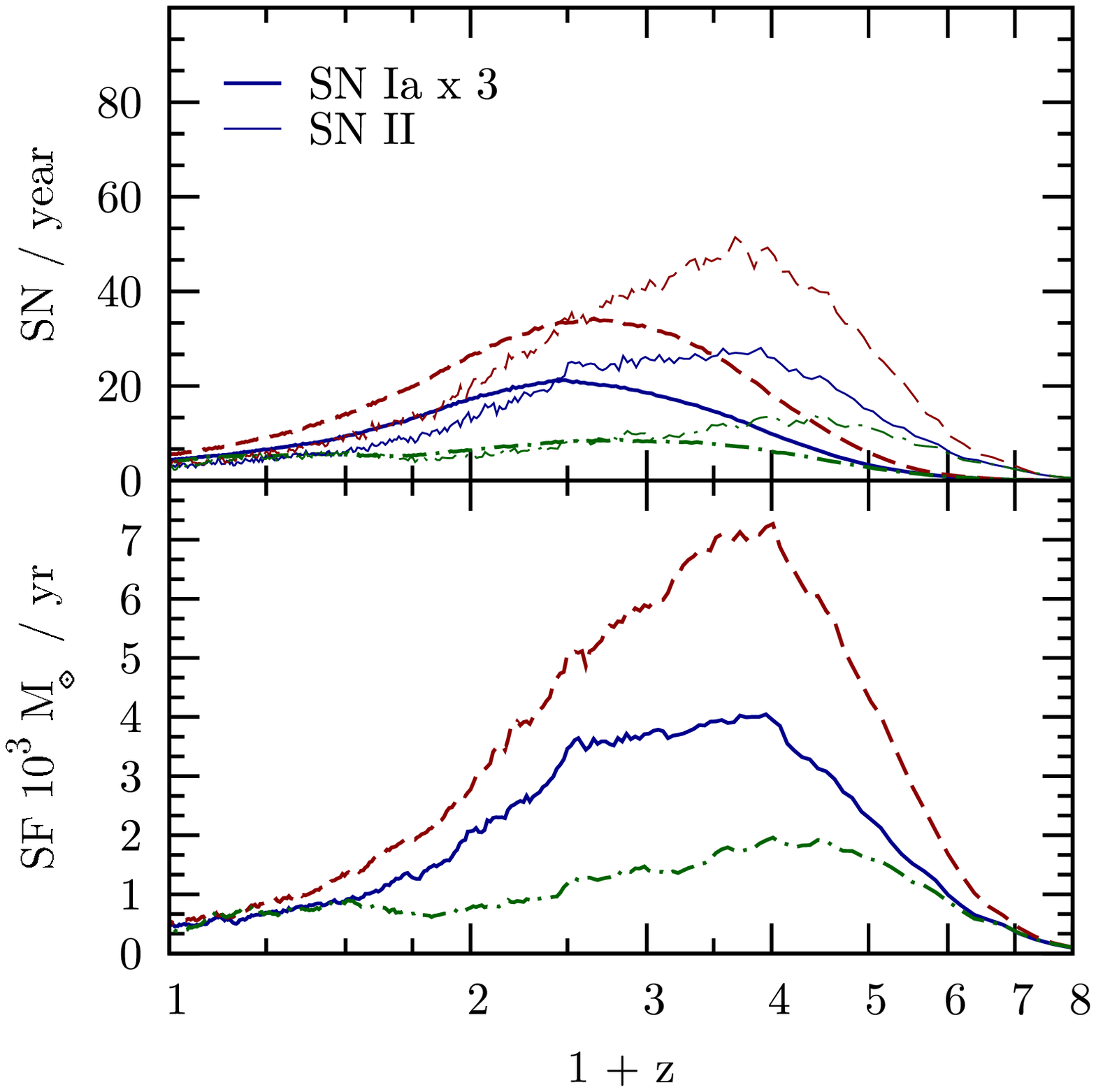}
\includegraphics[width=8.6cm]{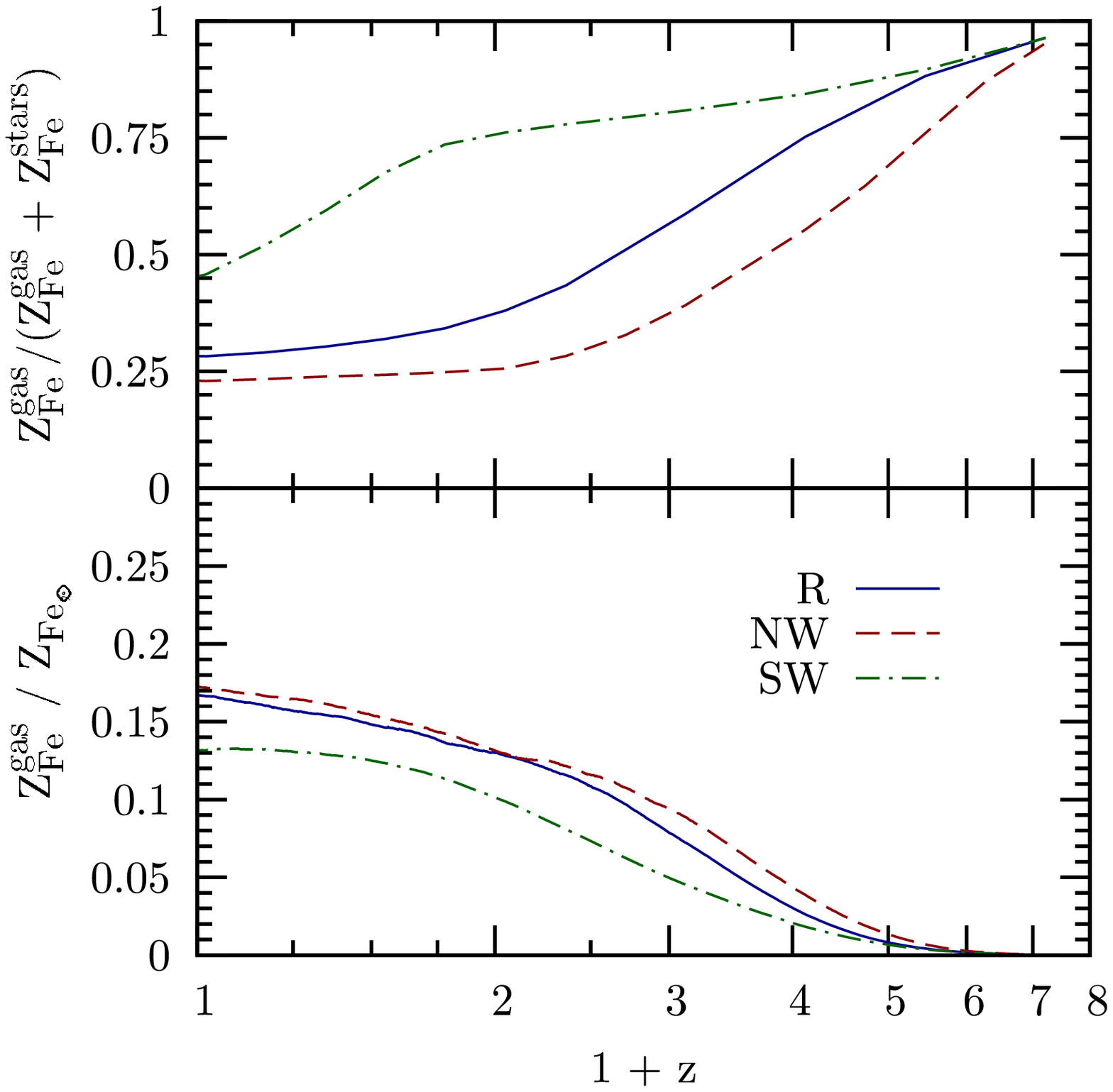} }}
\caption{The same as Figure \ref{fi:enr_ng}, but changing the feedback
  strength associated to galactic winds. The solid lines correspond to
  the reference run, while the short--dashed and the long--dashed
  lines are for the runs with strong winds (SW) and switching off
  winds (NW), respectively.}
\label{fi:enr_fdbk}
\end{figure*}

\begin{figure*}
\centerline{
\hbox{
\includegraphics[width=6cm]{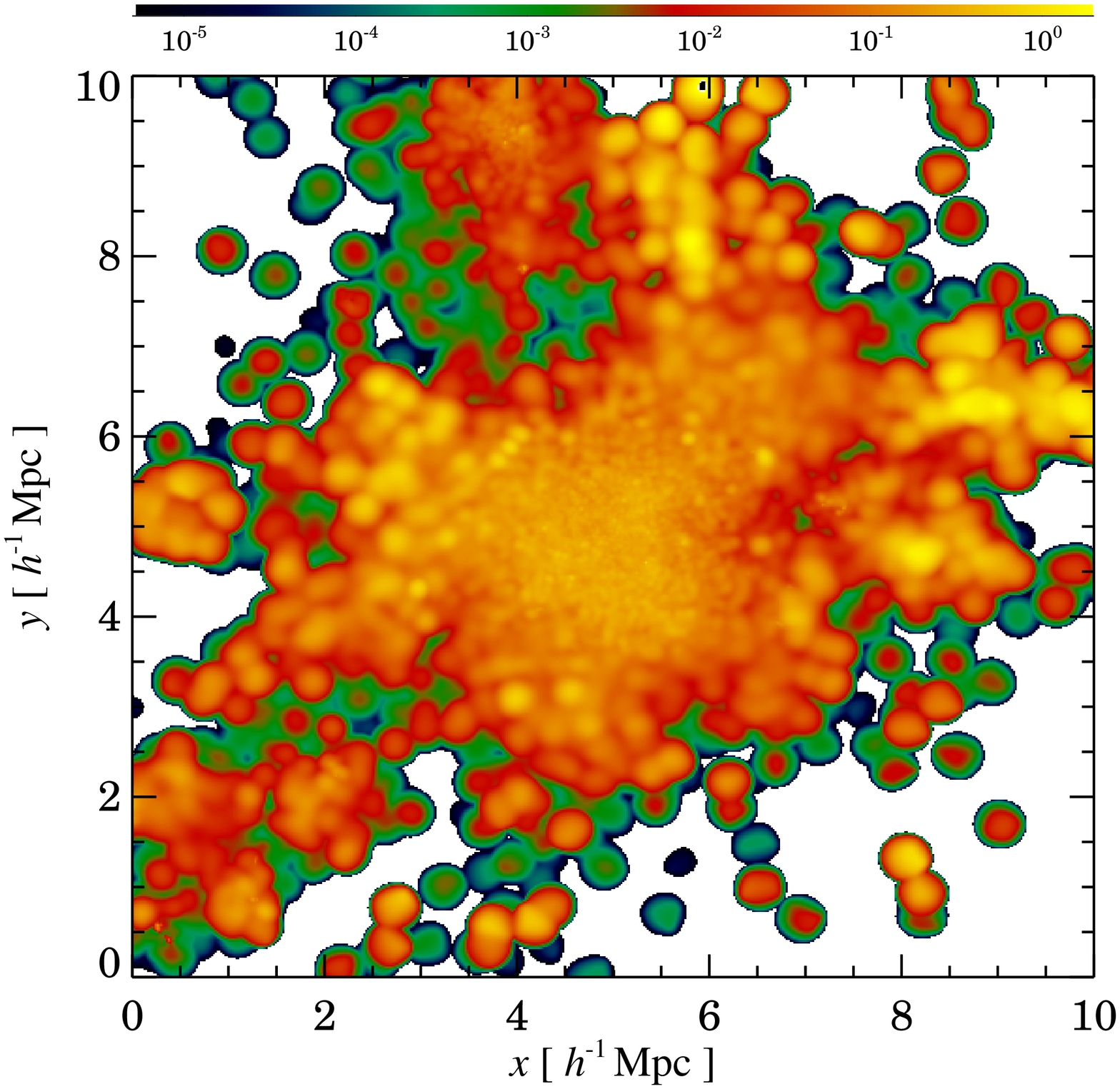}
\includegraphics[width=6cm]{Figs/Maps/map--z_0.00--Fe---5000x1000--512.R.eps} 
\includegraphics[width=6cm]{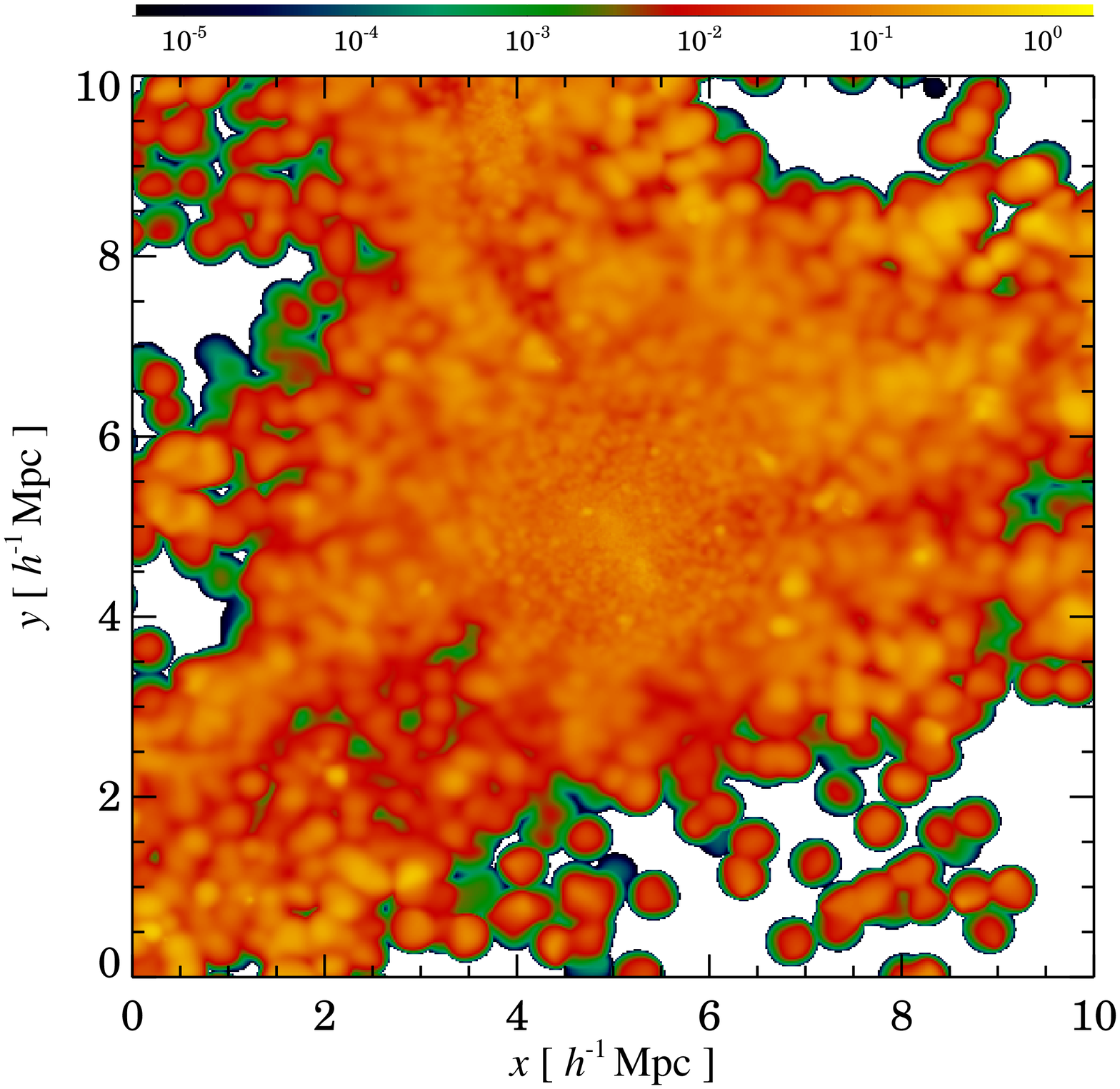} }
}

\caption{Maps of the Iron metallicity for the run with no winds (NW,
  left panel), for the reference run (R, central panel; $v_w=500\vel$)
  and for the run with strong winds (SW, right panel;
  $v_w=1000\vel$). Details of the maps are the same as in
  Fig.\ref{fi:maps_imf}.}
\label{fi:maps_feed}
\end{figure*}

\subsubsection{Changing the feedback strength}
Feedback from galactic winds has been originally implemented in the
{\small GADGET} code with the purpose of preventing overcooling
\citep{2003MNRAS.339..289S} in simulations of galaxy
formation. Besides producing a realistic star formation in
cosmological simulations \citep{2003MNRAS.339..312S}, galactic winds
are also expected to play a crucial role in the transport of metals
outside star--forming regions and, therefore, in the enrichment of the
diffuse gas. According to the implementation of galactic ejecta in the
{\small GADGET} code, gas particles to be uploaded in winds are chosen
among star forming particles which, by definition, are those more
metal rich. Recent numerical studies of the key role played by winds
in enriching diffuse cosmic baryons have been presented by
\cite{2006astro.ph..8268D}, who focused their analysis on the
high--redshift IGM, and by \cite{2006ApJ...650..560C}, who
concentrated on the study of the WHIM.

The left panel of Figure \ref{fi:glob_fdbk} clearly shows that
neglecting the effect of winds increases the star fraction in clusters
up to $\sim 35$ per cent. This value is possibly even an
underestimate, owing to the lack of numerical convergence of the star
fraction in the absence of winds \cite{2006MNRAS.367.1641B}. On the
other hand, increasing the wind speed to $1000\vel$ reduces the
fraction of cooled gas by almost a factor two. This reduction of the
star fraction inside the cluster virial region at $z=0$ is just the
consequence of the suppressed star formation due to the wind action
(see the left panel of Figure \ref{fi:enr_fdbk}). As shown in the
right panel of Figure \ref{fi:glob_fdbk}, an increasing wind
efficiency also provides a more efficient distribution of metals in
the gas. This is confirmed by the maps of Iron abundance shown in
Figure \ref{fi:maps_feed}. In the absence of winds, the pattern of the
Iron distribution is quite patchy, with lumps of high abundance
surrounding the star--forming regions. On the other hand, including
the effect of winds turns into a more diffused enrichment pattern, an
effect which increases with the winds' speed.

As shown in the right panel of Fig. \ref{fi:glob_fdbk}, this effect is
more pronounced for Oxygen, whose mass fraction in gas increases by a
factor of about 2.5, than for Iron, for which the same fraction
increases by about 50 per cent. As already discussed, Iron is more
efficiently distributed outside star--forming regions. This is due to
the fact that Iron is largely contributed by long--lived stars, which
explode when star particles possibly had the time to leave
star--forming regions. Therefore, these regions are expected to be
relatively more enriched in Oxygen. Since galactic winds are uploaded
with gas from star--forming particles, they are relatively more
enriched in Oxygen. As a consequence, increasing the feedback from
winds has the general effect of increasing the diffusion of metals, by
an amount which is larger for Oxygen than for Iron. This effect also
explains why the ICM metallicity is not suppressed in the same
proportion of star formation, when feedback from winds increases (see
left and central panel of Fig. \ref{fi:glob_fdbk}).

As shown in the right panel of Fig.\ref{fi:enr_fdbk}, the effect of
winds in unlocking metals from stars is more effective at high
redshift. This is quite expected, owing to the shallower potential
wells which are present at earlier times and make winds more effective
in transporting metals outside star--forming regions. This result
further illustrate the fundamental role that winds have played in
establishing the enrichment pattern of the high--redshift IGM
\citep{2006astro.ph..8268D}.

\begin{figure*}
\centerline{
\hbox{
\includegraphics[width=8.5cm]{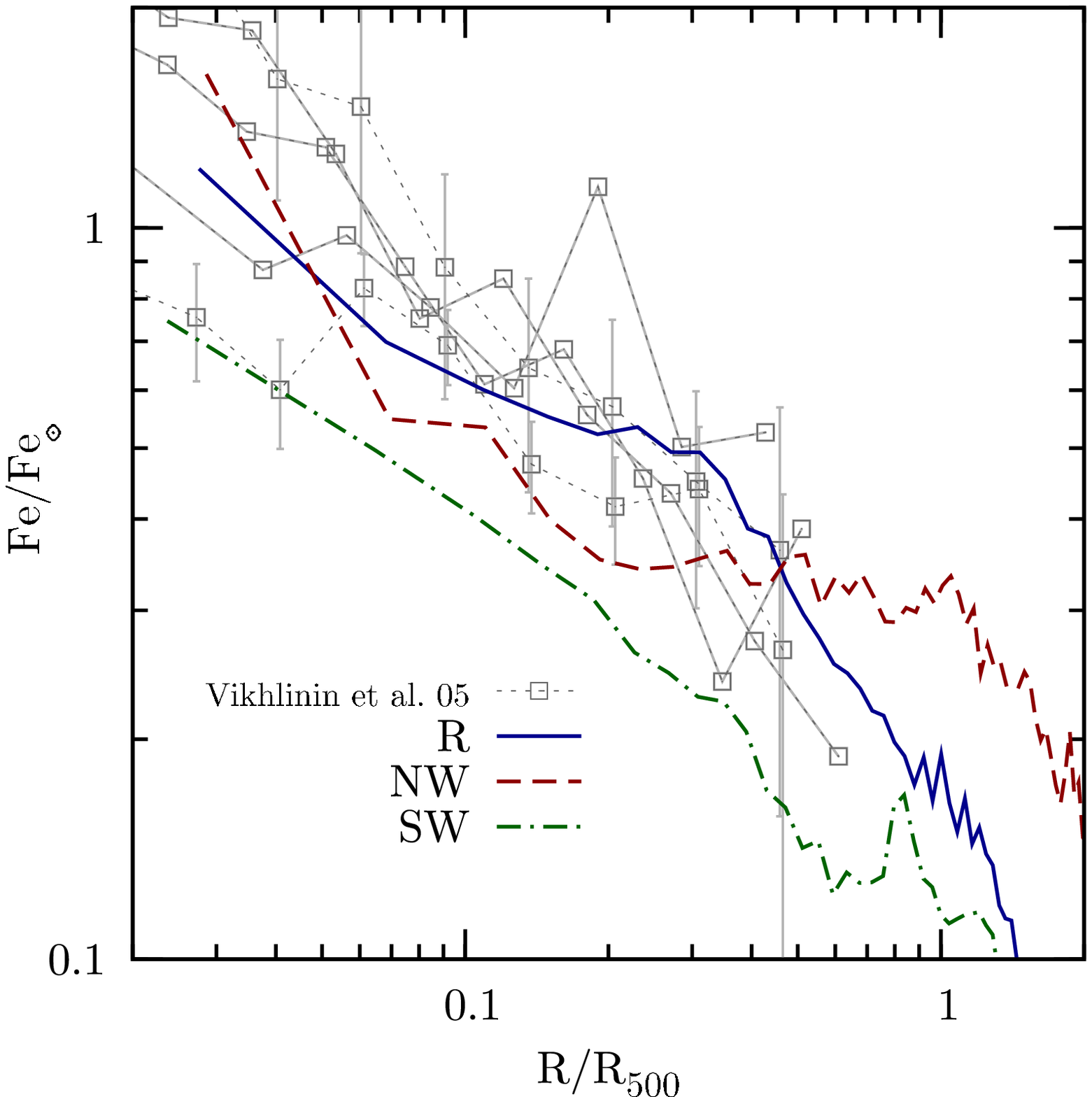} 
\includegraphics[width=8.8cm]{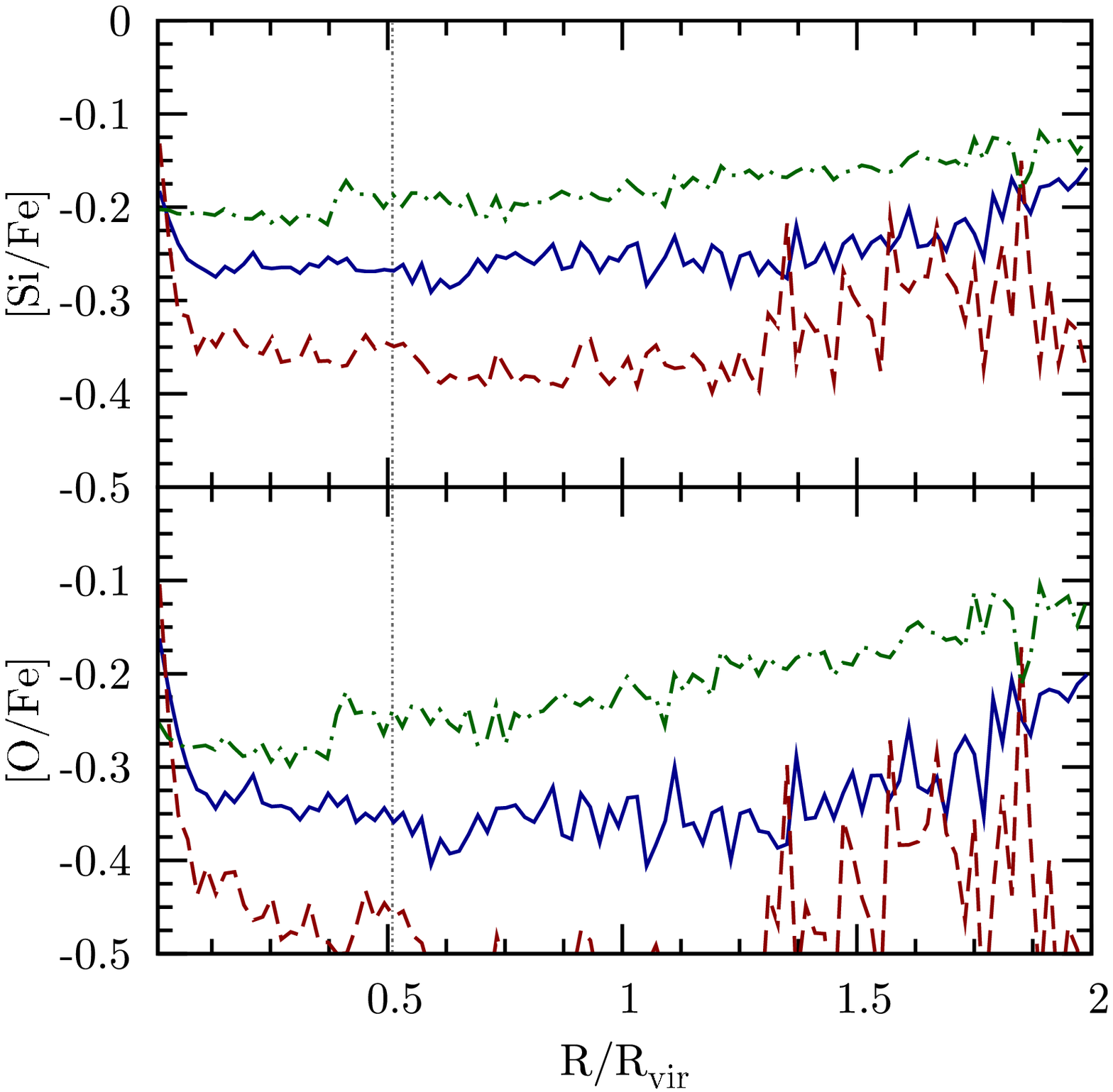}}} 
\caption{The effect of changing the feedback strength on the abundance
  profiles. The dot--dashed lines are for the run with strong winds
  (SW), while the dashes lines are for the run without winds (NW).
  Left panel: profiles of the mass--weighted Iron abundance. The data
  points are the same as in Figure \ref{fi:prof_ng}. Right panel: the
  relative abundance of Silicon (top) and Oxygen (bottom) with respect
  to Iron, for the same runs shown in the left panel.}
\label{fi:prof_fdbk}
\end{figure*}

The dependence of the abundance profiles on the feedback strength is
shown in Figure \ref{fi:prof_fdbk}. The effect of
increasing the wind speed from $500 \vel$ to $1000\vel$ is that of
suppressing the profile of Iron abundance, as a consequence of the
suppressed star formation, bringing it below the level indicated by
observational data. On the other hand, the effect of neglecting
galactic winds has a less direct interpretation. While for $R\magcir
0.3R_{500}$ the enrichment level is higher, due to the larger amount
of stars, it falls below the profile of the reference (R) run at
smaller radii, due to the reduced efficiency of metal transport. Only
in the innermost regions, the excess of star formation, taking place in
the absence of efficient feedback, causes a spike in the Iron
abundance. 

As for the profiles of the relative abundances of [O/Fe] and [Si/Fe]
(right panel of Fig.\ref{fi:prof_fdbk}), they confirm the different
effect that feedback has in determining the enrichment level for
different elements. Consistent with what shown in
Fig.\ref{fi:glob_fdbk}, Silicon and Oxygen becomes progressively
over--abundant with respect to Iron as the feedback strength
increases. In the innermost regions, the excess of low--redshift star
formation in the absence of efficient feedback manifests itself
through a sudden increase of the relative abundances. A comparison of
this figure with the right panel of Fig.\ref{fi:prof_imf} shows that
the effect of a stronger feedback is
similar to that induced by assuming a top--heavier IMF. This
highlights once more that using the observational determination of the
[$\alpha$/Fe] abundances to infer the IMF requires a good knowledge of
other gas--dynamical effects.

%\begin{figure}
%\centerline{
%\psfig{file=Figs/Enrich/z_0.00_mz.fdbk.eps,width=8.5cm}}
%\caption{The histograms showing the fraction of gas and
%  stars having a given level of metal enrichment, for different
%  feedback strengths. The solid lines correspond to
%  the reference run with $v_w=500\vel$, while the short--dashed and
%  the long--dashed lines are for $v_w=1000\vel$ and switching off
%  winds, respectively.}
%\label{fi:mz_phys}
%\end{figure}

% discussion and conclusions
%\input{sect.4}
\section{Discussion and conclusions}
We presented results from hydrodynamical simulations of galaxy
clusters, aimed at studying the metal enrichment of the Intra--Cluster
Medium (ICM). Our simulations were based on a version of the Tree-SPH
\gd code \citep{SP01.1,2005MNRAS.364.1105S}, which includes a detailed
treatment of the production of metals from different stellar
populations. In particular, we account for the delay times at which
different stellar populations release energy and metals from SNIa and
SNII explosions and stellar mass loss. The resulting chemo--dynamical
code is rather flexible since it allows one to change the assumed
stellar initial mass function (IMF), the life--time function and the
tables of stellar yields. Thanks to the efficient implementation of
the way in which the equations of the chemical evolution model are
solved, the corresponding computational cost is only of about 10 per
cent of the total cost of a simulation in a typical configuration.

The analyses presented in this paper have been carried out with the
twofold aim of assessing in detail the impact of numerical effects and
of changing the parameters defining the chemical evolution and
the feedback model on the
resulting pattern of ICM enrichment. For this reason, we have
concentrated our analysis only on two medium poor clusters, having
virial masses $M_{\rm vir}\simeq 2\times 10^{14}\msun$, which have
been simulated in a variety of conditions. For this reason, we have
preferred not to perform here any detailed comparison with
observational X--ray data. Such a comparison would require simulations
of a more extended set of clusters, so as to appreciate both the
cluster-to-cluster variation of the enrichment pattern and the
dependence on the cluster mass (Fabjan et al., in preparation).

The main results of our analysis can be summarized as follows.
\begin{description}
\item[(a)] Gas--dynamical effects, related to gas mixing, galactic
  winds and ram--pressure stripping, play an important role in
  determining the distribution and relative abundances of different
  chemical species. For instance, we find that the products of SNII
  are generally more concentrated around star--forming regions than
  those of low--mass stars, due to the different time--scales over
  which stars of different mass release metals. This emphasizes the
  relevance of properly including in the simulations a correct
  description of the lifetimes of different stellar populations.
\item[(b)] A comparison with observed metallicity profiles obtained by
  \cite{2005ApJ...628..655V} from Chandra data shows a reasonable and
  encouraging agreement in both shape and normalization, when using a
  \cite{1955ApJ...121..161S} IMF, within $\simeq 0.5R_{\rm vir}$. The
  [O/Fe] relative abundance is predicted to be undersolar, $\simeq
  -0.35$ over the cluster virial region, with a mild increase at
  larger radii and with a sudden increase in the core region, due to
  an excess of recent star formation.   
\item[(c)] For each gas particle we estimated the
  metallicity--weighted cosmic time of enrichment. We find that this
  typical age of enrichment of the hot gas corresponds to redshift
  $z\simeq 0.5$, over the whole cluster virial region. Only in the
  innermost region, which is heavily affected by the presence of the
  Brightest Cluster Galaxy (BCG), the enrichment took place at a
  significantly more recent epoch, due to an excess of low--redshift
  star formation.
\item[(d)] As expected, the IMF has a strong impact on the enrichment
  level of both the ICM and the stellar populations. Assuming a
  top--heavy IMF \citep{1987A&A...173...23A}, instead of the
  \cite{1955ApJ...121..161S} one, it turns into an increase of the
  Iron abundance and of the [O/Fe] relative abundance. Within the
  cluster virial regions, the Iron metallicity of the ICM increases by
  almost a factor of two, while Oxygen increases by about a factor of
  three. Using the IMF proposed by \cite{2001MNRAS.322..231K}, it
  turns into a general decrease of [O/Fe] with respect to the
  prediction of a Salpeter IMF, owing to the relatively smaller number
  of massive stars.
\item[(e)] Increasing the velocity of galactic ejecta from $500\vel$
  to $1000\vel$ provides a strong suppression of the star formation
  history and of the corresponding level of enrichment. The fraction
  of baryonic mass in stars within $R_{\rm vir}$ drops by about a
  factor of two, while the level of Iron enrichment drops only by
  about 40 per cent. This is due to the increasing efficiency of
  stronger winds to unlock metals from star--forming regions.
\end{description}

In order to judge the stability of our results against numerical
effects, we have considered several different prescriptions to
distribute metals around stars and the effect of progressively
increasing resolution.  For instance, we find that in the central
regions, $R\mincir 0.3R_{500}$, the pattern of ICM enrichment is quite
insensitive to the shape of the weighting kernel, to the number of
neighbours over which metals are distributed. The only significant
trend found is for the profiles of the Iron abundance to become higher
at larger radii as the number of neighbors used for the spreading
increases. As a general result, we can then conclude that our results
are not strongly affected by the details with which metals are
distributed around star--forming regions.  Furthermore, we find no
obvious trend in the Iron abundance profile with resolution within
$0.3R_{500}$. However, we do find a significant and systematic
increase of the enrichment level with resolution in the cluster
outskirts: at $R\magcir R_{\rm vir}$ the Iron abundance in the highest
resolution run is about 60 per cent higher that in the low resolution
run. These differences are not due to a lack of convergence of
resolution of the total amount of stars found at $z=0$.  Instead, they
are mainly due to the better--resolved star formation at high
redshift, which provides a more uniform enrichment of the
inter--galactic medium (IGM). This calls for the need of significantly
increasing the resolution to have numerically converged results on the
enrichment pattern in the cluster outskirts and in the Warm-Hot
Intergalactic Medium (WHIM).

The results presented in this paper and those from previous
chemo--dynamical models of the ICM \citep[e.g., see also
][]{2002MNRAS.330..821L,2003MNRAS.339.1117V,2005MNRAS.357..478S}
demonstrate that sophisticated models of chemical evolution can be
efficiently included in advanced codes for cosmological simulations of
galaxy clusters. This opens the possibility of accurately describing
the history of metal production through cosmic epochs, in a variety of
environments. While we have focused our analysis on the low--redshift
enrichment of the ICM, a number of other studies can be foreseen, from
the enrichment of the IGM at high redshift, $z\magcir 2$, to the metal
abundances in both elliptical and disk galaxies
\citep[e.g.,][]{2003MNRAS.340..908K,2004MNRAS.347..740K,2006MNRAS.371.1125S}.

Even restricting to the study of the ICM, a number of directions of
further developments and improvements with respect to the analysis
presented here can be devised.

As for the comparison with observational data, 
%the first step is
%that of simulating a statistically significant set of clusters,
%spanning a wide enough range of masses (Fabjan et al., in
%preparation), so as to probe the cluster by cluster variation of the
%enrichment pattern and any possible dependence of the ICM enrichment
%level on the cluster virial temperature \citep[e.g.,
%][]{2005ApJ...620..680B}, connecting them to the different dynamical
%histories of different objects. Another 
a fundamental step is represented by understanding in detail possible
observational biases in the measurement of the ICM metallicity. In the
analysis of observational data, the ICM metallicity is estimated by
fitting the X--ray spectrum to a plasma model. On the other hand, in
the analysis of simulations, metallicity is usually estimated by
either mass--weighting or emission--weighting the metallicity carried
by each gas particle. The best way of approaching this problem is by
performing mock observations of simulated clusters, which reproduce as
close as possible the observational setup (i.e., instrument response
and PSF, instrumental background, etc.; Rasia et al., in preparation),
and analyse them exactly in the same way as observational data.
%
%Indeed, software
%tools, which produce such mock observations, have been already used to
%analyze hydrodynamical simulations of galaxy clusters, with the aim of
%calibrating possible biases in the estimate of the temperature and,
%therefore, of the total collapsed mass \citep[e.g.,
%][]{2006MNRAS.369.2013R,2006ApJ...650..128K}. In fact, these analyses
%have shown that a thermally complex ICM leads to differences between
%the spectroscopic and the emission--weighted determination of the
%temperature of simulated clusters \citep[e.g.,
%][]{2004MNRAS.354...10M,2006ApJ...640..710V}. Therefore, it would come
%with no much surprise that non--negligible observational biases are
%also present in the determination of the ICM metallicity.
The preliminary comparison with observational data, presented in this
paper, shows an encouraging agreement with the profiles of the Iron
abundance as obtained from Chandra observations of a set of nearby
relaxed clusters \citep{2005ApJ...628..655V}. As a word of
caution, we point out that for a numerical model to be fully
successful, it should reproduce at the same time the properties of the
ICM and those of the cluster galaxy population. 
%Indeed, if simulations
%produce galaxies whose properties (e.g., luminosity function,
%color--magnitude relation) are grossly at variance with observations,
%there should be no surprise that it also produces the wrong pattern of
%ICM enrichment. 
However, a well known problem of simulations based on stellar
feedback, like those presented here, is that they produce central
cluster galaxies which are much bluer and more star--forming than observed
\citep[e.g., ][]{2006MNRAS.373..397S}. 
%This excess of
%star formation at small radii is also reflected in a central spike of
%the [O/Fe] relative abundance, which is not seen in observational data
%\citep{2002ApJ...572..160G,2002A&A...381...21F}. 
In this respect, AGN feedback is generally considered as the most
credible candidate to quench star formation at low redshift, thereby
accounting for the observed properties of both the brightest cluster
galaxies and of the ``cool cores''. Besides regulating star formation,
AGN feedback should also play an important role in distributing metals
from the core regions, through the generation of buoyant bubbles
created by collimated jets, which shock on the dense ambient gas
\citep[e.g., ][and references
therein]{2004MNRAS.355..995D,2006MNRAS.366..397S,2007MNRAS.375...15R}.

From the observational side, the sensitivity of the present generation
of X--ray satellites are allowing us to measure the distribution of
metals at most out to about $R_{500}$ for nearby, $z\sim 0.1$,
clusters, while global measurement of the ICM metallicity are possible
for distant cluster, out to $z\sim 1$ \citep[e.g.,
][]{2007A&A...462..429B}. X--ray telescopes of the next generation,
thanks to a much larger collecting area, improved spectral resolution
and lower instrumental background, will open a new window on our
knowledge of the chemo--dynamical history of the cosmic baryons.
%Doppler broadening and shifts of metal lines will
%allow us to measure for the first time gas motions in nearby clusters,
%while tracing the metallicity distribution out to their outskirts.  
Deep pointings of distant clusters will provide the signature of the
metal enrichment in the just--assembling ICM, out to $z\sim 2$, thus
bridging the gap between observations of the metal--enrichment of the
low--$z$ ICM and of the high--$z$ IGM. There is little doubt that
numerical simulations, like those presented here, will provide the
ideal tool for the interpretation of these observations within the
framework of cosmological models of structure formation.

\section*{Acknowledgments}
We are greatly indebted to Volker Springel for having provided us with
the non--public version of GADGET-2, and for his continuous advices on
the code whereabouts. We acknowledge useful discussions with Fabrizio
Brighenti, Francesco Calura, Cristina Chiappini, Sofia Cora, Claudio
Dalla Vecchia, Stefano Ettori, Giuseppe Murante, Laura Portinari,
Elena Rasia, Simone Recchi, Joop Schaye and Paolo Tozzi. We thank
Alexey Viklinin for having provided us with the observed metallicity
profiles. The simulations have been realized using the
super--computing facilities at the ``Centro Interuniversitario del
Nord-Est per il Calcolo Elettronico'' (CINECA, Bologna), with CPU time
assigned thanks to an INAF--CINECA grant and to an agreement between
CINECA and the University of Trieste. This work has been partially
supported by the INFN PD-51 grant.

\bibliographystyle{mn2e}
\bibliography{../Biblio/master}

\end{document}